\documentclass[aps,prb,amsfonts,amsmath,amssymb,nofootinbib,twocolumn,superscriptaddress]{revtex4-2}

\usepackage{braket}
\usepackage{graphics}
\usepackage{hyperref}
\usepackage{epsfig}
\usepackage{color}
\usepackage{verbatim}

\usepackage{ulem}
\usepackage{multirow}
\usepackage{bm}

\usepackage{pifont}
\usepackage{amssymb}
\usepackage{amsmath}

\begin{document}

\title{Coherent Phonon Pairs and Rotational Symmetry Breaking of Charge Density Wave Order in the Kagome Superconductor CsV$_3$Sb$_5$}

\author{Qinwen Deng}
\affiliation{Department of Physics and Astronomy, University of Pennsylvania, Philadelphia, Pennsylvania 19104, U.S.A.}
\author{Hengxin Tan}
\affiliation{Department of Condensed Matter Physics, Weizmann Institute of Science, Rehovot, Israel}
\author{Brenden R. Ortiz}
\affiliation{Materials Department, University of California Santa Barbara, Santa Barbara, California
93106, U.S.A.}
\affiliation{Materials Science and Technology Division, Oak Ridge National Laboratory, Oak Ridge, Tennessee 37831, U.S.A.}
\author{Andrea Capa Salinas}
\affiliation{Materials Department, University of California Santa Barbara, Santa Barbara, California
93106, U.S.A.}
\author{Stephen D. Wilson}
\affiliation{Materials Department, University of California Santa Barbara, Santa Barbara, California
93106, U.S.A.}
\author{Binghai Yan}
\affiliation{Department of Condensed Matter Physics, Weizmann Institute of Science, Rehovot, Israel}
\affiliation{Department of Physics, The Pennsylvania State University, University Park, Pennsylvania 16802, U.S.A.}
\author{Liang Wu}
\email{liangwu@sas.upenn.edu}
\affiliation{Department of Physics and Astronomy, University of Pennsylvania, Philadelphia, Pennsylvania 19104, U.S.A.}

\date{\today}

\begin{abstract}

In this work, we perform ultrafast time-resolved reflectivity measurements to study the symmetry breaking in the charge-density wave (CDW) phase of CsV$_3$Sb$_5$. By extracting the coherent phonon spectrum in the CDW phase of CsV$_3$Sb$_5$, we discover close phonon pairs near 1.3 THz and 3.1 THz, as well as a new mode at 1.84 THz. The 1.3 THz phonon pair and the 1.84 THz mode are observed up to the CDW transition temperature. Combining density-functional theory calculations, we point out these phonon pairs arise from the coexistence of Star-of-David and inverse Star-of-David distortions combined with six-fold rotational symmetry breaking. An anisotropy in the magnitude of transient reflectivity change is also revealed at the onset of CDW order. Our results thus indicate broken six-fold rotational symmetry in the charge-density wave state of CsV$_3$Sb$_5$, along with the absence of nematic fluctuation above T$_{\text{CDW}}$. Meanwhile, the measured coherent phonon spectrum in the CDW phase of CsV$_3$Sb$_{5-\text{x}}$Sn$_\text{x}$ with x = 0.03-0.04 matches with staggered inverse Star-of-David with interlayer $\pi$ phase shift. This CDW structure contrasts with undoped CsV$_3$Sb$_5$ and explains the evolution from phonon pair to a single mode at 1.3 THz by x = 0.03-0.04 Sn-doping. 
\end{abstract}

\maketitle 

The Kagome lattice, hosting a corner-sharing triangle network, has been the focus of extensive research for decades due to its interplay of inherent geometrical frustrations, nontrivial topology and unconventional correlation effects. As a recent example, the newly discovered vanadium-based Kagome superconductors AV$_3$Sb$_5$ (A = K, Rb or Cs) have attracted tremendous research interest due to their unconventional electronic landscape, including charge density wave (CDW) below T$_{\text{CDW}}$ $\approx$ 78-102 K and superconductivity with T$_c$ $\approx$ 0.9-2.5 K\cite{ortiz2019new, ortiz2020cs, ortiz2021superconductivity, yin2021superconductivity}. More intriguingly, additional exotic electronic instabilities emerge inside the CDW phase, including signatures of a possible time-reversal symmetry breaking loop current state\cite{mielke2022time, jiang2021unconventional, xing2024optical, guo2022switchable, xu2022three, park2021electronic, feng2021chiral, denner2021analysis, lin2021complex, guo2024correlated, christensen2022loop, tazai2024drastic}, electronic nematicity\cite{nie2022charge, li2023unidirectional, xiang2021twofold, wulferding2022emergent, asaba2024evidence, liu2024absence}, and pair density wave within the superconducting state\cite{chen2021roton}. Therefore, it is of paramount significance to clarify the structure and symmetry of the CDW order in this family of compounds, in order to understand the interaction between this plethora of charge orders and superconductivity. 

\begin{figure*}
    \centering
    \includegraphics[width=0.9\textwidth]{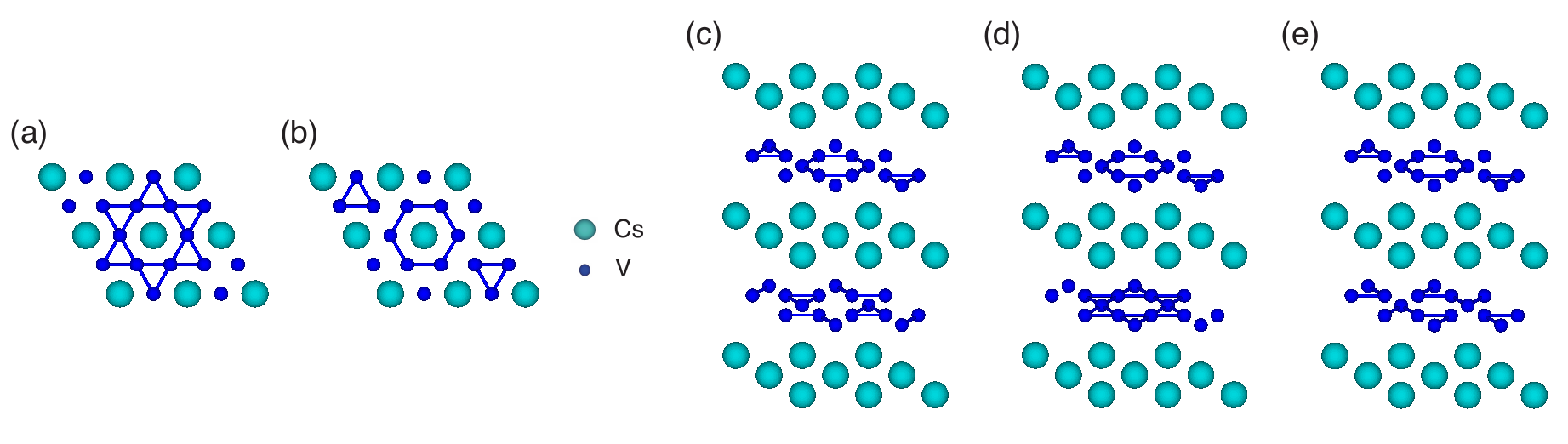}
    \caption{\textbf{CDW distortions in CsV$_3$Sb$_5$.} (a) 2 $\times$ 2 $\times$ 1 SD. (b) 2 $\times$ 2 $\times$ 1 ISD. (c) ISD + ISD with interlayer $\pi$-phase shift, (d) SD + ISD without interlayer phase shift, and (e) SD + ISD with interlayer $\pi$-phase shift. The Cs atoms are shown in cyan and V atoms are shown in blue. The Sb atoms are not shown for simplicity. Note that only the structure in (a, b, d) keeps the $D_{6h}$ symmetry. The lines connecting V atoms indicate shorter V-V bonds. }
    \label{fig1}
\end{figure*}

\begin{figure*}
    \centering
    \includegraphics[width=\textwidth]{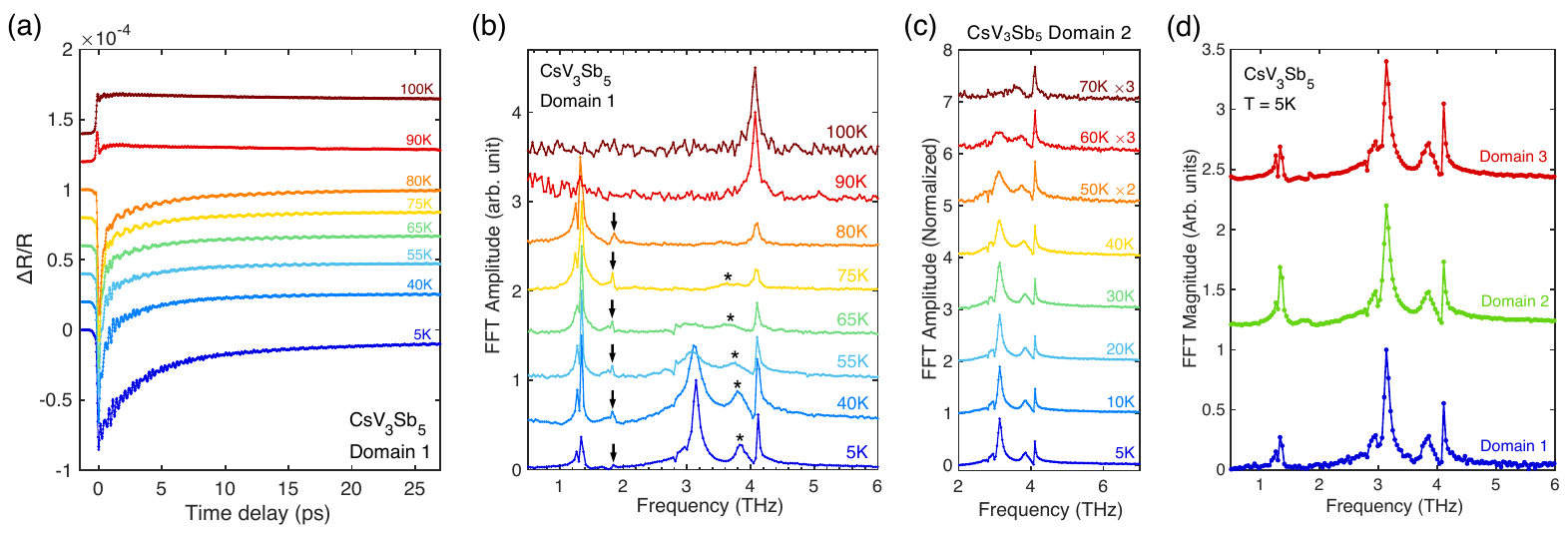}
    \caption{\textbf{Temperature-dependent coherent phonon spectroscopy in Kagome CsV$_3$Sb$_5$.}  (a) Time-resolved reflectivity curves at different temperatures above and below T$_{\text{CDW}}$ in a single birefringence domain. (b) Amplitudes of Fourier transforms of coherent phonon oscillations in (a) after subtracting a double-exponential background. The arrow marks the temperature evolution of the 1.84 THz mode, and the asterisk marks the temperature evolution of the 3.86 THz mode. (c) Same as (b) but measured in another birefringence domain and zoomed in around 4 THz to better show the two close modes near 3.1 THz. All datasets are normalized to its maximum amplitude. (d) Coherent phonon spectrum measured on all 3 different birefringence domains at 5 K. Curves are offset for clarity in all plots. 
 }
    \label{fig2}
\end{figure*}

An abundance of initial experiments reported a 3D CDW state in AV$_3$Sb$_5$\cite{liang2021three, li2021observation, ortiz2021fermi}. Among all members in this family of materials, CsV$_3$Sb$_5$ possesses the largest variety of reported 3D CDW phases with different interlayer stacking orders. These 3D CDW orders can be constructed from unstable phonon modes at M and L points in the momentum space which give rise to various 2 $\times$ 2 $\times$ 2 CDW patterns\cite{christensen2021theory, ritz2023impact, tan2021charge, ptok2022dynamical, deng2025revealing}. The in-plane 2 $\times$ 2 distortions on the Kagome lattice can be characterized by either a Star-of-David (SD) or inverse Star-of-David (ISD) pattern \cite{tan2021charge, christensen2021theory, park2021electronic} (Fig. \ref{fig1}a, b), with a well-defined phasing of 0 or $\pi$ between neighboring Kagome layers. The relative energy of different CDW structures is very close, and consensus has not been made on the precise ground state configuration despite extensive research. Different measurements mainly evidenced either the LLL phase with alternative SD + ISD without interlayer $\pi$ phase shift \cite{kang2022charge, subires2023order, hu2022coexistence, li2022coexistence, azoury2023direct} (Fig. \ref{fig1}d), in which the six-fold rotational symmetry $C_6$ is preserved, or the MLL phase with ISD + ISD with interlayer $\pi$ phase shift\cite{miao2021geometry, ratcliff2021coherent, xiao2023coexistence, wang2023structure, jin2024pi, xu2022three} (Fig. \ref{fig1}c), in which the six-fold rotational symmetry is broken down to two-fold. A further 2 $\times$ 2 $\times$ 4 supercell was also reported to coexist and compete with the 2 $\times$ 2 $\times$ 2 structures \cite{kautzsch2023structural, xiao2023coexistence, alkorta2025symmetry} as observed by X-ray diffraction, with contradicting results whether this 2 $\times$ 2 $\times$ 4 phase only has $C_2$\cite{ortiz2021fermi, stahl2022temperature, kautzsch2023structural, xiao2023coexistence}. This is possibly due to the phase coexistence, locally 
of multiple CDW ordered states as reported by recent dark-field X-ray results\cite{plumb2024phase}. Apart from whether the system breaks six-fold rotational symmetry, another major debate is at what temperature the six-fold symmetry breaks. Various experimental probes produced conflicting results regarding the onset temperature of rotational symmetry breaking, either at T$_{\text{CDW}}$\cite{xu2022three, kautzsch2023structural, wu2022simultaneous} or much lower than T$_{\text{CDW}}$\cite{nie2022charge, guo2022switchable, xiang2021twofold, zhao2021cascade, li2023unidirectional} via a distinct electronic nematic state. Therefore, whether the two-fold symmetry is directly related to the CDW has not yet been made clear. Another important question is whether fluctuations of these exotic electronic states emerge above T$_{\text{CDW}}$ in CsV$_3$Sb$_5$. Strong CDW fluctuations and nematic states have been reported above T$_{\text{CDW}}$\cite{chen2022charge, asaba2024evidence, zhong2024unveiling}, although recent strain dependent measurements showed no fluctuating vestigial nematicity\cite{liu2024absence}.  



In this work, we use ultrafast time-resolved reflectivity (TR-reflectivity) experiments to study the structural symmetry of the CDW phase of CsV$_3$Sb$_5$. The derived coherent phonon spectrum reveals close phonon pairs at 1.3 and 3.1 THz and a newly discovered mode at 1.84 THz. The 1.3 THz phonon pair and the 1.84 THz mode persist up to T$_{\text{CDW}}$. By comparing the calculated phonon frequencies using density functional theory (DFT), we demonstrate the close phonon pairs originate from the coexistence of SD and ISD pattern combined with six-fold rotational symmetry breaking. An anisotropy of the transient reflectivity change ($\Delta$R/R), defined as non-equal $\Delta$R/R values between two orthogonal polarization directions, also occurs upon entering CDW state. Our results thus unequivocally corroborate six-fold rotational symmetry breaking in the CDW phase of CsV$_3$Sb$_5$. Meanwhile, we pinpoint the CDW structure of CsV$_3$Sb$_{5-\text{x}}$Sn$_\text{x}$ with x = 0.03-0.04 to ISD + ISD with interlayer $\pi$-phase shift, which reduces the 1.3 THz phonon pair to a single phonon mode.

CsV$_3$Sb$_5$ single crystals are synthesized using the self-flux method\cite{ortiz2019new, ortiz2021superconductivity}. We perform TR-reflectivity measurements on freshly cleaved (001) surface of CsV$_3$Sb$_5$ single crystals. For all pump-probe experiments, pump wavelength is at 1560 nm and probe wavelength is at 780 nm, with repetition rate of 80 MHz and pulse duration of 100 fs. To reduce sample heating, both beams have a fluence less than 10 $\mu$J/cm$^2$. Both beams are at normal incidence and focused by an objective lens to achieve a spot size of $\sim$10 $\mu$m in diameter. In our previous experiments\cite{xu2022three}, we pinpointed the three-state birefringence domains in CsV$_3$Sb$_5$ below T$_{\text{CDW}}$, with the domain size being $\sim$100 $\mu$m. We can thus probe the transient reflectivity change in a single birefringence domain. Here, we use an xyz-stage to carefully adjust the light spot position onto a selected birefringence domain. The pump beam intensity is modulated at a frequency of 84 kHz using a photo-elastic modulator. 

The coherent phonon generation via the pump pulse modulates the refractive index by the ion motion\cite{shen1984principles, shen1965theory, giordmaine1966light} which causes transient reflectivity changes. Fig. \ref{fig2}a shows the transient reflectivity change ($\Delta$R/R) versus the pump-probe time delay measured inside a single birefringence domain of CsV$_3$Sb$_5$ at different temperatures. A single phonon oscillation
can be observed above the CDW transition temperature T$_{\text{CDW}}$ $\sim$90 K. In contrast, the sign of the transient reflectivity traces flips below T$_{\text{CDW}}$, with the emergence of multiple phonon modes, as can be observed from the complex oscillation pattern. To better understand the evolution of coherent phonon modes as a function of temperature in the CDW phase, a double exponential decay background ($A_0 + A_1\exp{(-t/\tau_1)} + A_2\exp{(-t/\tau_2)}$) is fitted to all time traces and then subtracted to show the oscillation parts, which are then Fourier transformed to reveal the coherent phonon modes (Fig. \ref{fig2}b). Above T$_{\text{CDW}}$, only one mode at 4.1 THz exists. This 4.1 THz mode persists at all temperatures and should be assigned as a main lattice mode. Indeed, it matches with a 
4.10 THz $A_{1g}$ mode in the pristine phase of CsV$_3$Sb$_5$ (Table \ref{table-1}). In contrast, at T = 5 K, another two intense modes centered at 1.3 and 3.1 THz are conspicuously present, similar to previous pump-probe reflectivity studies\cite{ratcliff2021coherent, PhysRevB.104.165110}. As temperature increases, the 1.3 THz mode exhibits an increase in amplitude but an abrupt disappearance at T $\approx$ 90 K, matching with reported T$_{\text{CDW}}$ = 94 K\cite{uykur2021low, ortiz2020cs}. We note modest local laser heating may cause a slight decrease of the measured T$_{\text{CDW}}$. Little frequency softening is seen for the 1.3 THz mode when increasing the temperature, more consistent with a zone-folded phonon mode by CDW\cite{joshi2019short}. This explains its disappearance in the phonon spectrum above T$_{\text{CDW}}$. In contrast, the amplitude of 3.1 THz mode decreases during warm-up and vanishes at $\sim$65 K (Fig. \ref{fig2}b, c). Same behavior is also captured in previous calculations\cite{alkorta2025symmetry}. Previous Raman studies \cite{liu2022observation, he2024anharmonic} showed this 3.1 THz mode exhibits significant weakening and broadening upon warming toward T$_{\text{CDW}}$, becoming overdamped between 60 and 90 K with the linewidth as large as 50 cm$^{-1}$ right below T$_{\text{CDW}}$. This may explain why this 3.1 THz mode vanishes in our detection above $\sim$65 K. Apart from these dominant modes, we also find a 3.86 THz mode and a weaker 1.84 THz mode that is newly-discovered by time-resolved reflectivity (Fig. \ref{fig2}b). Similar to the 1.3 THz mode, the 1.84 THz mode also shows little frequency change and persists up to T$_{\text{CDW}}$. We do not observe other modes up to 15 THz.  

\begin{figure}
    \centering
    \includegraphics[width=8.6cm]{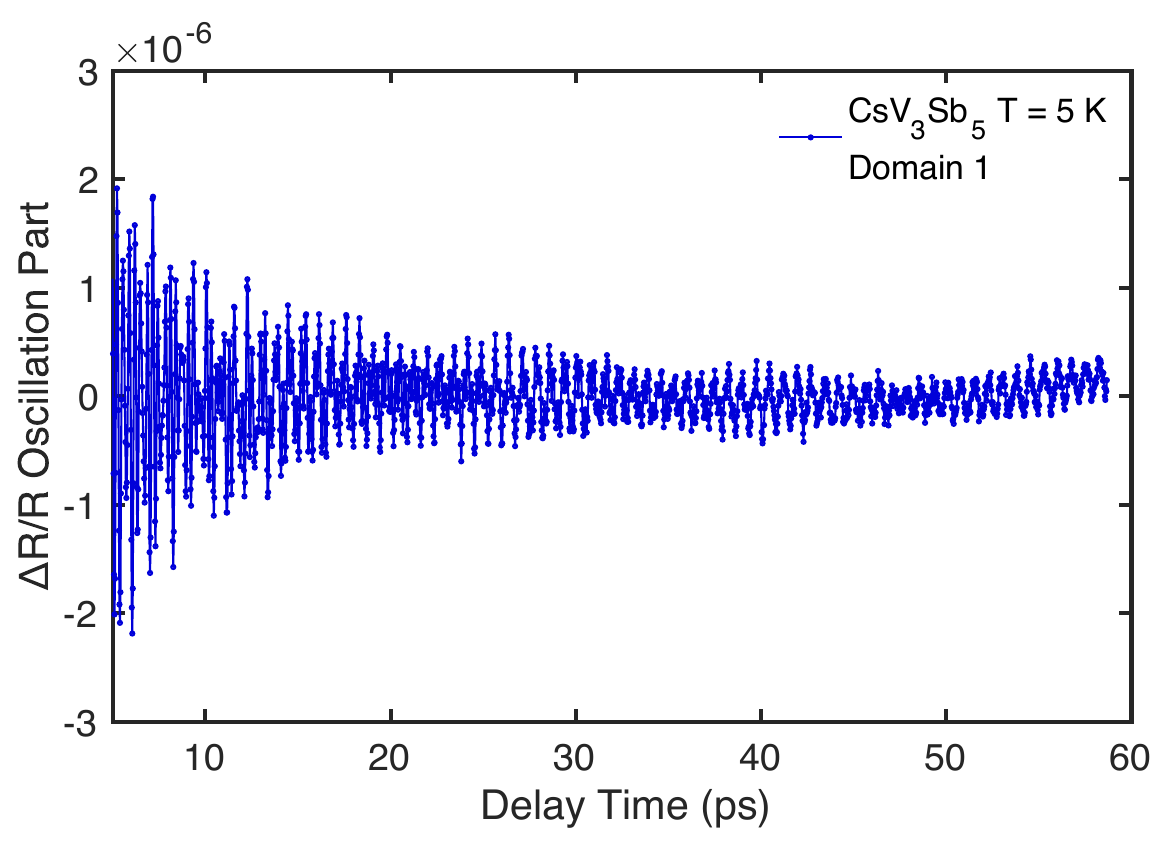}
    \caption{Oscillation part of the TR-reflectivity curve corresponding to the phonon spectrum of domain 1 in Fig. \ref{fig2}. Clear beatings can be observed in the oscillation pattern which corroborates the dual modes near 1.3 THz. The data is taken at T = 5 K. }
    \label{fig3}
\end{figure}

\begin{figure}
    \centering
    \includegraphics[width=8.5cm]{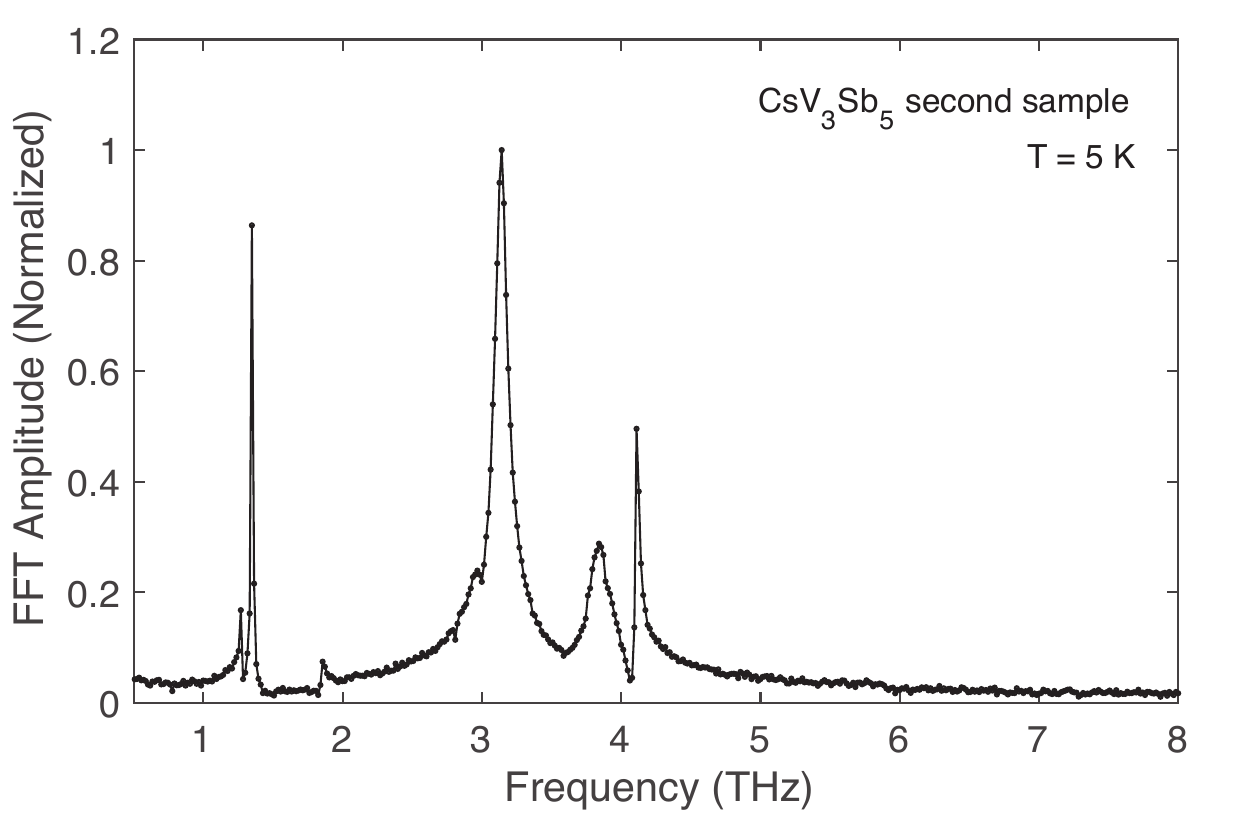}
    \caption{Coherent phonon spectrum measured by TR-reflectivity on a second CsV$_3$Sb$_5$ sample at T = 5 K. It also shows the dual modes at 1.3 THz and 3.1 THz, as well as the newly discovered weaker mode near 1.8 THz. }
    \label{fig4}
\end{figure}

\begin{figure}[]
    \centering
    \includegraphics[width=8.6cm]{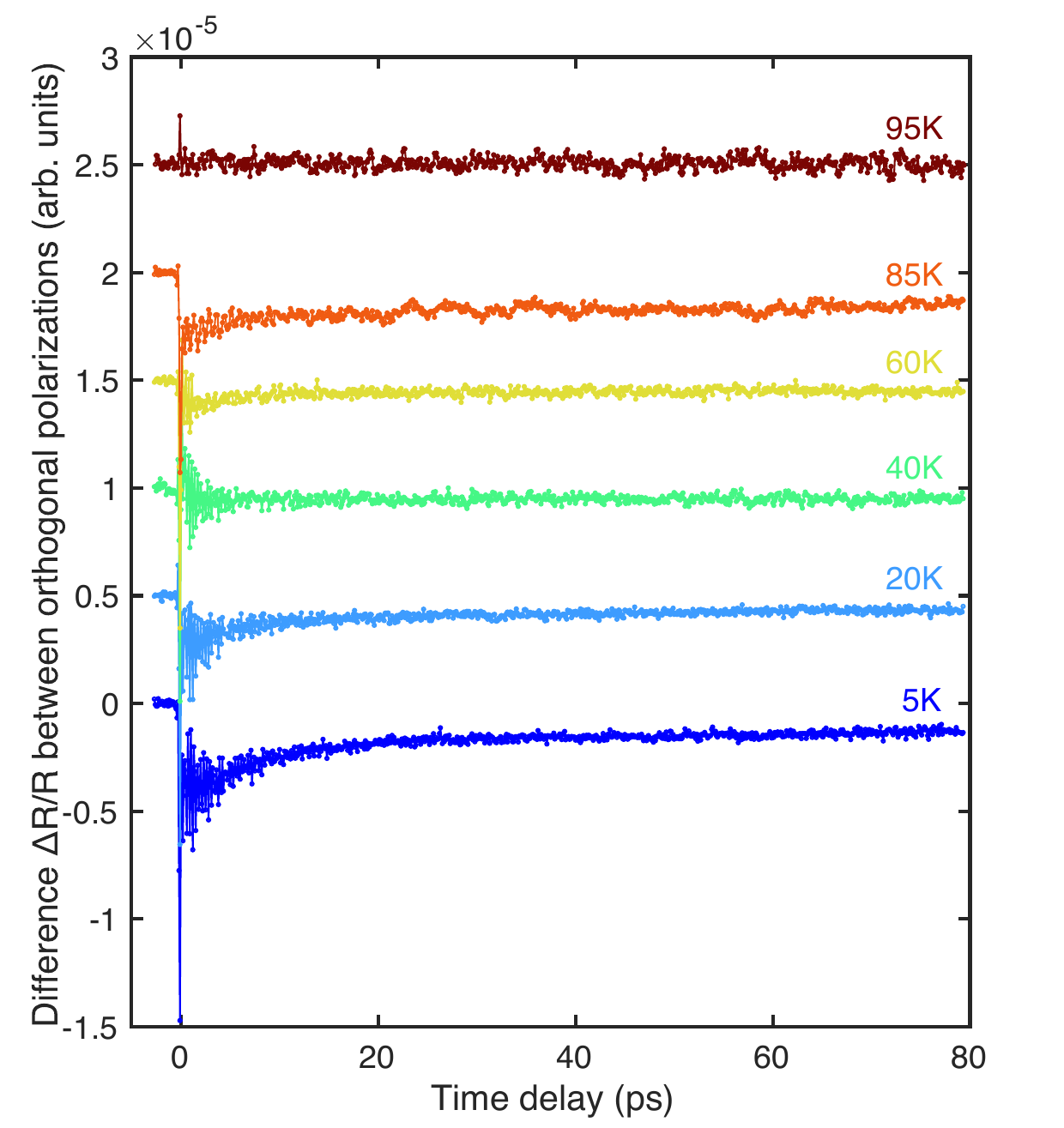}
    \caption{Differential time-resolved reflectivity curves between orthogonal incident probe polarizations at different temperatures across T$_{\text{CDW}}$. The curves are offset for clarity. The light spot position is inside a single birefringence domain. Nonzero differential values below T$_{\text{CDW}}$ indicates in-plane anisotropy, suggesting $C_6$ is broken. However, no difference is observed beyond noise level above T$_{\text{CDW}}$, indicating the preserve of $C_6$ and absence of nematic fluctuations. }
    \label{fig5}
\end{figure}

Contrary to previously reported pump-probe experiments on CsV$_3$Sb$_5$\cite{ratcliff2021coherent, PhysRevB.104.165110, wu2022simultaneous}, we observe clear close phonon pairs near both 1.3 THz and 3.1 THz in our phonon spectrum. At T = 5 K, the 1.3 THz mode consists of two close phonon modes as shown in Fig. \ref{fig2}b with two peaks centered at 1.26 THz and 1.33 THz, respectively. These dual modes persists upon warming, up to T$_{\text{CDW}}$ where it vanishes. Additionally, two close phonon peaks correspond to beatings in the time domain. Indeed, the beating waveform arising from these dual modes near 1.3 THz can be resolved in longer time scans (Fig. \ref{fig3}) with delay time difference of $\sim$14 ps between wave nodes, in accordance with the frequency splitting ($\sim$ 0.07 THz) between these two peaks. Repeated measurements on another CsV$_3$Sb$_5$ sample also shows this 1.3 THz dual modes (Fig. \ref{fig4}). Meanwhile, same experiments on a second birefringence domain (Fig. \ref{fig2}c and Domain 2 in Fig. \ref{fig2}d) visualizes two close phonon modes near 3.1 THz, with its two peaks centered at 2.96 THz and 3.14 THz at T = 5 K. These two close modes persist up to 60 K when the 3.1 THz mode fades away in our detection. Moreover, both 1.3 and 3.1 THz phonon pairs are visible in the coherent phonon spectrum extracted from all three birefringence domains (Fig. \ref{fig2}d). 

\begin{figure}
    \centering
    \includegraphics[width=8.6cm]{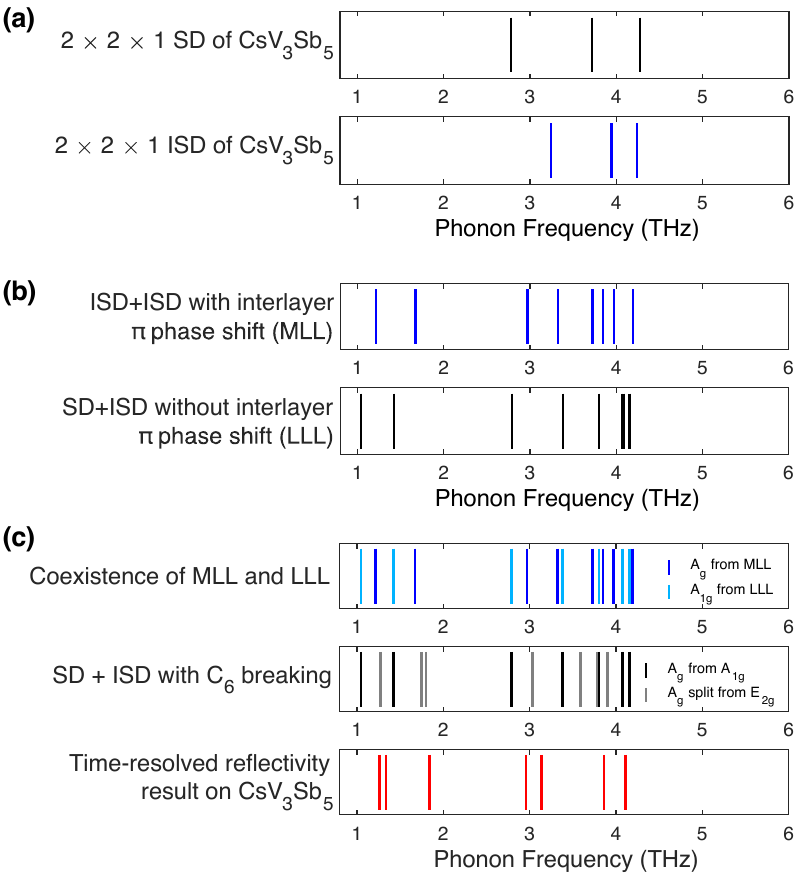}
    \caption{\textbf{Comparison of the DFT results and the measured phonon modes by TR-reflectivity on CsV$_3$Sb$_5$ at T = 5 K.} The vertical lines denote the frequency of the $A_{1g}$($A_g$) modes in DFT calculation results or the measured phonon frequency. For DFT-calculated phonons, only the modes below 4.5 THz are shown. (a) The calculated $A_{1g}$ mode frequencies in SD and ISD CDW phases in CsV$_3$Sb$_5$. (b) The calculated $A_g$ ($A_{1g}$) mode frequencies in the ISD + ISD with interlayer $\pi$ phase shift (SD + ISD without interlayer $\pi$ phase shift) CDW order. (c) The calculated fully-symmetric mode frequencies in the CDW order with the coexistence of MLL and LLL, and the calculated $A_g$ mode frequencies in SD + ISD with $C_6$ breaking CDW phases in CsV$_3$Sb$_5$ and their comparison to the observed phonon modes in our TR-reflectivity results. }
    \label{fig6}
\end{figure}

Furthermore, we reveal evidence of $C_6$ breaking from the anisotropy of transient reflectivity response by varying probe polarizations, which manifests breaking of rotational symmetry and has been used to study the nematic order in Fe-based superconductors\cite{stojchevska2012doping, thewalt2018imaging, liu2018transient}. This probe polarization dependent measurement has also been successfully used to probe the breaking of $C_6$ in Kagome CDW material CsCr$_3$Sb$_5$\cite{liu2024charge}. By varying the probe polarization within the (001) surface, we observe the anisotropic transient reflectivity signal inside the Kagome plane at T = 5 K, indicating broken $C_6$. To highlight this anisotropy, we subtract the signals from orthogonal probe polarization states that correspond to maximum and minimum signal size respectively. As shown in Fig. \ref{fig5}, this anisotropy persists in the CDW phase but vanishes above T$_{\text{CDW}}$, corroborating $C_6$ breaking onsets at the CDW transition. Further probe polarization dependent measurements are shown in Appendix B. This is different from the low-temperature electronic nematic order seen by NMR\cite{nie2022charge} and STM quasiparticle interference measurement\cite{li2023unidirectional, zhao2021cascade} 
that onsets at $\sim$ 35 K far below T$_{\text{CDW}}$. Our observed isotropic response of transient reflectivity above T$_{\text{CDW}}$ demonstrates $C_6$ symmetry, sharing the conclusion of Liu et al.\cite{liu2024absence} suggesting no vestigial nematic order associated with the $C_6$-breaking CDW. 

\begin{table*}
\small 
\centering
\resizebox{14cm}{!}{%
\begin{tabular}{l|lllllllll}
\hline\hline
Pristine                                                                                                  & \multicolumn{1}{l|}{\begin{tabular}[c]{@{}l@{}}$A_{1g}$ \\ 4.10\end{tabular}} & \begin{tabular}[c]{@{}l@{}}$E_{2g}$ \\ 3.86\end{tabular}     &                                                                                &                                                           &                                                           &                                                           &                                                           &                                                           &                                                           \\ \hline
2 $\times$ 2 $\times$ 1 SD                                                                                & \begin{tabular}[c]{@{}l@{}}$A_{1g}$\\ 2.78\end{tabular}                       & \begin{tabular}[c]{@{}l@{}}$A_{1g}$\\ 3.72\end{tabular} & \multicolumn{1}{l|}{\begin{tabular}[c]{@{}l@{}}$A_{1g}$\\ 4.27\end{tabular}} & \begin{tabular}[c]{@{}l@{}}$E_{2g}$\\ 1.49\end{tabular} & \begin{tabular}[c]{@{}l@{}}$E_{2g}$\\ 1.65\end{tabular} & \begin{tabular}[c]{@{}l@{}}$E_{2g}$\\ 3.58\end{tabular} & \begin{tabular}[c]{@{}l@{}}$E_{2g}$\\ 3.88\end{tabular} &                                                           &                                                           \\ \hline
2 $\times$ 2 $\times$ 1 ISD                                                                               & \begin{tabular}[c]{@{}l@{}}$A_{1g}$\\ 3.24\end{tabular}                       & \begin{tabular}[c]{@{}l@{}}$A_{1g}$\\ 3.95\end{tabular} & \multicolumn{1}{l|}{\begin{tabular}[c]{@{}l@{}}$A_{1g}$\\ 4.24\end{tabular}} & \begin{tabular}[c]{@{}l@{}}$E_{2g}$\\ 1.73\end{tabular} & \begin{tabular}[c]{@{}l@{}}$E_{2g}$\\ 2.97\end{tabular} & \begin{tabular}[c]{@{}l@{}}$E_{2g}$\\ 3.74\end{tabular} & \begin{tabular}[c]{@{}l@{}}$E_{2g}$\\ 3.87\end{tabular} &                                                           &                                                           \\ \hline
\begin{tabular}[c]{@{}l@{}}ISD + ISD with inter-\\layer $\pi$ phase shift\end{tabular}                    & \begin{tabular}[c]{@{}l@{}}$A_g$\\ 1.21\end{tabular}                          & \begin{tabular}[c]{@{}l@{}}$A_g$\\ 1.67\end{tabular}    & \begin{tabular}[c]{@{}l@{}}$A_g$\\ 2.97\end{tabular}                         & \begin{tabular}[c]{@{}l@{}}$A_g$\\ 3.32\end{tabular}    & \begin{tabular}[c]{@{}l@{}}$A_g$\\ 3.73\end{tabular}    & \begin{tabular}[c]{@{}l@{}}$A_g$\\ 3.85\end{tabular}    & \begin{tabular}[c]{@{}l@{}}$A_g$\\ 3.97\end{tabular}    & \begin{tabular}[c]{@{}l@{}}$A_g$\\ 4.19\end{tabular}    &                                                           \\ \hline
\multirow{2}{*}{\begin{tabular}[c]{@{}l@{}}SD + ISD without inter-~~~\\layer $\pi$ phase shift\end{tabular}} & \begin{tabular}[c]{@{}l@{}}$A_{1g}$\\ 1.04\end{tabular}                       & \begin{tabular}[c]{@{}l@{}}$A_{1g}$\\ 1.42\end{tabular} & \begin{tabular}[c]{@{}l@{}}$A_{1g}$\\ 2.79\end{tabular}                      & \begin{tabular}[c]{@{}l@{}}$A_{1g}$\\ 3.38\end{tabular} & \begin{tabular}[c]{@{}l@{}}$A_{1g}$\\ 3.80\end{tabular} & \begin{tabular}[c]{@{}l@{}}$A_{1g}$\\ 4.07\end{tabular} & \begin{tabular}[c]{@{}l@{}}$A_{1g}$\\ 4.08\end{tabular} & \begin{tabular}[c]{@{}l@{}}$A_{1g}$\\ 4.16\end{tabular} &                                                           \\ \cline{2-10} 
                                                                                                          & \begin{tabular}[c]{@{}l@{}}$E_{2g}$\\ 0.31~~~\end{tabular}                       & \begin{tabular}[c]{@{}l@{}}$E_{2g}$\\ 1.27~~~\end{tabular} & \begin{tabular}[c]{@{}l@{}}$E_{2g}$\\ 1.75~~~\end{tabular}                      & \begin{tabular}[c]{@{}l@{}}$E_{2g}$\\ 1.80~~~\end{tabular} & \begin{tabular}[c]{@{}l@{}}$E_{2g}$\\ 3.03~~~\end{tabular} & \begin{tabular}[c]{@{}l@{}}$E_{2g}$\\ 3.59~~~\end{tabular} & \begin{tabular}[c]{@{}l@{}}$E_{2g}$\\ 3.78~~~\end{tabular} & \begin{tabular}[c]{@{}l@{}}$E_{2g}$\\ 3.89~~~\end{tabular} & \begin{tabular}[c]{@{}l@{}}$E_{2g}$\\ 3.90~~~\end{tabular} \\ \hline\hline
\end{tabular}%
}
\caption{Frequency (unit: THz) of the selected Raman-active modes in the pristine and CDW phases of CsV$_3$Sb$_5$ calculated by DFT. Only relevant $A_{1g}$, $E_{2g}$ and $A_g$ modes with frequencies below 4.5 THz are included. When interlayer $\pi$ phase shift between the SD and ISD layer is included, the $E_{2g}$ modes in the SD + ISD without interlayer $\pi$ phase shift (LLL) phase will split to induce $A_g$ modes that can be detected in TR-reflectivity. }
\label{table-1}
\end{table*}

To explain the observed close phonon modes and determine the CDW structure of CsV$_3$Sb$_5$, we perform DFT calculations of phonon frequencies and compare with the phonon spectrum detected by TR-reflectivity. In TR-reflectivity measurements, the femtosecond pump pulse selectively excites Raman-active phonons coherently\cite{merlin1997generating, stevens2002coherent}. On the theory side, coherent phonon excitation is often described either as impulsive stimulated Raman scattering (ISRS)\cite{dhar1994time, yan1985impulsive, gray2024time} or as displacive excitation of coherent phonons (DECP)\cite{cheng1991mechanism, zeiger1992theory}. In ISRS, there is no restrictions on the symmetry of the Raman-active modes\cite{dhar1994time}, and ISRS does not require absorption in the material\cite{de1985femtosecond, ruhman1987time}. Meanwhile, DECP requires absorption at the pump frequency in order to disturb the electronic energy distribution in the material, and only fully symmetric modes can be observed\cite{zeiger1992theory, cheng1991mechanism} in DECP mechanism. Fully symmetric phonon modes host $\Gamma_1^{+}$ symmetry, such as $A_{1g}$ modes in $D_{6h}$ point group and $A_g$ modes in $D_{2h}$ point group. 

To determine the coherent phonon excitation mechanism in our pump-probe measurement, we first compare our detected phonon modes with previous Raman spectroscopy results on CsV$_3$Sb$_5$\cite{wu2022charge, liu2022observation}. Above T$_{\text{CDW}}$, Raman spectroscopy detected one $A_{1g}$ mode at 4.1 THz and one $E_{2g}$ mode at 3.6 THz with comparable amplitudes. This main 3.6 THz $E_{2g}$ mode persists across T$_{\text{CDW}}$ at all temperatures and has the maximum amplitude among all non-fully-symmetric Raman-active modes, with minimal frequency change of less than 1 cm$^{-1}$ ($\rightarrow$ 0.030 THz) and increasing amplitude as temperature increases\cite{wu2022charge, liu2022observation}. However, in our TR-reflectivity measurements, we only detect the 4.1 THz $A_{1g}$ mode above T$_{\text{CDW}}$. Below T$_{\text{CDW}}$, we do not observe this strongest 3.6 THz $E_{2g}$ mode either. The closest detected mode is the 3.86 THz mode, but it quickly weakens and softens as temperature increases, in contrast to the 3.6 THz $E_{2g}$ mode. Since we do not observe this dominant 3.6 THz mode at any temperature, 
we can rule out the detection of non-fully-symmetric modes in our TR-reflectivity measurements. Meanwhile, our pump photon energy ($\sim$ 0.79 eV) is much larger than the partial CDW gap opening in CsV$_3$Sb$_5$\cite{nakayama2021multiple, azoury2023direct, zhou2021origin, wang2021distinctive} indicating strong absorption at pump frequency. Further confirmation of DECP mechanism is shown in pump polarization dependent measurements detailed in Appendix C. We confirm therein the observed coherent phonons in the CDW state are all fully symmetric. Thus, we use the DECP mechanism to interpret our data, and we assume all the detected phonon modes in our time-resolved reflectivity measurements are fully symmetric modes. This also matches with previous pump-probe studies on CsV$_3$Sb$_5$\cite{ratcliff2021coherent}.

We first examine the phonon spectra of the two $C_6$-symmetric 2 $\times$ 2 $\times$ 1 CDW orders, SD and ISD, to see if they match with our observed phonon spectrum. Fig. \ref{fig6}a shows the DFT calculated $A_{1g}$ phonon spectrum of SD and ISD phases in the relevant frequency region. Both phases lack the observed 1.3 and 3.1 THz phonon pairs and the 1.84 THz mode from our TR-reflectivity measurements.

\begin{figure*}[t]
    \centering
    \includegraphics[width=\textwidth]{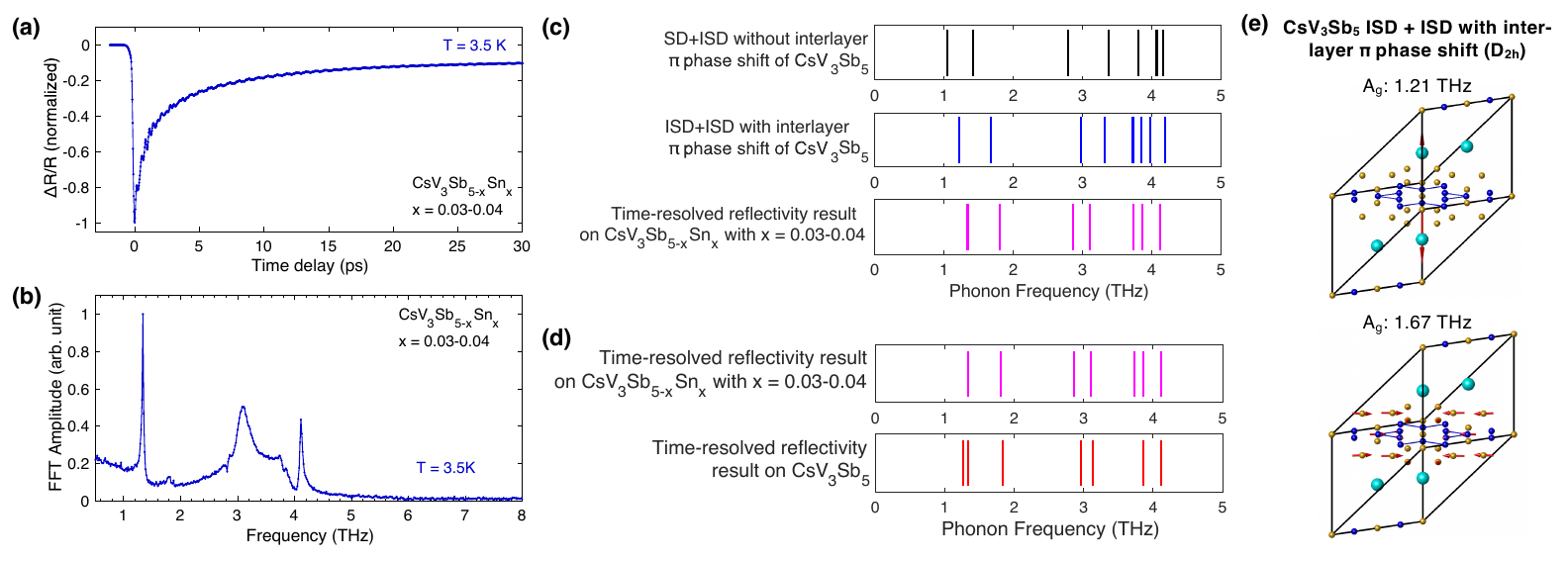}
    \caption{\textbf{Coherent phonon spectrum in CsV$_3$Sb$_{5-\text{x}}$Sn$_\text{x}$ with x = 0.03-0.04.} (a) TR-reflectivity time trace on CsV$_3$Sb$_{5-\text{x}}$Sn$_\text{x}$ with x = 0.03-0.04 at T = 3.5 K. (b) Coherent phonon spectrum measured from TR-reflectivity at T = 3.5 K. (c) The calculated $A_{1g}$ ($A_g$) mode frequencies in the SD + ISD without interlayer $\pi$ phase shift (ISD + ISD with interlayer $\pi$ phase shift) CDW state in CsV$_3$Sb$_5$ and their comparison to our TR-reflectivity results in CsV$_3$Sb$_{5-\text{x}}$Sn$_\text{x}$ with x = 0.03-0.04. (d) Comparison of the measured coherent phonon spectrum by time-resolved reflectivity in CsV$_3$Sb$_5$ and CsV$_3$Sb$_{5-\text{x}}$Sn$_\text{x}$ with x = 0.03-0.04. In (c-d), The phonon spectrum of CsV$_3$Sb$_{5-\text{x}}$Sn$_\text{x}$ with x = 0.03-0.04 measured from TR-reflectivity is taken at T = 3.5 K. In (d), the phonon spectrum for CsV$_3$Sb$_5$ is taken at T = 5 K. (e) The DFT-calculated oscillation pattern of the fully-symmetric Raman active $A_g$ modes in ISD + ISD with interlayer $\pi$ phase shift (MLL) CDW state in CsV$_3$Sb$_5$ that are near our detected 1.34 and 1.80 THz modes in CsV$_3$Sb$_{5-\text{x}}$Sn$_\text{x}$ with x = 0.03-0.04 via time-resolved reflectivity, respectively. The 1.21 and 1.67 THz $A_g$ modes are the lowest and second-lowest frequency mode in the $A_g$ phonon spectrum of MLL phase in (c).}
    \label{fig7}
\end{figure*}

Thus, we need to consider a modulation along c axis of the CDW order, which is consistent with previous X-ray diffraction studies\cite{li2021observation, kautzsch2023structural}. We first examine the two most reported 2 $\times$ 2 $\times$ 2 CDW states (Fig. \ref{fig6}b): either ISD + ISD with interlayer $\pi$ phase shift (MLL), where the $C_6$ rotational symmetry is broken, or SD + ISD without interlayer $\pi$ phase shift (LLL), where $C_6$ is preserved. The calculated fully-symmetric $A_g$ ($A_{1g}$) mode frequencies in the ISD + ISD with interlayer $\pi$ phase shift (SD + ISD without interlayer $\pi$ phase shift) CDW order has been listed in Table \ref{table-1} and plotted in Fig. \ref{fig6}b. The MLL phase agrees with our TR-reflectivity results better, with the frequency difference of the two $A_g$ modes near 3.1 THz closer to the observed value. The MLL phase also hosts an $A_g$ mode at 1.67 THz that matches the observed 1.84 THz phonon. This $A_g$ mode comes from the $C_6$-breaking induced split of an $E_{2g}$ mode at 1.73 THz in 2 $\times$ 2 $\times$ 1 ISD phase (Table \ref{table-1}) ($E_{2g}$ $\rightarrow$ $A_g + B_{1g}$). However, MLL only shows one $A_g$ mode near 1.3 THz, which fails to explain the observed two close modes near 1.3 THz. For the LLL phase, although it has two $A_{1g}$ phonon pairs near 1.3 THz and 3.1 THz, their frequency differences are too large to match the measured values of our observed 1.3 THz and 3.1 THz phonon pairs. Also, no $A_{1g}$ mode in LLL matches well with our observed 1.84 THz mode. Finally, our previous birefringence measurement indicates $C_6$ symmetry breaking\cite{xu2022three}, contradictory to the preserved $C_6$ in LLL. In short, either MLL or LLL cannot explain our observed coherent phonon spectrum in CsV$_3$Sb$_5$.

We note a recent time-resolved X-ray diffraction (XRD) measurement that demonstrated the coexistence of MLL and LLL CDW orders\cite{ning2024dynamical} which breaks $C_6$. By comparing the combined phonon spectrum of these two CDW orders, we point out this coexistence configuration can explain our experimentally detected phonon spectrum (Fig. \ref{fig6}c) by showing fully-symmetric phonon pairs near 1.3 and 3.1 THz and a fully-symmetric mode near 1.8 THz, explaining our measured phonon spectrum better than merely MLL or LLL. This coexistence of MLL and LLL phase will also be energetically favorable as it contains the reported two energetically favorable CDW phases MLL and LLL\cite{tan2021charge, christensen2021theory}. Further evidence of the coexistence of MLL and LLL CDW orders comes from our pump fluence dependent TR-reflectivity measurements, as shown in Appendix D. Herein, we pinpoint the lower and higher frequency peak of the 1.3 THz dual modes originates from LLL and MLL phase respectively. The observed 1.84 THz phonon is then explained by the 1.67 THz $A_g$ mode in MLL phase, which comes from $C_6$-breaking induced split of an $E_{2g}$ mode at 1.73 THz in 2 $\times$ 2 $\times$ 1 ISD phase (Table \ref{table-1}). While there are more fully-symmetric modes in calculation than observed, some modes could be missing in our observation, possibly due to their weaker modulations on the refractive index. Moreover, some calculated modes between 3.5 - 4 THz may be too close in frequency to be resolved individually in the measurements.


Besides, if $C_6$ is broken by interlayer $\pi$ phase shift between the SD and ISD layer (Fig. \ref{fig1}e), $E_{2g}$ modes in the $C_6$-symmetric LLL phase will split to give rise to a fully symmetric $A_g$ mode that can be detected in TR-reflectivity. $A_{1g}$ modes in LLL will also evolve into $A_g$ modes. We list the $A_g$ modes in this $C_6$-breaking SD + ISD state in Fig. \ref{fig6}c. In this phase, there also exists close 1.3 THz and 3.1 THz $A_g$ pairs, along with $A_g$ modes near 1.8 THz that are split from $E_{2g}$ modes, i.e. the 1.75 and 1.80 THz $E_{2g}$ mode in LLL (Table \ref{table-1}). However, this SD + ISD with interlayer $\pi$-phase shift phase may not be among the most energetically stable phases\cite{ratcliff2021coherent, tan2021charge, christensen2021theory}. Considering our pump fluence dependence data, recent time-resolved XRD results\cite{ning2024dynamical} and energy stability of these phases, the coexistence of MLL and LLL phase is more likely, but we can't fully rule out SD + ISD with interlayer $\pi$-phase shift. 

In either case, our observed two close phonon pairs near 1.3 and 3.1 THz strongly corroborate C$_6$-breaking in the CDW phase of CsV$_3$Sb$_5$ and the coexistence of SD and ISD patterns. Meanwhile, to explain our observed 1.84 THz mode, $C_6$-breaking is also required to give rise to a fully symmetric $A_g$ mode near 1.8 THz by splitting the corresponding $E_{2g}$ mode. Thus, the persistence of two close modes near 1.3 THz and the 1.84 THz mode up to T$_{\text{CDW}}$ in our measurements marks the $C_6$ breaking happens simultaneously with the onset of CDW, contrary to the electronic nematicity revealed by NMR\cite{nie2022charge} and STM\cite{li2023unidirectional, zhao2021cascade} that only onset at temperatures far below T$_{\text{CDW}}$. 


There is also a recent Raman measurement reporting two close modes near 1.3 THz\cite{jin2024pi} 
but only above $\sim$ 65 K. The temperature-dependent evolution of the intensities of these two modes indicated a spectral transformation between them and the higher frequency mode became weaker above $\sim$ 65 K. This is different from our TR-reflectivity results where the two close modes near 1.3 THz are always present from T$_{\text{CDW}}$ down to 5 K, and our measured intensity of the higher frequency mode keeps increasing from 5 K to 80 K. 
We also note that an alternative interpretation based on calculation analysis has been proposed in \cite{alkorta2025symmetry}, showing a 2 $\times$ 2 $\times$ 4 ISD-stacking phase coexisting with MLL provides a 1.4 THz mode which can combine with the 1.21 THz $A_g$ mode in MLL to explain the observed dual 1.3 THz modes.

In contrast to pure CsV$_3$Sb$_5$, the investigation of TR-reflectivity on CsV$_3$Sb$_{5-\text{x}}$Sn$_\text{x}$ with x = 0.03-0.04 uncovers a different CDW structural reconstruction from CsV$_3$Sb$_5$. Fig. \ref{fig7}b shows the coherent phonon spectrum extracted from $\Delta$R/R vs. time delay (Fig. \ref{fig7}a) in this x = 0.03-0.04 Sn-doped sample. Similar to CsV$_3$Sb$_5$, at T = 3.5 K, the most prominent modes are at 1.34, 3.11 and 4.18 THz, along with weaker peaks around 1.80 and 3.8 THz. However, in contrast to dual modes at 1.3 THz in CsV$_3$Sb$_5$, we observe only one mode at 1.34 THz in this Sn-doped sample, with its frequency matching the higher frequency peak in the 1.3 THz dual mode in undoped CsV$_3$Sb$_5$ (Fig. \ref{fig7}d). Meanwhile, the 3.11 THz mode is weaker and broader compared to CsV$_3$Sb$_5$, making its dual mode feature weaker. We then compare our detected phonon modes with the DFT-calculated fully-symmetric phonon spectrum in various CDW phases (Fig. \ref{fig7}c). Overall, the phonon spectrum of ISD + ISD with interlayer $\pi$-phase shift matches the experimental results better, since the observed 1.34, 1.80 and 3.11 THz modes are better matched in frequency. We thus conclude the actual CDW structure of this x = 0.03-0.04 sample to be ISD + ISD with interlayer $\pi$-phase shift (MLL), breaking $C_6$ rotational symmetry as well. This matches with recent ARPES\cite{kang2022charge} and X-ray diffraction\cite{kautzsch2023incommensurate} results. Thus, our observed 1.34 THz mode corresponds to the calculated 1.21 THz $A_g$ mode which is a Cs-related mode, and our observed 1.80 THz mode corresponds to the calculated 1.67 THz $A_g$ mode which involves V and Sb oscillations (Fig. \ref{fig7}e). The CDW structure in this x = 0.03-0.04 sample contrasts with the coexistence of SD and ISD distortion in undoped CsV$_3$Sb$_5$, which explains the evolution from dual modes to single mode at 1.3 THz by a slight x = 0.03-0.04 Sn-doping. The frequency match between the 1.34 THz mode in this x = 0.03-0.04 Sn-doped sample and the higher frequency peak in the 1.3 THz dual modes in undoped CsV$_3$Sb$_5$ indicates this higher frequency peak in the 1.3 THz dual modes in undoped CsV$_3$Sb$_5$ comes from MLL, matching with the conclusion in Appendix D and further confirms the coexistence of MLL and LLL CDW phase in undoped CsV$_3$Sb$_5$.

In conclusion, we have studied coherent phonon excitation in CsV$_3$Sb$_5$ via time-resolved reflectivity to determine the structural configuration of its CDW phase. The phonon spectra confirm six-fold rotational symmetry breaking and the coexistence of SD and ISD distortion in the CDW phase of CsV$_3$Sb$_5$. From our measurements and DFT calculations, there are two possibilities: coexistence of MLL and LLL, and SD + ISD with interlayer $\pi$-phase shift. Considering our fluence dependence measurements, we think coexistence of MLL and LLL phase is more likely, but we can't fully rule out SD + ISD with interlayer $\pi$-phase shift. Meanwhile, an x = 0.03-0.04 Sn-doping changes the CDW structure to MLL with only in-plane ISD distortion, reducing 1.3 THz dual modes in undoped CsV$_3$Sb$_5$ to a single mode. This observation provides deeper insights for comprehending the interplay with other electronic phases in this system, such as the in-plane 1 $\times$ 4 unidirectional charge order\cite{wang2021electronic, zhao2021cascade, li2023unidirectional}, electronic nematicity, and superconductivity. We also envision our methodology will promote the understanding of the CDW phase diagram in doped AV$_3$Sb$_5$\cite{oey2022fermi, zhong2023nodeless, kautzsch2023incommensurate} and other Kagome systems such as FeGe\cite{teng2022discovery, oh2024tunability}, and stimulate further research on ultrafast manipulation of these symmetry-breaking states.

\section*{Acknowledgement}
The construction of the pump-probe setup was supported by the Air Force Office of Scientific Research under award no. FA9550-22-1-0410. Q.D. was mainly supported by the Vagelos Institute of Energy Science and Technology graduate fellowship and also partly supported by the Air Force Office of Scientific Research under award no. FA9550-22-1-0410 and the National Science Foundation EPM program under grant no. DMR-2213891. S.D.W. and B.R.O. gratefully acknowledge support via the UC Santa Barbara National Science Foundation Quantum Foundry funded via the Q-AMASE-i program under award DMR-1906325. B.R.O. thanks support from the U.S. Department of Energy (DOE), Office of Science, Basic Energy Sciences (BES), Materials Sciences and Engineering Division. B.Y. acknowledges the financial support by the Israel Science Foundation (ISF: 2932/21, 2974/23), German Research Foundation (DFG, CRC-183, A02), and by a research grant from the Estate of Gerald Alexander.  L.W. acknowledges support from the Sloan Foundation under the award FG-2025-25036.

\bibliography{CsV3Sb5_Pure_TRR_refs}

\begin{thebibliography}{92}%
\makeatletter
\providecommand \@ifxundefined [1]{%
 \@ifx{#1\undefined}
}%
\providecommand \@ifnum [1]{%
 \ifnum #1\expandafter \@firstoftwo
 \else \expandafter \@secondoftwo
 \fi
}%
\providecommand \@ifx [1]{%
 \ifx #1\expandafter \@firstoftwo
 \else \expandafter \@secondoftwo
 \fi
}%
\providecommand \natexlab [1]{#1}%
\providecommand \enquote  [1]{``#1''}%
\providecommand \bibnamefont  [1]{#1}%
\providecommand \bibfnamefont [1]{#1}%
\providecommand \citenamefont [1]{#1}%
\providecommand \href@noop [0]{\@secondoftwo}%
\providecommand \href [0]{\begingroup \@sanitize@url \@href}%
\providecommand \@href[1]{\@@startlink{#1}\@@href}%
\providecommand \@@href[1]{\endgroup#1\@@endlink}%
\providecommand \@sanitize@url [0]{\catcode `\\12\catcode `\$12\catcode
  `\&12\catcode `\#12\catcode `\^12\catcode `\_12\catcode `\%12\relax}%
\providecommand \@@startlink[1]{}%
\providecommand \@@endlink[0]{}%
\providecommand \url  [0]{\begingroup\@sanitize@url \@url }%
\providecommand \@url [1]{\endgroup\@href {#1}{\urlprefix }}%
\providecommand \urlprefix  [0]{URL }%
\providecommand \Eprint [0]{\href }%
\providecommand \doibase [0]{https://doi.org/}%
\providecommand \selectlanguage [0]{\@gobble}%
\providecommand \bibinfo  [0]{\@secondoftwo}%
\providecommand \bibfield  [0]{\@secondoftwo}%
\providecommand \translation [1]{[#1]}%
\providecommand \BibitemOpen [0]{}%
\providecommand \bibitemStop [0]{}%
\providecommand \bibitemNoStop [0]{.\EOS\space}%
\providecommand \EOS [0]{\spacefactor3000\relax}%
\providecommand \BibitemShut  [1]{\csname bibitem#1\endcsname}%
\let\auto@bib@innerbib\@empty
\bibitem [{\citenamefont {Ortiz}\ \emph {et~al.}(2019)\citenamefont {Ortiz},
  \citenamefont {Gomes}, \citenamefont {Morey}, \citenamefont {Winiarski},
  \citenamefont {Bordelon}, \citenamefont {Mangum}, \citenamefont {Oswald},
  \citenamefont {Rodriguez-Rivera}, \citenamefont {Neilson}, \citenamefont
  {Wilson} \emph {et~al.}}]{ortiz2019new}%
  \BibitemOpen
  \bibfield  {author} {\bibinfo {author} {\bibfnamefont {B.~R.}\ \bibnamefont
  {Ortiz}}, \bibinfo {author} {\bibfnamefont {L.~C.}\ \bibnamefont {Gomes}},
  \bibinfo {author} {\bibfnamefont {J.~R.}\ \bibnamefont {Morey}}, \bibinfo
  {author} {\bibfnamefont {M.}~\bibnamefont {Winiarski}}, \bibinfo {author}
  {\bibfnamefont {M.}~\bibnamefont {Bordelon}}, \bibinfo {author}
  {\bibfnamefont {J.~S.}\ \bibnamefont {Mangum}}, \bibinfo {author}
  {\bibfnamefont {I.~W.}\ \bibnamefont {Oswald}}, \bibinfo {author}
  {\bibfnamefont {J.~A.}\ \bibnamefont {Rodriguez-Rivera}}, \bibinfo {author}
  {\bibfnamefont {J.~R.}\ \bibnamefont {Neilson}}, \bibinfo {author}
  {\bibfnamefont {S.~D.}\ \bibnamefont {Wilson}}, \emph {et~al.},\ }\bibfield
  {title} {\bibinfo {title} {New kagome prototype materials: discovery of kv 3
  sb 5, rbv 3 sb 5, and csv 3 sb 5},\ }\href@noop {} {\bibfield  {journal}
  {\bibinfo  {journal} {Physical Review Materials}\ }\textbf {\bibinfo {volume}
  {3}},\ \bibinfo {pages} {094407} (\bibinfo {year} {2019})}\BibitemShut
  {NoStop}%
\bibitem [{\citenamefont {Ortiz}\ \emph {et~al.}(2020)\citenamefont {Ortiz},
  \citenamefont {Teicher}, \citenamefont {Hu}, \citenamefont {Zuo},
  \citenamefont {Sarte}, \citenamefont {Schueller}, \citenamefont {Abeykoon},
  \citenamefont {Krogstad}, \citenamefont {Rosenkranz}, \citenamefont {Osborn}
  \emph {et~al.}}]{ortiz2020cs}%
  \BibitemOpen
  \bibfield  {author} {\bibinfo {author} {\bibfnamefont {B.~R.}\ \bibnamefont
  {Ortiz}}, \bibinfo {author} {\bibfnamefont {S.~M.}\ \bibnamefont {Teicher}},
  \bibinfo {author} {\bibfnamefont {Y.}~\bibnamefont {Hu}}, \bibinfo {author}
  {\bibfnamefont {J.~L.}\ \bibnamefont {Zuo}}, \bibinfo {author} {\bibfnamefont
  {P.~M.}\ \bibnamefont {Sarte}}, \bibinfo {author} {\bibfnamefont {E.~C.}\
  \bibnamefont {Schueller}}, \bibinfo {author} {\bibfnamefont {A.~M.}\
  \bibnamefont {Abeykoon}}, \bibinfo {author} {\bibfnamefont {M.~J.}\
  \bibnamefont {Krogstad}}, \bibinfo {author} {\bibfnamefont {S.}~\bibnamefont
  {Rosenkranz}}, \bibinfo {author} {\bibfnamefont {R.}~\bibnamefont {Osborn}},
  \emph {et~al.},\ }\bibfield  {title} {\bibinfo {title} {Cs v 3 sb 5: A z 2
  topological kagome metal with a superconducting ground state},\ }\href@noop
  {} {\bibfield  {journal} {\bibinfo  {journal} {Physical Review Letters}\
  }\textbf {\bibinfo {volume} {125}},\ \bibinfo {pages} {247002} (\bibinfo
  {year} {2020})}\BibitemShut {NoStop}%
\bibitem [{\citenamefont {Ortiz}\ \emph
  {et~al.}(2021{\natexlab{a}})\citenamefont {Ortiz}, \citenamefont {Sarte},
  \citenamefont {Kenney}, \citenamefont {Graf}, \citenamefont {Teicher},
  \citenamefont {Seshadri},\ and\ \citenamefont
  {Wilson}}]{ortiz2021superconductivity}%
  \BibitemOpen
  \bibfield  {author} {\bibinfo {author} {\bibfnamefont {B.~R.}\ \bibnamefont
  {Ortiz}}, \bibinfo {author} {\bibfnamefont {P.~M.}\ \bibnamefont {Sarte}},
  \bibinfo {author} {\bibfnamefont {E.~M.}\ \bibnamefont {Kenney}}, \bibinfo
  {author} {\bibfnamefont {M.~J.}\ \bibnamefont {Graf}}, \bibinfo {author}
  {\bibfnamefont {S.~M.}\ \bibnamefont {Teicher}}, \bibinfo {author}
  {\bibfnamefont {R.}~\bibnamefont {Seshadri}},\ and\ \bibinfo {author}
  {\bibfnamefont {S.~D.}\ \bibnamefont {Wilson}},\ }\bibfield  {title}
  {\bibinfo {title} {Superconductivity in the z 2 kagome metal kv 3 sb 5},\
  }\href@noop {} {\bibfield  {journal} {\bibinfo  {journal} {Physical Review
  Materials}\ }\textbf {\bibinfo {volume} {5}},\ \bibinfo {pages} {034801}
  (\bibinfo {year} {2021}{\natexlab{a}})}\BibitemShut {NoStop}%
\bibitem [{\citenamefont {Yin}\ \emph {et~al.}(2021)\citenamefont {Yin},
  \citenamefont {Tu}, \citenamefont {Gong}, \citenamefont {Fu}, \citenamefont
  {Yan},\ and\ \citenamefont {Lei}}]{yin2021superconductivity}%
  \BibitemOpen
  \bibfield  {author} {\bibinfo {author} {\bibfnamefont {Q.}~\bibnamefont
  {Yin}}, \bibinfo {author} {\bibfnamefont {Z.}~\bibnamefont {Tu}}, \bibinfo
  {author} {\bibfnamefont {C.}~\bibnamefont {Gong}}, \bibinfo {author}
  {\bibfnamefont {Y.}~\bibnamefont {Fu}}, \bibinfo {author} {\bibfnamefont
  {S.}~\bibnamefont {Yan}},\ and\ \bibinfo {author} {\bibfnamefont
  {H.}~\bibnamefont {Lei}},\ }\bibfield  {title} {\bibinfo {title}
  {Superconductivity and normal-state properties of kagome metal rbv3sb5 single
  crystals},\ }\href@noop {} {\bibfield  {journal} {\bibinfo  {journal}
  {Chinese Physics Letters}\ }\textbf {\bibinfo {volume} {38}},\ \bibinfo
  {pages} {037403} (\bibinfo {year} {2021})}\BibitemShut {NoStop}%
\bibitem [{\citenamefont {Mielke~III}\ \emph {et~al.}(2022)\citenamefont
  {Mielke~III}, \citenamefont {Das}, \citenamefont {Yin}, \citenamefont {Liu},
  \citenamefont {Gupta}, \citenamefont {Jiang}, \citenamefont {Medarde},
  \citenamefont {Wu}, \citenamefont {Lei}, \citenamefont {Chang} \emph
  {et~al.}}]{mielke2022time}%
  \BibitemOpen
  \bibfield  {author} {\bibinfo {author} {\bibfnamefont {C.}~\bibnamefont
  {Mielke~III}}, \bibinfo {author} {\bibfnamefont {D.}~\bibnamefont {Das}},
  \bibinfo {author} {\bibfnamefont {J.-X.}\ \bibnamefont {Yin}}, \bibinfo
  {author} {\bibfnamefont {H.}~\bibnamefont {Liu}}, \bibinfo {author}
  {\bibfnamefont {R.}~\bibnamefont {Gupta}}, \bibinfo {author} {\bibfnamefont
  {Y.-X.}\ \bibnamefont {Jiang}}, \bibinfo {author} {\bibfnamefont
  {M.}~\bibnamefont {Medarde}}, \bibinfo {author} {\bibfnamefont
  {X.}~\bibnamefont {Wu}}, \bibinfo {author} {\bibfnamefont {H.~C.}\
  \bibnamefont {Lei}}, \bibinfo {author} {\bibfnamefont {J.}~\bibnamefont
  {Chang}}, \emph {et~al.},\ }\bibfield  {title} {\bibinfo {title}
  {Time-reversal symmetry-breaking charge order in a kagome superconductor},\
  }\href@noop {} {\bibfield  {journal} {\bibinfo  {journal} {Nature}\ }\textbf
  {\bibinfo {volume} {602}},\ \bibinfo {pages} {245} (\bibinfo {year}
  {2022})}\BibitemShut {NoStop}%
\bibitem [{\citenamefont {Jiang}\ \emph {et~al.}(2021)\citenamefont {Jiang},
  \citenamefont {Yin}, \citenamefont {Denner}, \citenamefont {Shumiya},
  \citenamefont {Ortiz}, \citenamefont {Xu}, \citenamefont {Guguchia},
  \citenamefont {He}, \citenamefont {Hossain}, \citenamefont {Liu} \emph
  {et~al.}}]{jiang2021unconventional}%
  \BibitemOpen
  \bibfield  {author} {\bibinfo {author} {\bibfnamefont {Y.-X.}\ \bibnamefont
  {Jiang}}, \bibinfo {author} {\bibfnamefont {J.-X.}\ \bibnamefont {Yin}},
  \bibinfo {author} {\bibfnamefont {M.~M.}\ \bibnamefont {Denner}}, \bibinfo
  {author} {\bibfnamefont {N.}~\bibnamefont {Shumiya}}, \bibinfo {author}
  {\bibfnamefont {B.~R.}\ \bibnamefont {Ortiz}}, \bibinfo {author}
  {\bibfnamefont {G.}~\bibnamefont {Xu}}, \bibinfo {author} {\bibfnamefont
  {Z.}~\bibnamefont {Guguchia}}, \bibinfo {author} {\bibfnamefont
  {J.}~\bibnamefont {He}}, \bibinfo {author} {\bibfnamefont {M.~S.}\
  \bibnamefont {Hossain}}, \bibinfo {author} {\bibfnamefont {X.}~\bibnamefont
  {Liu}}, \emph {et~al.},\ }\bibfield  {title} {\bibinfo {title}
  {Unconventional chiral charge order in kagome superconductor kv3sb5},\
  }\href@noop {} {\bibfield  {journal} {\bibinfo  {journal} {Nature materials}\
  }\textbf {\bibinfo {volume} {20}},\ \bibinfo {pages} {1353} (\bibinfo {year}
  {2021})}\BibitemShut {NoStop}%
\bibitem [{\citenamefont {Xing}\ \emph {et~al.}(2024)\citenamefont {Xing},
  \citenamefont {Bae}, \citenamefont {Ritz}, \citenamefont {Yang},
  \citenamefont {Birol}, \citenamefont {Capa~Salinas}, \citenamefont {Ortiz},
  \citenamefont {Wilson}, \citenamefont {Wang}, \citenamefont {Fernandes} \emph
  {et~al.}}]{xing2024optical}%
  \BibitemOpen
  \bibfield  {author} {\bibinfo {author} {\bibfnamefont {Y.}~\bibnamefont
  {Xing}}, \bibinfo {author} {\bibfnamefont {S.}~\bibnamefont {Bae}}, \bibinfo
  {author} {\bibfnamefont {E.}~\bibnamefont {Ritz}}, \bibinfo {author}
  {\bibfnamefont {F.}~\bibnamefont {Yang}}, \bibinfo {author} {\bibfnamefont
  {T.}~\bibnamefont {Birol}}, \bibinfo {author} {\bibfnamefont {A.~N.}\
  \bibnamefont {Capa~Salinas}}, \bibinfo {author} {\bibfnamefont {B.~R.}\
  \bibnamefont {Ortiz}}, \bibinfo {author} {\bibfnamefont {S.~D.}\ \bibnamefont
  {Wilson}}, \bibinfo {author} {\bibfnamefont {Z.}~\bibnamefont {Wang}},
  \bibinfo {author} {\bibfnamefont {R.~M.}\ \bibnamefont {Fernandes}}, \emph
  {et~al.},\ }\bibfield  {title} {\bibinfo {title} {Optical manipulation of the
  charge-density-wave state in rbv3sb5},\ }\href@noop {} {\bibfield  {journal}
  {\bibinfo  {journal} {Nature}\ }\textbf {\bibinfo {volume} {631}},\ \bibinfo
  {pages} {60} (\bibinfo {year} {2024})}\BibitemShut {NoStop}%
\bibitem [{\citenamefont {Guo}\ \emph {et~al.}(2022)\citenamefont {Guo},
  \citenamefont {Putzke}, \citenamefont {Konyzheva}, \citenamefont {Huang},
  \citenamefont {Gutierrez-Amigo}, \citenamefont {Errea}, \citenamefont {Chen},
  \citenamefont {Vergniory}, \citenamefont {Felser}, \citenamefont {Fischer}
  \emph {et~al.}}]{guo2022switchable}%
  \BibitemOpen
  \bibfield  {author} {\bibinfo {author} {\bibfnamefont {C.}~\bibnamefont
  {Guo}}, \bibinfo {author} {\bibfnamefont {C.}~\bibnamefont {Putzke}},
  \bibinfo {author} {\bibfnamefont {S.}~\bibnamefont {Konyzheva}}, \bibinfo
  {author} {\bibfnamefont {X.}~\bibnamefont {Huang}}, \bibinfo {author}
  {\bibfnamefont {M.}~\bibnamefont {Gutierrez-Amigo}}, \bibinfo {author}
  {\bibfnamefont {I.}~\bibnamefont {Errea}}, \bibinfo {author} {\bibfnamefont
  {D.}~\bibnamefont {Chen}}, \bibinfo {author} {\bibfnamefont {M.~G.}\
  \bibnamefont {Vergniory}}, \bibinfo {author} {\bibfnamefont {C.}~\bibnamefont
  {Felser}}, \bibinfo {author} {\bibfnamefont {M.~H.}\ \bibnamefont {Fischer}},
  \emph {et~al.},\ }\bibfield  {title} {\bibinfo {title} {Switchable chiral
  transport in charge-ordered kagome metal csv3sb5},\ }\href@noop {} {\bibfield
   {journal} {\bibinfo  {journal} {Nature}\ }\textbf {\bibinfo {volume}
  {611}},\ \bibinfo {pages} {461} (\bibinfo {year} {2022})}\BibitemShut
  {NoStop}%
\bibitem [{\citenamefont {Xu}\ \emph {et~al.}(2022)\citenamefont {Xu},
  \citenamefont {Ni}, \citenamefont {Liu}, \citenamefont {Ortiz}, \citenamefont
  {Deng}, \citenamefont {Wilson}, \citenamefont {Yan}, \citenamefont
  {Balents},\ and\ \citenamefont {Wu}}]{xu2022three}%
  \BibitemOpen
  \bibfield  {author} {\bibinfo {author} {\bibfnamefont {Y.}~\bibnamefont
  {Xu}}, \bibinfo {author} {\bibfnamefont {Z.}~\bibnamefont {Ni}}, \bibinfo
  {author} {\bibfnamefont {Y.}~\bibnamefont {Liu}}, \bibinfo {author}
  {\bibfnamefont {B.~R.}\ \bibnamefont {Ortiz}}, \bibinfo {author}
  {\bibfnamefont {Q.}~\bibnamefont {Deng}}, \bibinfo {author} {\bibfnamefont
  {S.~D.}\ \bibnamefont {Wilson}}, \bibinfo {author} {\bibfnamefont
  {B.}~\bibnamefont {Yan}}, \bibinfo {author} {\bibfnamefont {L.}~\bibnamefont
  {Balents}},\ and\ \bibinfo {author} {\bibfnamefont {L.}~\bibnamefont {Wu}},\
  }\bibfield  {title} {\bibinfo {title} {Three-state nematicity and
  magneto-optical kerr effect in the charge density waves in kagome
  superconductors},\ }\href@noop {} {\bibfield  {journal} {\bibinfo  {journal}
  {Nature physics}\ }\textbf {\bibinfo {volume} {18}},\ \bibinfo {pages} {1470}
  (\bibinfo {year} {2022})}\BibitemShut {NoStop}%
\bibitem [{\citenamefont {Park}\ \emph {et~al.}(2021)\citenamefont {Park},
  \citenamefont {Ye},\ and\ \citenamefont {Balents}}]{park2021electronic}%
  \BibitemOpen
  \bibfield  {author} {\bibinfo {author} {\bibfnamefont {T.}~\bibnamefont
  {Park}}, \bibinfo {author} {\bibfnamefont {M.}~\bibnamefont {Ye}},\ and\
  \bibinfo {author} {\bibfnamefont {L.}~\bibnamefont {Balents}},\ }\bibfield
  {title} {\bibinfo {title} {Electronic instabilities of kagome metals: saddle
  points and landau theory},\ }\href@noop {} {\bibfield  {journal} {\bibinfo
  {journal} {Physical Review B}\ }\textbf {\bibinfo {volume} {104}},\ \bibinfo
  {pages} {035142} (\bibinfo {year} {2021})}\BibitemShut {NoStop}%
\bibitem [{\citenamefont {Feng}\ \emph {et~al.}(2021)\citenamefont {Feng},
  \citenamefont {Jiang}, \citenamefont {Wang},\ and\ \citenamefont
  {Hu}}]{feng2021chiral}%
  \BibitemOpen
  \bibfield  {author} {\bibinfo {author} {\bibfnamefont {X.}~\bibnamefont
  {Feng}}, \bibinfo {author} {\bibfnamefont {K.}~\bibnamefont {Jiang}},
  \bibinfo {author} {\bibfnamefont {Z.}~\bibnamefont {Wang}},\ and\ \bibinfo
  {author} {\bibfnamefont {J.}~\bibnamefont {Hu}},\ }\bibfield  {title}
  {\bibinfo {title} {Chiral flux phase in the kagome superconductor av3sb5},\
  }\href@noop {} {\bibfield  {journal} {\bibinfo  {journal} {Science bulletin}\
  }\textbf {\bibinfo {volume} {66}},\ \bibinfo {pages} {1384} (\bibinfo {year}
  {2021})}\BibitemShut {NoStop}%
\bibitem [{\citenamefont {Denner}\ \emph {et~al.}(2021)\citenamefont {Denner},
  \citenamefont {Thomale},\ and\ \citenamefont {Neupert}}]{denner2021analysis}%
  \BibitemOpen
  \bibfield  {author} {\bibinfo {author} {\bibfnamefont {M.~M.}\ \bibnamefont
  {Denner}}, \bibinfo {author} {\bibfnamefont {R.}~\bibnamefont {Thomale}},\
  and\ \bibinfo {author} {\bibfnamefont {T.}~\bibnamefont {Neupert}},\
  }\bibfield  {title} {\bibinfo {title} {Analysis of charge order in the kagome
  metal a v 3 sb 5 (a= k, rb, cs)},\ }\href@noop {} {\bibfield  {journal}
  {\bibinfo  {journal} {Physical Review Letters}\ }\textbf {\bibinfo {volume}
  {127}},\ \bibinfo {pages} {217601} (\bibinfo {year} {2021})}\BibitemShut
  {NoStop}%
\bibitem [{\citenamefont {Lin}\ and\ \citenamefont
  {Nandkishore}(2021)}]{lin2021complex}%
  \BibitemOpen
  \bibfield  {author} {\bibinfo {author} {\bibfnamefont {Y.-P.}\ \bibnamefont
  {Lin}}\ and\ \bibinfo {author} {\bibfnamefont {R.~M.}\ \bibnamefont
  {Nandkishore}},\ }\bibfield  {title} {\bibinfo {title} {Complex charge
  density waves at van hove singularity on hexagonal lattices: Haldane-model
  phase diagram and potential realization in the kagome metals a v 3 sb 5 (a=
  k, rb, cs)},\ }\href@noop {} {\bibfield  {journal} {\bibinfo  {journal}
  {Physical Review B}\ }\textbf {\bibinfo {volume} {104}},\ \bibinfo {pages}
  {045122} (\bibinfo {year} {2021})}\BibitemShut {NoStop}%
\bibitem [{\citenamefont {Guo}\ \emph {et~al.}(2024)\citenamefont {Guo},
  \citenamefont {Wagner}, \citenamefont {Putzke}, \citenamefont {Chen},
  \citenamefont {Wang}, \citenamefont {Zhang}, \citenamefont {Gutierrez-Amigo},
  \citenamefont {Errea}, \citenamefont {Vergniory}, \citenamefont {Felser}
  \emph {et~al.}}]{guo2024correlated}%
  \BibitemOpen
  \bibfield  {author} {\bibinfo {author} {\bibfnamefont {C.}~\bibnamefont
  {Guo}}, \bibinfo {author} {\bibfnamefont {G.}~\bibnamefont {Wagner}},
  \bibinfo {author} {\bibfnamefont {C.}~\bibnamefont {Putzke}}, \bibinfo
  {author} {\bibfnamefont {D.}~\bibnamefont {Chen}}, \bibinfo {author}
  {\bibfnamefont {K.}~\bibnamefont {Wang}}, \bibinfo {author} {\bibfnamefont
  {L.}~\bibnamefont {Zhang}}, \bibinfo {author} {\bibfnamefont
  {M.}~\bibnamefont {Gutierrez-Amigo}}, \bibinfo {author} {\bibfnamefont
  {I.}~\bibnamefont {Errea}}, \bibinfo {author} {\bibfnamefont {M.~G.}\
  \bibnamefont {Vergniory}}, \bibinfo {author} {\bibfnamefont {C.}~\bibnamefont
  {Felser}}, \emph {et~al.},\ }\bibfield  {title} {\bibinfo {title} {Correlated
  order at the tipping point in the kagome metal csv3sb5},\ }\href@noop {}
  {\bibfield  {journal} {\bibinfo  {journal} {Nature Physics}\ }\textbf
  {\bibinfo {volume} {20}},\ \bibinfo {pages} {579} (\bibinfo {year}
  {2024})}\BibitemShut {NoStop}%
\bibitem [{\citenamefont {Christensen}\ \emph {et~al.}(2022)\citenamefont
  {Christensen}, \citenamefont {Birol}, \citenamefont {Andersen},\ and\
  \citenamefont {Fernandes}}]{christensen2022loop}%
  \BibitemOpen
  \bibfield  {author} {\bibinfo {author} {\bibfnamefont {M.~H.}\ \bibnamefont
  {Christensen}}, \bibinfo {author} {\bibfnamefont {T.}~\bibnamefont {Birol}},
  \bibinfo {author} {\bibfnamefont {B.~M.}\ \bibnamefont {Andersen}},\ and\
  \bibinfo {author} {\bibfnamefont {R.~M.}\ \bibnamefont {Fernandes}},\
  }\bibfield  {title} {\bibinfo {title} {Loop currents in a v 3 sb 5 kagome
  metals: Multipolar and toroidal magnetic orders},\ }\href@noop {} {\bibfield
  {journal} {\bibinfo  {journal} {Physical Review B}\ }\textbf {\bibinfo
  {volume} {106}},\ \bibinfo {pages} {144504} (\bibinfo {year}
  {2022})}\BibitemShut {NoStop}%
\bibitem [{\citenamefont {Tazai}\ \emph {et~al.}(2024)\citenamefont {Tazai},
  \citenamefont {Yamakawa},\ and\ \citenamefont {Kontani}}]{tazai2024drastic}%
  \BibitemOpen
  \bibfield  {author} {\bibinfo {author} {\bibfnamefont {R.}~\bibnamefont
  {Tazai}}, \bibinfo {author} {\bibfnamefont {Y.}~\bibnamefont {Yamakawa}},\
  and\ \bibinfo {author} {\bibfnamefont {H.}~\bibnamefont {Kontani}},\
  }\bibfield  {title} {\bibinfo {title} {Drastic magnetic-field-induced chiral
  current order and emergent current-bond-field interplay in kagome metals},\
  }\href@noop {} {\bibfield  {journal} {\bibinfo  {journal} {Proceedings of the
  National Academy of Sciences}\ }\textbf {\bibinfo {volume} {121}},\ \bibinfo
  {pages} {e2303476121} (\bibinfo {year} {2024})}\BibitemShut {NoStop}%
\bibitem [{\citenamefont {Nie}\ \emph {et~al.}(2022)\citenamefont {Nie},
  \citenamefont {Sun}, \citenamefont {Ma}, \citenamefont {Song}, \citenamefont
  {Zheng}, \citenamefont {Liang}, \citenamefont {Wu}, \citenamefont {Yu},
  \citenamefont {Li}, \citenamefont {Shan} \emph {et~al.}}]{nie2022charge}%
  \BibitemOpen
  \bibfield  {author} {\bibinfo {author} {\bibfnamefont {L.}~\bibnamefont
  {Nie}}, \bibinfo {author} {\bibfnamefont {K.}~\bibnamefont {Sun}}, \bibinfo
  {author} {\bibfnamefont {W.}~\bibnamefont {Ma}}, \bibinfo {author}
  {\bibfnamefont {D.}~\bibnamefont {Song}}, \bibinfo {author} {\bibfnamefont
  {L.}~\bibnamefont {Zheng}}, \bibinfo {author} {\bibfnamefont
  {Z.}~\bibnamefont {Liang}}, \bibinfo {author} {\bibfnamefont
  {P.}~\bibnamefont {Wu}}, \bibinfo {author} {\bibfnamefont {F.}~\bibnamefont
  {Yu}}, \bibinfo {author} {\bibfnamefont {J.}~\bibnamefont {Li}}, \bibinfo
  {author} {\bibfnamefont {M.}~\bibnamefont {Shan}}, \emph {et~al.},\
  }\bibfield  {title} {\bibinfo {title} {Charge-density-wave-driven electronic
  nematicity in a kagome superconductor},\ }\href@noop {} {\bibfield  {journal}
  {\bibinfo  {journal} {Nature}\ }\textbf {\bibinfo {volume} {604}},\ \bibinfo
  {pages} {59} (\bibinfo {year} {2022})}\BibitemShut {NoStop}%
\bibitem [{\citenamefont {Li}\ \emph {et~al.}(2023)\citenamefont {Li},
  \citenamefont {Zhao}, \citenamefont {Ortiz}, \citenamefont {Oey},
  \citenamefont {Wang}, \citenamefont {Wilson},\ and\ \citenamefont
  {Zeljkovic}}]{li2023unidirectional}%
  \BibitemOpen
  \bibfield  {author} {\bibinfo {author} {\bibfnamefont {H.}~\bibnamefont
  {Li}}, \bibinfo {author} {\bibfnamefont {H.}~\bibnamefont {Zhao}}, \bibinfo
  {author} {\bibfnamefont {B.~R.}\ \bibnamefont {Ortiz}}, \bibinfo {author}
  {\bibfnamefont {Y.}~\bibnamefont {Oey}}, \bibinfo {author} {\bibfnamefont
  {Z.}~\bibnamefont {Wang}}, \bibinfo {author} {\bibfnamefont {S.~D.}\
  \bibnamefont {Wilson}},\ and\ \bibinfo {author} {\bibfnamefont
  {I.}~\bibnamefont {Zeljkovic}},\ }\bibfield  {title} {\bibinfo {title}
  {Unidirectional coherent quasiparticles in the high-temperature rotational
  symmetry broken phase of a v3sb5 kagome superconductors},\ }\href@noop {}
  {\bibfield  {journal} {\bibinfo  {journal} {Nature Physics}\ }\textbf
  {\bibinfo {volume} {19}},\ \bibinfo {pages} {637} (\bibinfo {year}
  {2023})}\BibitemShut {NoStop}%
\bibitem [{\citenamefont {Xiang}\ \emph {et~al.}(2021)\citenamefont {Xiang},
  \citenamefont {Li}, \citenamefont {Li}, \citenamefont {Xie}, \citenamefont
  {Yang}, \citenamefont {Wang}, \citenamefont {Yao},\ and\ \citenamefont
  {Wen}}]{xiang2021twofold}%
  \BibitemOpen
  \bibfield  {author} {\bibinfo {author} {\bibfnamefont {Y.}~\bibnamefont
  {Xiang}}, \bibinfo {author} {\bibfnamefont {Q.}~\bibnamefont {Li}}, \bibinfo
  {author} {\bibfnamefont {Y.}~\bibnamefont {Li}}, \bibinfo {author}
  {\bibfnamefont {W.}~\bibnamefont {Xie}}, \bibinfo {author} {\bibfnamefont
  {H.}~\bibnamefont {Yang}}, \bibinfo {author} {\bibfnamefont {Z.}~\bibnamefont
  {Wang}}, \bibinfo {author} {\bibfnamefont {Y.}~\bibnamefont {Yao}},\ and\
  \bibinfo {author} {\bibfnamefont {H.-H.}\ \bibnamefont {Wen}},\ }\bibfield
  {title} {\bibinfo {title} {Twofold symmetry of c-axis resistivity in
  topological kagome superconductor csv3sb5 with in-plane rotating magnetic
  field},\ }\href@noop {} {\bibfield  {journal} {\bibinfo  {journal} {Nature
  communications}\ }\textbf {\bibinfo {volume} {12}},\ \bibinfo {pages} {6727}
  (\bibinfo {year} {2021})}\BibitemShut {NoStop}%
\bibitem [{\citenamefont {Wulferding}\ \emph {et~al.}(2022)\citenamefont
  {Wulferding}, \citenamefont {Lee}, \citenamefont {Choi}, \citenamefont {Yin},
  \citenamefont {Tu}, \citenamefont {Gong}, \citenamefont {Lei}, \citenamefont
  {Yousuf}, \citenamefont {Song}, \citenamefont {Lee} \emph
  {et~al.}}]{wulferding2022emergent}%
  \BibitemOpen
  \bibfield  {author} {\bibinfo {author} {\bibfnamefont {D.}~\bibnamefont
  {Wulferding}}, \bibinfo {author} {\bibfnamefont {S.}~\bibnamefont {Lee}},
  \bibinfo {author} {\bibfnamefont {Y.}~\bibnamefont {Choi}}, \bibinfo {author}
  {\bibfnamefont {Q.}~\bibnamefont {Yin}}, \bibinfo {author} {\bibfnamefont
  {Z.}~\bibnamefont {Tu}}, \bibinfo {author} {\bibfnamefont {C.}~\bibnamefont
  {Gong}}, \bibinfo {author} {\bibfnamefont {H.}~\bibnamefont {Lei}}, \bibinfo
  {author} {\bibfnamefont {S.}~\bibnamefont {Yousuf}}, \bibinfo {author}
  {\bibfnamefont {J.}~\bibnamefont {Song}}, \bibinfo {author} {\bibfnamefont
  {H.}~\bibnamefont {Lee}}, \emph {et~al.},\ }\bibfield  {title} {\bibinfo
  {title} {Emergent nematicity and intrinsic versus extrinsic electronic
  scattering processes in the kagome metal csv 3 sb 5},\ }\href@noop {}
  {\bibfield  {journal} {\bibinfo  {journal} {Physical Review Research}\
  }\textbf {\bibinfo {volume} {4}},\ \bibinfo {pages} {023215} (\bibinfo {year}
  {2022})}\BibitemShut {NoStop}%
\bibitem [{\citenamefont {Asaba}\ \emph {et~al.}(2024)\citenamefont {Asaba},
  \citenamefont {Onishi}, \citenamefont {Kageyama}, \citenamefont {Kiyosue},
  \citenamefont {Ohtsuka}, \citenamefont {Suetsugu}, \citenamefont {Kohsaka},
  \citenamefont {Gaggl}, \citenamefont {Kasahara}, \citenamefont {Murayama}
  \emph {et~al.}}]{asaba2024evidence}%
  \BibitemOpen
  \bibfield  {author} {\bibinfo {author} {\bibfnamefont {T.}~\bibnamefont
  {Asaba}}, \bibinfo {author} {\bibfnamefont {A.}~\bibnamefont {Onishi}},
  \bibinfo {author} {\bibfnamefont {Y.}~\bibnamefont {Kageyama}}, \bibinfo
  {author} {\bibfnamefont {T.}~\bibnamefont {Kiyosue}}, \bibinfo {author}
  {\bibfnamefont {K.}~\bibnamefont {Ohtsuka}}, \bibinfo {author} {\bibfnamefont
  {S.}~\bibnamefont {Suetsugu}}, \bibinfo {author} {\bibfnamefont
  {Y.}~\bibnamefont {Kohsaka}}, \bibinfo {author} {\bibfnamefont
  {T.}~\bibnamefont {Gaggl}}, \bibinfo {author} {\bibfnamefont
  {Y.}~\bibnamefont {Kasahara}}, \bibinfo {author} {\bibfnamefont
  {H.}~\bibnamefont {Murayama}}, \emph {et~al.},\ }\bibfield  {title} {\bibinfo
  {title} {Evidence for an odd-parity nematic phase above the
  charge-density-wave transition in a kagome metal},\ }\href@noop {} {\bibfield
   {journal} {\bibinfo  {journal} {Nature Physics}\ }\textbf {\bibinfo {volume}
  {20}},\ \bibinfo {pages} {40} (\bibinfo {year} {2024})}\BibitemShut {NoStop}%
\bibitem [{\citenamefont {Liu}\ \emph {et~al.}(2024{\natexlab{a}})\citenamefont
  {Liu}, \citenamefont {Shi}, \citenamefont {Jiang}, \citenamefont {Rosenberg},
  \citenamefont {DeStefano}, \citenamefont {Liu}, \citenamefont {Hu},
  \citenamefont {Zhao}, \citenamefont {Wang}, \citenamefont {Yao} \emph
  {et~al.}}]{liu2024absence}%
  \BibitemOpen
  \bibfield  {author} {\bibinfo {author} {\bibfnamefont {Z.}~\bibnamefont
  {Liu}}, \bibinfo {author} {\bibfnamefont {Y.}~\bibnamefont {Shi}}, \bibinfo
  {author} {\bibfnamefont {Q.}~\bibnamefont {Jiang}}, \bibinfo {author}
  {\bibfnamefont {E.~W.}\ \bibnamefont {Rosenberg}}, \bibinfo {author}
  {\bibfnamefont {J.~M.}\ \bibnamefont {DeStefano}}, \bibinfo {author}
  {\bibfnamefont {J.}~\bibnamefont {Liu}}, \bibinfo {author} {\bibfnamefont
  {C.}~\bibnamefont {Hu}}, \bibinfo {author} {\bibfnamefont {Y.}~\bibnamefont
  {Zhao}}, \bibinfo {author} {\bibfnamefont {Z.}~\bibnamefont {Wang}}, \bibinfo
  {author} {\bibfnamefont {Y.}~\bibnamefont {Yao}}, \emph {et~al.},\ }\bibfield
   {title} {\bibinfo {title} {Absence of e 2 g nematic instability and dominant
  a 1 g response in the kagome metal csv 3 sb 5},\ }\href@noop {} {\bibfield
  {journal} {\bibinfo  {journal} {Physical Review X}\ }\textbf {\bibinfo
  {volume} {14}},\ \bibinfo {pages} {031015} (\bibinfo {year}
  {2024}{\natexlab{a}})}\BibitemShut {NoStop}%
\bibitem [{\citenamefont {Chen}\ \emph {et~al.}(2021)\citenamefont {Chen},
  \citenamefont {Yang}, \citenamefont {Hu}, \citenamefont {Zhao}, \citenamefont
  {Yuan}, \citenamefont {Xing}, \citenamefont {Qian}, \citenamefont {Huang},
  \citenamefont {Li}, \citenamefont {Ye} \emph {et~al.}}]{chen2021roton}%
  \BibitemOpen
  \bibfield  {author} {\bibinfo {author} {\bibfnamefont {H.}~\bibnamefont
  {Chen}}, \bibinfo {author} {\bibfnamefont {H.}~\bibnamefont {Yang}}, \bibinfo
  {author} {\bibfnamefont {B.}~\bibnamefont {Hu}}, \bibinfo {author}
  {\bibfnamefont {Z.}~\bibnamefont {Zhao}}, \bibinfo {author} {\bibfnamefont
  {J.}~\bibnamefont {Yuan}}, \bibinfo {author} {\bibfnamefont {Y.}~\bibnamefont
  {Xing}}, \bibinfo {author} {\bibfnamefont {G.}~\bibnamefont {Qian}}, \bibinfo
  {author} {\bibfnamefont {Z.}~\bibnamefont {Huang}}, \bibinfo {author}
  {\bibfnamefont {G.}~\bibnamefont {Li}}, \bibinfo {author} {\bibfnamefont
  {Y.}~\bibnamefont {Ye}}, \emph {et~al.},\ }\bibfield  {title} {\bibinfo
  {title} {Roton pair density wave in a strong-coupling kagome
  superconductor},\ }\href@noop {} {\bibfield  {journal} {\bibinfo  {journal}
  {Nature}\ }\textbf {\bibinfo {volume} {599}},\ \bibinfo {pages} {222}
  (\bibinfo {year} {2021})}\BibitemShut {NoStop}%
\bibitem [{\citenamefont {Liang}\ \emph {et~al.}(2021)\citenamefont {Liang},
  \citenamefont {Hou}, \citenamefont {Zhang}, \citenamefont {Ma}, \citenamefont
  {Wu}, \citenamefont {Zhang}, \citenamefont {Yu}, \citenamefont {Ying},
  \citenamefont {Jiang}, \citenamefont {Shan} \emph {et~al.}}]{liang2021three}%
  \BibitemOpen
  \bibfield  {author} {\bibinfo {author} {\bibfnamefont {Z.}~\bibnamefont
  {Liang}}, \bibinfo {author} {\bibfnamefont {X.}~\bibnamefont {Hou}}, \bibinfo
  {author} {\bibfnamefont {F.}~\bibnamefont {Zhang}}, \bibinfo {author}
  {\bibfnamefont {W.}~\bibnamefont {Ma}}, \bibinfo {author} {\bibfnamefont
  {P.}~\bibnamefont {Wu}}, \bibinfo {author} {\bibfnamefont {Z.}~\bibnamefont
  {Zhang}}, \bibinfo {author} {\bibfnamefont {F.}~\bibnamefont {Yu}}, \bibinfo
  {author} {\bibfnamefont {J.-J.}\ \bibnamefont {Ying}}, \bibinfo {author}
  {\bibfnamefont {K.}~\bibnamefont {Jiang}}, \bibinfo {author} {\bibfnamefont
  {L.}~\bibnamefont {Shan}}, \emph {et~al.},\ }\bibfield  {title} {\bibinfo
  {title} {Three-dimensional charge density wave and surface-dependent
  vortex-core states in a kagome superconductor csv 3 sb 5},\ }\href@noop {}
  {\bibfield  {journal} {\bibinfo  {journal} {Physical Review X}\ }\textbf
  {\bibinfo {volume} {11}},\ \bibinfo {pages} {031026} (\bibinfo {year}
  {2021})}\BibitemShut {NoStop}%
\bibitem [{\citenamefont {Li}\ \emph {et~al.}(2021)\citenamefont {Li},
  \citenamefont {Zhang}, \citenamefont {Yilmaz}, \citenamefont {Pai},
  \citenamefont {Marvinney}, \citenamefont {Said}, \citenamefont {Yin},
  \citenamefont {Gong}, \citenamefont {Tu}, \citenamefont {Vescovo} \emph
  {et~al.}}]{li2021observation}%
  \BibitemOpen
  \bibfield  {author} {\bibinfo {author} {\bibfnamefont {H.}~\bibnamefont
  {Li}}, \bibinfo {author} {\bibfnamefont {T.}~\bibnamefont {Zhang}}, \bibinfo
  {author} {\bibfnamefont {T.}~\bibnamefont {Yilmaz}}, \bibinfo {author}
  {\bibfnamefont {Y.}~\bibnamefont {Pai}}, \bibinfo {author} {\bibfnamefont
  {C.}~\bibnamefont {Marvinney}}, \bibinfo {author} {\bibfnamefont
  {A.}~\bibnamefont {Said}}, \bibinfo {author} {\bibfnamefont {Q.}~\bibnamefont
  {Yin}}, \bibinfo {author} {\bibfnamefont {C.}~\bibnamefont {Gong}}, \bibinfo
  {author} {\bibfnamefont {Z.}~\bibnamefont {Tu}}, \bibinfo {author}
  {\bibfnamefont {E.}~\bibnamefont {Vescovo}}, \emph {et~al.},\ }\bibfield
  {title} {\bibinfo {title} {Observation of unconventional charge density wave
  without acoustic phonon anomaly in kagome superconductors a v 3 sb 5 (a= rb,
  cs)},\ }\href@noop {} {\bibfield  {journal} {\bibinfo  {journal} {Physical
  Review X}\ }\textbf {\bibinfo {volume} {11}},\ \bibinfo {pages} {031050}
  (\bibinfo {year} {2021})}\BibitemShut {NoStop}%
\bibitem [{\citenamefont {Ortiz}\ \emph
  {et~al.}(2021{\natexlab{b}})\citenamefont {Ortiz}, \citenamefont {Teicher},
  \citenamefont {Kautzsch}, \citenamefont {Sarte}, \citenamefont {Ratcliff},
  \citenamefont {Harter}, \citenamefont {Ruff}, \citenamefont {Seshadri},\ and\
  \citenamefont {Wilson}}]{ortiz2021fermi}%
  \BibitemOpen
  \bibfield  {author} {\bibinfo {author} {\bibfnamefont {B.~R.}\ \bibnamefont
  {Ortiz}}, \bibinfo {author} {\bibfnamefont {S.~M.}\ \bibnamefont {Teicher}},
  \bibinfo {author} {\bibfnamefont {L.}~\bibnamefont {Kautzsch}}, \bibinfo
  {author} {\bibfnamefont {P.~M.}\ \bibnamefont {Sarte}}, \bibinfo {author}
  {\bibfnamefont {N.}~\bibnamefont {Ratcliff}}, \bibinfo {author}
  {\bibfnamefont {J.}~\bibnamefont {Harter}}, \bibinfo {author} {\bibfnamefont
  {J.~P.}\ \bibnamefont {Ruff}}, \bibinfo {author} {\bibfnamefont
  {R.}~\bibnamefont {Seshadri}},\ and\ \bibinfo {author} {\bibfnamefont
  {S.~D.}\ \bibnamefont {Wilson}},\ }\bibfield  {title} {\bibinfo {title}
  {Fermi surface mapping and the nature of charge-density-wave order in the
  kagome superconductor csv 3 sb 5},\ }\href@noop {} {\bibfield  {journal}
  {\bibinfo  {journal} {Physical Review X}\ }\textbf {\bibinfo {volume} {11}},\
  \bibinfo {pages} {041030} (\bibinfo {year} {2021}{\natexlab{b}})}\BibitemShut
  {NoStop}%
\bibitem [{\citenamefont {Christensen}\ \emph {et~al.}(2021)\citenamefont
  {Christensen}, \citenamefont {Birol}, \citenamefont {Andersen},\ and\
  \citenamefont {Fernandes}}]{christensen2021theory}%
  \BibitemOpen
  \bibfield  {author} {\bibinfo {author} {\bibfnamefont {M.~H.}\ \bibnamefont
  {Christensen}}, \bibinfo {author} {\bibfnamefont {T.}~\bibnamefont {Birol}},
  \bibinfo {author} {\bibfnamefont {B.~M.}\ \bibnamefont {Andersen}},\ and\
  \bibinfo {author} {\bibfnamefont {R.~M.}\ \bibnamefont {Fernandes}},\
  }\bibfield  {title} {\bibinfo {title} {Theory of the charge density wave in a
  v 3 sb 5 kagome metals},\ }\href@noop {} {\bibfield  {journal} {\bibinfo
  {journal} {Physical Review B}\ }\textbf {\bibinfo {volume} {104}},\ \bibinfo
  {pages} {214513} (\bibinfo {year} {2021})}\BibitemShut {NoStop}%
\bibitem [{\citenamefont {Ritz}\ \emph {et~al.}(2023)\citenamefont {Ritz},
  \citenamefont {Fernandes},\ and\ \citenamefont {Birol}}]{ritz2023impact}%
  \BibitemOpen
  \bibfield  {author} {\bibinfo {author} {\bibfnamefont {E.~T.}\ \bibnamefont
  {Ritz}}, \bibinfo {author} {\bibfnamefont {R.~M.}\ \bibnamefont
  {Fernandes}},\ and\ \bibinfo {author} {\bibfnamefont {T.}~\bibnamefont
  {Birol}},\ }\bibfield  {title} {\bibinfo {title} {Impact of sb degrees of
  freedom on the charge density wave phase diagram of the kagome metal csv 3 sb
  5},\ }\href@noop {} {\bibfield  {journal} {\bibinfo  {journal} {Physical
  Review B}\ }\textbf {\bibinfo {volume} {107}},\ \bibinfo {pages} {205131}
  (\bibinfo {year} {2023})}\BibitemShut {NoStop}%
\bibitem [{\citenamefont {Tan}\ \emph {et~al.}(2021)\citenamefont {Tan},
  \citenamefont {Liu}, \citenamefont {Wang},\ and\ \citenamefont
  {Yan}}]{tan2021charge}%
  \BibitemOpen
  \bibfield  {author} {\bibinfo {author} {\bibfnamefont {H.}~\bibnamefont
  {Tan}}, \bibinfo {author} {\bibfnamefont {Y.}~\bibnamefont {Liu}}, \bibinfo
  {author} {\bibfnamefont {Z.}~\bibnamefont {Wang}},\ and\ \bibinfo {author}
  {\bibfnamefont {B.}~\bibnamefont {Yan}},\ }\bibfield  {title} {\bibinfo
  {title} {Charge density waves and electronic properties of superconducting
  kagome metals},\ }\href@noop {} {\bibfield  {journal} {\bibinfo  {journal}
  {Physical review letters}\ }\textbf {\bibinfo {volume} {127}},\ \bibinfo
  {pages} {046401} (\bibinfo {year} {2021})}\BibitemShut {NoStop}%
\bibitem [{\citenamefont {Ptok}\ \emph {et~al.}(2022)\citenamefont {Ptok},
  \citenamefont {Kobia{\l}ka}, \citenamefont {Sternik}, \citenamefont
  {{\L}a{\.z}ewski}, \citenamefont {Jochym}, \citenamefont {Ole{\'s}},\ and\
  \citenamefont {Piekarz}}]{ptok2022dynamical}%
  \BibitemOpen
  \bibfield  {author} {\bibinfo {author} {\bibfnamefont {A.}~\bibnamefont
  {Ptok}}, \bibinfo {author} {\bibfnamefont {A.}~\bibnamefont {Kobia{\l}ka}},
  \bibinfo {author} {\bibfnamefont {M.}~\bibnamefont {Sternik}}, \bibinfo
  {author} {\bibfnamefont {J.}~\bibnamefont {{\L}a{\.z}ewski}}, \bibinfo
  {author} {\bibfnamefont {P.~T.}\ \bibnamefont {Jochym}}, \bibinfo {author}
  {\bibfnamefont {A.~M.}\ \bibnamefont {Ole{\'s}}},\ and\ \bibinfo {author}
  {\bibfnamefont {P.}~\bibnamefont {Piekarz}},\ }\bibfield  {title} {\bibinfo
  {title} {Dynamical study of the origin of the charge density wave in av 3 sb
  5 (a= k, rb, cs) compounds},\ }\href@noop {} {\bibfield  {journal} {\bibinfo
  {journal} {Physical Review B}\ }\textbf {\bibinfo {volume} {105}},\ \bibinfo
  {pages} {235134} (\bibinfo {year} {2022})}\BibitemShut {NoStop}%
\bibitem [{\citenamefont {Deng}\ \emph {et~al.}(2025)\citenamefont {Deng},
  \citenamefont {Tan}, \citenamefont {Ortiz}, \citenamefont {Wilson},
  \citenamefont {Yan},\ and\ \citenamefont {Wu}}]{deng2025revealing}%
  \BibitemOpen
  \bibfield  {author} {\bibinfo {author} {\bibfnamefont {Q.}~\bibnamefont
  {Deng}}, \bibinfo {author} {\bibfnamefont {H.}~\bibnamefont {Tan}}, \bibinfo
  {author} {\bibfnamefont {B.~R.}\ \bibnamefont {Ortiz}}, \bibinfo {author}
  {\bibfnamefont {S.~D.}\ \bibnamefont {Wilson}}, \bibinfo {author}
  {\bibfnamefont {B.}~\bibnamefont {Yan}},\ and\ \bibinfo {author}
  {\bibfnamefont {L.}~\bibnamefont {Wu}},\ }\bibfield  {title} {\bibinfo
  {title} {Revealing rotational symmetry breaking charge density wave order in
  the kagome superconductor (rb, k) v 3 sb 5 by ultrafast pump-probe
  experiments},\ }\href@noop {} {\bibfield  {journal} {\bibinfo  {journal}
  {Physical Review B}\ }\textbf {\bibinfo {volume} {111}},\ \bibinfo {pages}
  {165134} (\bibinfo {year} {2025})}\BibitemShut {NoStop}%
\bibitem [{\citenamefont {Kang}\ \emph {et~al.}(2022)\citenamefont {Kang},
  \citenamefont {Fang}, \citenamefont {Yoo}, \citenamefont {Ortiz},
  \citenamefont {Oey}, \citenamefont {Choi}, \citenamefont {Ryu}, \citenamefont
  {Kim}, \citenamefont {Jozwiak}, \citenamefont {Bostwick} \emph
  {et~al.}}]{kang2022charge}%
  \BibitemOpen
  \bibfield  {author} {\bibinfo {author} {\bibfnamefont {M.}~\bibnamefont
  {Kang}}, \bibinfo {author} {\bibfnamefont {S.}~\bibnamefont {Fang}}, \bibinfo
  {author} {\bibfnamefont {J.}~\bibnamefont {Yoo}}, \bibinfo {author}
  {\bibfnamefont {B.~R.}\ \bibnamefont {Ortiz}}, \bibinfo {author}
  {\bibfnamefont {Y.~M.}\ \bibnamefont {Oey}}, \bibinfo {author} {\bibfnamefont
  {J.}~\bibnamefont {Choi}}, \bibinfo {author} {\bibfnamefont {S.~H.}\
  \bibnamefont {Ryu}}, \bibinfo {author} {\bibfnamefont {J.}~\bibnamefont
  {Kim}}, \bibinfo {author} {\bibfnamefont {C.}~\bibnamefont {Jozwiak}},
  \bibinfo {author} {\bibfnamefont {A.}~\bibnamefont {Bostwick}}, \emph
  {et~al.},\ }\bibfield  {title} {\bibinfo {title} {Charge order landscape and
  competition with superconductivity in kagome metals},\ }\href@noop {}
  {\bibfield  {journal} {\bibinfo  {journal} {Nature Materials}\ }\textbf
  {\bibinfo {volume} {22}},\ \bibinfo {pages} {186} (\bibinfo {year}
  {2022})}\BibitemShut {NoStop}%
\bibitem [{\citenamefont {Subires}\ \emph {et~al.}(2023)\citenamefont
  {Subires}, \citenamefont {Korshunov}, \citenamefont {Said}, \citenamefont
  {S{\'a}nchez}, \citenamefont {Ortiz}, \citenamefont {Wilson}, \citenamefont
  {Bosak},\ and\ \citenamefont {Blanco-Canosa}}]{subires2023order}%
  \BibitemOpen
  \bibfield  {author} {\bibinfo {author} {\bibfnamefont {D.}~\bibnamefont
  {Subires}}, \bibinfo {author} {\bibfnamefont {A.}~\bibnamefont {Korshunov}},
  \bibinfo {author} {\bibfnamefont {A.}~\bibnamefont {Said}}, \bibinfo {author}
  {\bibfnamefont {L.}~\bibnamefont {S{\'a}nchez}}, \bibinfo {author}
  {\bibfnamefont {B.~R.}\ \bibnamefont {Ortiz}}, \bibinfo {author}
  {\bibfnamefont {S.~D.}\ \bibnamefont {Wilson}}, \bibinfo {author}
  {\bibfnamefont {A.}~\bibnamefont {Bosak}},\ and\ \bibinfo {author}
  {\bibfnamefont {S.}~\bibnamefont {Blanco-Canosa}},\ }\bibfield  {title}
  {\bibinfo {title} {Order-disorder charge density wave instability in the
  kagome metal (cs, rb) v3sb5},\ }\href@noop {} {\bibfield  {journal} {\bibinfo
   {journal} {Nature Communications}\ }\textbf {\bibinfo {volume} {14}},\
  \bibinfo {pages} {1015} (\bibinfo {year} {2023})}\BibitemShut {NoStop}%
\bibitem [{\citenamefont {Hu}\ \emph {et~al.}(2022)\citenamefont {Hu},
  \citenamefont {Wu}, \citenamefont {Ortiz}, \citenamefont {Han}, \citenamefont
  {Plumb}, \citenamefont {Wilson}, \citenamefont {Schnyder}, \citenamefont
  {Shi} \emph {et~al.}}]{hu2022coexistence}%
  \BibitemOpen
  \bibfield  {author} {\bibinfo {author} {\bibfnamefont {Y.}~\bibnamefont
  {Hu}}, \bibinfo {author} {\bibfnamefont {X.}~\bibnamefont {Wu}}, \bibinfo
  {author} {\bibfnamefont {B.~R.}\ \bibnamefont {Ortiz}}, \bibinfo {author}
  {\bibfnamefont {X.}~\bibnamefont {Han}}, \bibinfo {author} {\bibfnamefont
  {N.~C.}\ \bibnamefont {Plumb}}, \bibinfo {author} {\bibfnamefont {S.~D.}\
  \bibnamefont {Wilson}}, \bibinfo {author} {\bibfnamefont {A.~P.}\
  \bibnamefont {Schnyder}}, \bibinfo {author} {\bibfnamefont {M.}~\bibnamefont
  {Shi}}, \emph {et~al.},\ }\bibfield  {title} {\bibinfo {title} {Coexistence
  of trihexagonal and star-of-david pattern in the charge density wave of the
  kagome superconductor a v 3 sb 5},\ }\href@noop {} {\bibfield  {journal}
  {\bibinfo  {journal} {Physical Review B}\ }\textbf {\bibinfo {volume}
  {106}},\ \bibinfo {pages} {L241106} (\bibinfo {year} {2022})}\BibitemShut
  {NoStop}%
\bibitem [{\citenamefont {Li}\ \emph {et~al.}(2022)\citenamefont {Li},
  \citenamefont {Wu}, \citenamefont {Liu}, \citenamefont {Polley},
  \citenamefont {Guo}, \citenamefont {Wang}, \citenamefont {Han}, \citenamefont
  {Dendzik}, \citenamefont {Berntsen}, \citenamefont {Thiagarajan} \emph
  {et~al.}}]{li2022coexistence}%
  \BibitemOpen
  \bibfield  {author} {\bibinfo {author} {\bibfnamefont {C.}~\bibnamefont
  {Li}}, \bibinfo {author} {\bibfnamefont {X.}~\bibnamefont {Wu}}, \bibinfo
  {author} {\bibfnamefont {H.}~\bibnamefont {Liu}}, \bibinfo {author}
  {\bibfnamefont {C.}~\bibnamefont {Polley}}, \bibinfo {author} {\bibfnamefont
  {Q.}~\bibnamefont {Guo}}, \bibinfo {author} {\bibfnamefont {Y.}~\bibnamefont
  {Wang}}, \bibinfo {author} {\bibfnamefont {X.}~\bibnamefont {Han}}, \bibinfo
  {author} {\bibfnamefont {M.}~\bibnamefont {Dendzik}}, \bibinfo {author}
  {\bibfnamefont {M.~H.}\ \bibnamefont {Berntsen}}, \bibinfo {author}
  {\bibfnamefont {B.}~\bibnamefont {Thiagarajan}}, \emph {et~al.},\ }\bibfield
  {title} {\bibinfo {title} {Coexistence of two intertwined charge density
  waves in a kagome system},\ }\href@noop {} {\bibfield  {journal} {\bibinfo
  {journal} {Physical Review Research}\ }\textbf {\bibinfo {volume} {4}},\
  \bibinfo {pages} {033072} (\bibinfo {year} {2022})}\BibitemShut {NoStop}%
\bibitem [{\citenamefont {Azoury}\ \emph {et~al.}(2023)\citenamefont {Azoury},
  \citenamefont {von Hoegen}, \citenamefont {Su}, \citenamefont {Oh},
  \citenamefont {Holder}, \citenamefont {Tan}, \citenamefont {Ortiz},
  \citenamefont {Capa~Salinas}, \citenamefont {Wilson}, \citenamefont {Yan}
  \emph {et~al.}}]{azoury2023direct}%
  \BibitemOpen
  \bibfield  {author} {\bibinfo {author} {\bibfnamefont {D.}~\bibnamefont
  {Azoury}}, \bibinfo {author} {\bibfnamefont {A.}~\bibnamefont {von Hoegen}},
  \bibinfo {author} {\bibfnamefont {Y.}~\bibnamefont {Su}}, \bibinfo {author}
  {\bibfnamefont {K.~H.}\ \bibnamefont {Oh}}, \bibinfo {author} {\bibfnamefont
  {T.}~\bibnamefont {Holder}}, \bibinfo {author} {\bibfnamefont
  {H.}~\bibnamefont {Tan}}, \bibinfo {author} {\bibfnamefont {B.~R.}\
  \bibnamefont {Ortiz}}, \bibinfo {author} {\bibfnamefont {A.}~\bibnamefont
  {Capa~Salinas}}, \bibinfo {author} {\bibfnamefont {S.~D.}\ \bibnamefont
  {Wilson}}, \bibinfo {author} {\bibfnamefont {B.}~\bibnamefont {Yan}}, \emph
  {et~al.},\ }\bibfield  {title} {\bibinfo {title} {Direct observation of the
  collective modes of the charge density wave in the kagome metal csv3sb5},\
  }\href@noop {} {\bibfield  {journal} {\bibinfo  {journal} {Proceedings of the
  National Academy of Sciences}\ }\textbf {\bibinfo {volume} {120}},\ \bibinfo
  {pages} {e2308588120} (\bibinfo {year} {2023})}\BibitemShut {NoStop}%
\bibitem [{\citenamefont {Miao}\ \emph {et~al.}(2021)\citenamefont {Miao},
  \citenamefont {Li}, \citenamefont {Meier}, \citenamefont {Huon},
  \citenamefont {Lee}, \citenamefont {Said}, \citenamefont {Lei}, \citenamefont
  {Ortiz}, \citenamefont {Wilson}, \citenamefont {Yin} \emph
  {et~al.}}]{miao2021geometry}%
  \BibitemOpen
  \bibfield  {author} {\bibinfo {author} {\bibfnamefont {H.}~\bibnamefont
  {Miao}}, \bibinfo {author} {\bibfnamefont {H.~X.}\ \bibnamefont {Li}},
  \bibinfo {author} {\bibfnamefont {W.}~\bibnamefont {Meier}}, \bibinfo
  {author} {\bibfnamefont {A.}~\bibnamefont {Huon}}, \bibinfo {author}
  {\bibfnamefont {H.~N.}\ \bibnamefont {Lee}}, \bibinfo {author} {\bibfnamefont
  {A.}~\bibnamefont {Said}}, \bibinfo {author} {\bibfnamefont {H.}~\bibnamefont
  {Lei}}, \bibinfo {author} {\bibfnamefont {B.}~\bibnamefont {Ortiz}}, \bibinfo
  {author} {\bibfnamefont {S.}~\bibnamefont {Wilson}}, \bibinfo {author}
  {\bibfnamefont {J.}~\bibnamefont {Yin}}, \emph {et~al.},\ }\bibfield  {title}
  {\bibinfo {title} {Geometry of the charge density wave in the kagome metal a
  v 3 sb 5},\ }\href@noop {} {\bibfield  {journal} {\bibinfo  {journal}
  {Physical Review B}\ }\textbf {\bibinfo {volume} {104}},\ \bibinfo {pages}
  {195132} (\bibinfo {year} {2021})}\BibitemShut {NoStop}%
\bibitem [{\citenamefont {Ratcliff}\ \emph {et~al.}(2021)\citenamefont
  {Ratcliff}, \citenamefont {Hallett}, \citenamefont {Ortiz}, \citenamefont
  {Wilson},\ and\ \citenamefont {Harter}}]{ratcliff2021coherent}%
  \BibitemOpen
  \bibfield  {author} {\bibinfo {author} {\bibfnamefont {N.}~\bibnamefont
  {Ratcliff}}, \bibinfo {author} {\bibfnamefont {L.}~\bibnamefont {Hallett}},
  \bibinfo {author} {\bibfnamefont {B.~R.}\ \bibnamefont {Ortiz}}, \bibinfo
  {author} {\bibfnamefont {S.~D.}\ \bibnamefont {Wilson}},\ and\ \bibinfo
  {author} {\bibfnamefont {J.~W.}\ \bibnamefont {Harter}},\ }\bibfield  {title}
  {\bibinfo {title} {Coherent phonon spectroscopy and interlayer modulation of
  charge density wave order in the kagome metal csv 3 sb 5},\ }\href@noop {}
  {\bibfield  {journal} {\bibinfo  {journal} {Physical Review Materials}\
  }\textbf {\bibinfo {volume} {5}},\ \bibinfo {pages} {L111801} (\bibinfo
  {year} {2021})}\BibitemShut {NoStop}%
\bibitem [{\citenamefont {Xiao}\ \emph {et~al.}(2023)\citenamefont {Xiao},
  \citenamefont {Lin}, \citenamefont {Li}, \citenamefont {Zheng}, \citenamefont
  {Francoual}, \citenamefont {Plueckthun}, \citenamefont {Xia}, \citenamefont
  {Qiu}, \citenamefont {Zhang}, \citenamefont {Guo} \emph
  {et~al.}}]{xiao2023coexistence}%
  \BibitemOpen
  \bibfield  {author} {\bibinfo {author} {\bibfnamefont {Q.}~\bibnamefont
  {Xiao}}, \bibinfo {author} {\bibfnamefont {Y.}~\bibnamefont {Lin}}, \bibinfo
  {author} {\bibfnamefont {Q.}~\bibnamefont {Li}}, \bibinfo {author}
  {\bibfnamefont {X.}~\bibnamefont {Zheng}}, \bibinfo {author} {\bibfnamefont
  {S.}~\bibnamefont {Francoual}}, \bibinfo {author} {\bibfnamefont
  {C.}~\bibnamefont {Plueckthun}}, \bibinfo {author} {\bibfnamefont
  {W.}~\bibnamefont {Xia}}, \bibinfo {author} {\bibfnamefont {Q.}~\bibnamefont
  {Qiu}}, \bibinfo {author} {\bibfnamefont {S.}~\bibnamefont {Zhang}}, \bibinfo
  {author} {\bibfnamefont {Y.}~\bibnamefont {Guo}}, \emph {et~al.},\ }\bibfield
   {title} {\bibinfo {title} {Coexistence of multiple stacking charge density
  waves in kagome superconductor csv 3 sb 5},\ }\href@noop {} {\bibfield
  {journal} {\bibinfo  {journal} {Physical Review Research}\ }\textbf {\bibinfo
  {volume} {5}},\ \bibinfo {pages} {L012032} (\bibinfo {year}
  {2023})}\BibitemShut {NoStop}%
\bibitem [{\citenamefont {Wang}\ \emph {et~al.}(2023)\citenamefont {Wang},
  \citenamefont {Wu}, \citenamefont {Li}, \citenamefont {Jiang},\ and\
  \citenamefont {Hu}}]{wang2023structure}%
  \BibitemOpen
  \bibfield  {author} {\bibinfo {author} {\bibfnamefont {Y.}~\bibnamefont
  {Wang}}, \bibinfo {author} {\bibfnamefont {T.}~\bibnamefont {Wu}}, \bibinfo
  {author} {\bibfnamefont {Z.}~\bibnamefont {Li}}, \bibinfo {author}
  {\bibfnamefont {K.}~\bibnamefont {Jiang}},\ and\ \bibinfo {author}
  {\bibfnamefont {J.}~\bibnamefont {Hu}},\ }\bibfield  {title} {\bibinfo
  {title} {Structure of the kagome superconductor csv 3 sb 5 in the charge
  density wave state},\ }\href@noop {} {\bibfield  {journal} {\bibinfo
  {journal} {Physical Review B}\ }\textbf {\bibinfo {volume} {107}},\ \bibinfo
  {pages} {184106} (\bibinfo {year} {2023})}\BibitemShut {NoStop}%
\bibitem [{\citenamefont {Jin}\ \emph {et~al.}(2024)\citenamefont {Jin},
  \citenamefont {Ren}, \citenamefont {Tan}, \citenamefont {Xie}, \citenamefont
  {Lu}, \citenamefont {Zhang}, \citenamefont {Ji},\ and\ \citenamefont
  {Zhang}}]{jin2024pi}%
  \BibitemOpen
  \bibfield  {author} {\bibinfo {author} {\bibfnamefont {F.}~\bibnamefont
  {Jin}}, \bibinfo {author} {\bibfnamefont {W.}~\bibnamefont {Ren}}, \bibinfo
  {author} {\bibfnamefont {M.}~\bibnamefont {Tan}}, \bibinfo {author}
  {\bibfnamefont {M.}~\bibnamefont {Xie}}, \bibinfo {author} {\bibfnamefont
  {B.}~\bibnamefont {Lu}}, \bibinfo {author} {\bibfnamefont {Z.}~\bibnamefont
  {Zhang}}, \bibinfo {author} {\bibfnamefont {J.}~\bibnamefont {Ji}},\ and\
  \bibinfo {author} {\bibfnamefont {Q.}~\bibnamefont {Zhang}},\ }\bibfield
  {title} {\bibinfo {title} {$\pi$ phase interlayer shift and stacking fault in
  the kagome superconductor csv 3 sb 5},\ }\href@noop {} {\bibfield  {journal}
  {\bibinfo  {journal} {Physical Review Letters}\ }\textbf {\bibinfo {volume}
  {132}},\ \bibinfo {pages} {066501} (\bibinfo {year} {2024})}\BibitemShut
  {NoStop}%
\bibitem [{\citenamefont {Kautzsch}\ \emph
  {et~al.}(2023{\natexlab{a}})\citenamefont {Kautzsch}, \citenamefont {Ortiz},
  \citenamefont {Mallayya}, \citenamefont {Plumb}, \citenamefont {Pokharel},
  \citenamefont {Ruff}, \citenamefont {Islam}, \citenamefont {Kim},
  \citenamefont {Seshadri},\ and\ \citenamefont
  {Wilson}}]{kautzsch2023structural}%
  \BibitemOpen
  \bibfield  {author} {\bibinfo {author} {\bibfnamefont {L.}~\bibnamefont
  {Kautzsch}}, \bibinfo {author} {\bibfnamefont {B.~R.}\ \bibnamefont {Ortiz}},
  \bibinfo {author} {\bibfnamefont {K.}~\bibnamefont {Mallayya}}, \bibinfo
  {author} {\bibfnamefont {J.}~\bibnamefont {Plumb}}, \bibinfo {author}
  {\bibfnamefont {G.}~\bibnamefont {Pokharel}}, \bibinfo {author}
  {\bibfnamefont {J.~P.}\ \bibnamefont {Ruff}}, \bibinfo {author}
  {\bibfnamefont {Z.}~\bibnamefont {Islam}}, \bibinfo {author} {\bibfnamefont
  {E.-A.}\ \bibnamefont {Kim}}, \bibinfo {author} {\bibfnamefont
  {R.}~\bibnamefont {Seshadri}},\ and\ \bibinfo {author} {\bibfnamefont
  {S.~D.}\ \bibnamefont {Wilson}},\ }\bibfield  {title} {\bibinfo {title}
  {Structural evolution of the kagome superconductors a v 3 sb 5 (a= k, rb, and
  cs) through charge density wave order},\ }\href@noop {} {\bibfield  {journal}
  {\bibinfo  {journal} {Physical Review Materials}\ }\textbf {\bibinfo {volume}
  {7}},\ \bibinfo {pages} {024806} (\bibinfo {year}
  {2023}{\natexlab{a}})}\BibitemShut {NoStop}%
\bibitem [{\citenamefont {Alkorta}\ \emph {et~al.}(2025)\citenamefont
  {Alkorta}, \citenamefont {Gutierrez-Amigo}, \citenamefont {Guo},
  \citenamefont {Moll}, \citenamefont {Vergniory},\ and\ \citenamefont
  {Errea}}]{alkorta2025symmetry}%
  \BibitemOpen
  \bibfield  {author} {\bibinfo {author} {\bibfnamefont {M.}~\bibnamefont
  {Alkorta}}, \bibinfo {author} {\bibfnamefont {M.}~\bibnamefont
  {Gutierrez-Amigo}}, \bibinfo {author} {\bibfnamefont {C.}~\bibnamefont
  {Guo}}, \bibinfo {author} {\bibfnamefont {P.~J.}\ \bibnamefont {Moll}},
  \bibinfo {author} {\bibfnamefont {M.~G.}\ \bibnamefont {Vergniory}},\ and\
  \bibinfo {author} {\bibfnamefont {I.}~\bibnamefont {Errea}},\ }\bibfield
  {title} {\bibinfo {title} {Symmetry-broken charge-ordered ground state in csv
  $ \_3 $ sb $ \_5 $ kagome metal},\ }\href@noop {} {\bibfield  {journal}
  {\bibinfo  {journal} {arXiv preprint arXiv:2505.19686}\ } (\bibinfo {year}
  {2025})}\BibitemShut {NoStop}%
\bibitem [{\citenamefont {Stahl}\ \emph {et~al.}(2022)\citenamefont {Stahl},
  \citenamefont {Chen}, \citenamefont {Ritschel}, \citenamefont {Shekhar},
  \citenamefont {Sadrollahi}, \citenamefont {Rahn}, \citenamefont {Ivashko},
  \citenamefont {Zimmermann}, \citenamefont {Felser},\ and\ \citenamefont
  {Geck}}]{stahl2022temperature}%
  \BibitemOpen
  \bibfield  {author} {\bibinfo {author} {\bibfnamefont {Q.}~\bibnamefont
  {Stahl}}, \bibinfo {author} {\bibfnamefont {D.}~\bibnamefont {Chen}},
  \bibinfo {author} {\bibfnamefont {T.}~\bibnamefont {Ritschel}}, \bibinfo
  {author} {\bibfnamefont {C.}~\bibnamefont {Shekhar}}, \bibinfo {author}
  {\bibfnamefont {E.}~\bibnamefont {Sadrollahi}}, \bibinfo {author}
  {\bibfnamefont {M.}~\bibnamefont {Rahn}}, \bibinfo {author} {\bibfnamefont
  {O.}~\bibnamefont {Ivashko}}, \bibinfo {author} {\bibfnamefont {M.~v.}\
  \bibnamefont {Zimmermann}}, \bibinfo {author} {\bibfnamefont
  {C.}~\bibnamefont {Felser}},\ and\ \bibinfo {author} {\bibfnamefont
  {J.}~\bibnamefont {Geck}},\ }\bibfield  {title} {\bibinfo {title}
  {Temperature-driven reorganization of electronic order in csv 3 sb 5},\
  }\href@noop {} {\bibfield  {journal} {\bibinfo  {journal} {Physical Review
  B}\ }\textbf {\bibinfo {volume} {105}},\ \bibinfo {pages} {195136} (\bibinfo
  {year} {2022})}\BibitemShut {NoStop}%
\bibitem [{\citenamefont {Plumb}\ \emph {et~al.}(2024)\citenamefont {Plumb},
  \citenamefont {Salinas}, \citenamefont {Mallayya}, \citenamefont {Kisiel},
  \citenamefont {Carneiro}, \citenamefont {Gomez}, \citenamefont {Pokharel},
  \citenamefont {Kim}, \citenamefont {Sarker}, \citenamefont {Islam} \emph
  {et~al.}}]{plumb2024phase}%
  \BibitemOpen
  \bibfield  {author} {\bibinfo {author} {\bibfnamefont {J.}~\bibnamefont
  {Plumb}}, \bibinfo {author} {\bibfnamefont {A.~C.}\ \bibnamefont {Salinas}},
  \bibinfo {author} {\bibfnamefont {K.}~\bibnamefont {Mallayya}}, \bibinfo
  {author} {\bibfnamefont {E.}~\bibnamefont {Kisiel}}, \bibinfo {author}
  {\bibfnamefont {F.~B.}\ \bibnamefont {Carneiro}}, \bibinfo {author}
  {\bibfnamefont {R.}~\bibnamefont {Gomez}}, \bibinfo {author} {\bibfnamefont
  {G.}~\bibnamefont {Pokharel}}, \bibinfo {author} {\bibfnamefont {E.-A.}\
  \bibnamefont {Kim}}, \bibinfo {author} {\bibfnamefont {S.}~\bibnamefont
  {Sarker}}, \bibinfo {author} {\bibfnamefont {Z.}~\bibnamefont {Islam}}, \emph
  {et~al.},\ }\bibfield  {title} {\bibinfo {title} {Phase-separated charge
  order and twinning across length scales in csv 3 sb 5},\ }\href@noop {}
  {\bibfield  {journal} {\bibinfo  {journal} {Physical Review Materials}\
  }\textbf {\bibinfo {volume} {8}},\ \bibinfo {pages} {093601} (\bibinfo {year}
  {2024})}\BibitemShut {NoStop}%
\bibitem [{\citenamefont {Wu}\ \emph {et~al.}(2022{\natexlab{a}})\citenamefont
  {Wu}, \citenamefont {Wang}, \citenamefont {Liu}, \citenamefont {Li},
  \citenamefont {Xu}, \citenamefont {Yin}, \citenamefont {Gong}, \citenamefont
  {Tu}, \citenamefont {Lei}, \citenamefont {Dong} \emph
  {et~al.}}]{wu2022simultaneous}%
  \BibitemOpen
  \bibfield  {author} {\bibinfo {author} {\bibfnamefont {Q.}~\bibnamefont
  {Wu}}, \bibinfo {author} {\bibfnamefont {Z.}~\bibnamefont {Wang}}, \bibinfo
  {author} {\bibfnamefont {Q.}~\bibnamefont {Liu}}, \bibinfo {author}
  {\bibfnamefont {R.}~\bibnamefont {Li}}, \bibinfo {author} {\bibfnamefont
  {S.}~\bibnamefont {Xu}}, \bibinfo {author} {\bibfnamefont {Q.}~\bibnamefont
  {Yin}}, \bibinfo {author} {\bibfnamefont {C.}~\bibnamefont {Gong}}, \bibinfo
  {author} {\bibfnamefont {Z.}~\bibnamefont {Tu}}, \bibinfo {author}
  {\bibfnamefont {H.}~\bibnamefont {Lei}}, \bibinfo {author} {\bibfnamefont
  {T.}~\bibnamefont {Dong}}, \emph {et~al.},\ }\bibfield  {title} {\bibinfo
  {title} {Simultaneous formation of two-fold rotation symmetry with charge
  order in the kagome superconductor csv 3 sb 5 by optical polarization
  rotation measurement},\ }\href@noop {} {\bibfield  {journal} {\bibinfo
  {journal} {Physical Review B}\ }\textbf {\bibinfo {volume} {106}},\ \bibinfo
  {pages} {205109} (\bibinfo {year} {2022}{\natexlab{a}})}\BibitemShut
  {NoStop}%
\bibitem [{\citenamefont {Zhao}\ \emph {et~al.}(2021)\citenamefont {Zhao},
  \citenamefont {Li}, \citenamefont {Ortiz}, \citenamefont {Teicher},
  \citenamefont {Park}, \citenamefont {Ye}, \citenamefont {Wang}, \citenamefont
  {Balents}, \citenamefont {Wilson},\ and\ \citenamefont
  {Zeljkovic}}]{zhao2021cascade}%
  \BibitemOpen
  \bibfield  {author} {\bibinfo {author} {\bibfnamefont {H.}~\bibnamefont
  {Zhao}}, \bibinfo {author} {\bibfnamefont {H.}~\bibnamefont {Li}}, \bibinfo
  {author} {\bibfnamefont {B.~R.}\ \bibnamefont {Ortiz}}, \bibinfo {author}
  {\bibfnamefont {S.~M.}\ \bibnamefont {Teicher}}, \bibinfo {author}
  {\bibfnamefont {T.}~\bibnamefont {Park}}, \bibinfo {author} {\bibfnamefont
  {M.}~\bibnamefont {Ye}}, \bibinfo {author} {\bibfnamefont {Z.}~\bibnamefont
  {Wang}}, \bibinfo {author} {\bibfnamefont {L.}~\bibnamefont {Balents}},
  \bibinfo {author} {\bibfnamefont {S.~D.}\ \bibnamefont {Wilson}},\ and\
  \bibinfo {author} {\bibfnamefont {I.}~\bibnamefont {Zeljkovic}},\ }\bibfield
  {title} {\bibinfo {title} {Cascade of correlated electron states in the
  kagome superconductor csv3sb5},\ }\href@noop {} {\bibfield  {journal}
  {\bibinfo  {journal} {Nature}\ }\textbf {\bibinfo {volume} {599}},\ \bibinfo
  {pages} {216} (\bibinfo {year} {2021})}\BibitemShut {NoStop}%
\bibitem [{\citenamefont {Chen}\ \emph {et~al.}(2022)\citenamefont {Chen},
  \citenamefont {Chen}, \citenamefont {Schnelle}, \citenamefont {Felser},\ and\
  \citenamefont {Gaulin}}]{chen2022charge}%
  \BibitemOpen
  \bibfield  {author} {\bibinfo {author} {\bibfnamefont {Q.}~\bibnamefont
  {Chen}}, \bibinfo {author} {\bibfnamefont {D.}~\bibnamefont {Chen}}, \bibinfo
  {author} {\bibfnamefont {W.}~\bibnamefont {Schnelle}}, \bibinfo {author}
  {\bibfnamefont {C.}~\bibnamefont {Felser}},\ and\ \bibinfo {author}
  {\bibfnamefont {B.}~\bibnamefont {Gaulin}},\ }\bibfield  {title} {\bibinfo
  {title} {Charge density wave order and fluctuations above t cdw and below
  superconducting t c in the kagome metal csv 3 sb 5},\ }\href@noop {}
  {\bibfield  {journal} {\bibinfo  {journal} {Physical Review Letters}\
  }\textbf {\bibinfo {volume} {129}},\ \bibinfo {pages} {056401} (\bibinfo
  {year} {2022})}\BibitemShut {NoStop}%
\bibitem [{\citenamefont {Zhong}\ \emph {et~al.}(2024)\citenamefont {Zhong},
  \citenamefont {Suzuki}, \citenamefont {Liu}, \citenamefont {Liu},
  \citenamefont {Nie}, \citenamefont {Shi}, \citenamefont {Meng}, \citenamefont
  {Lv}, \citenamefont {Ding}, \citenamefont {Kanai} \emph
  {et~al.}}]{zhong2024unveiling}%
  \BibitemOpen
  \bibfield  {author} {\bibinfo {author} {\bibfnamefont {Y.}~\bibnamefont
  {Zhong}}, \bibinfo {author} {\bibfnamefont {T.}~\bibnamefont {Suzuki}},
  \bibinfo {author} {\bibfnamefont {H.}~\bibnamefont {Liu}}, \bibinfo {author}
  {\bibfnamefont {K.}~\bibnamefont {Liu}}, \bibinfo {author} {\bibfnamefont
  {Z.}~\bibnamefont {Nie}}, \bibinfo {author} {\bibfnamefont {Y.}~\bibnamefont
  {Shi}}, \bibinfo {author} {\bibfnamefont {S.}~\bibnamefont {Meng}}, \bibinfo
  {author} {\bibfnamefont {B.}~\bibnamefont {Lv}}, \bibinfo {author}
  {\bibfnamefont {H.}~\bibnamefont {Ding}}, \bibinfo {author} {\bibfnamefont
  {T.}~\bibnamefont {Kanai}}, \emph {et~al.},\ }\bibfield  {title} {\bibinfo
  {title} {Unveiling van hove singularity modulation and fluctuated charge
  order in kagome superconductor cs v 3 s b 5 via time-resolved arpes},\
  }\href@noop {} {\bibfield  {journal} {\bibinfo  {journal} {Physical Review
  Research}\ }\textbf {\bibinfo {volume} {6}},\ \bibinfo {pages} {043328}
  (\bibinfo {year} {2024})}\BibitemShut {NoStop}%
\bibitem [{\citenamefont {Shen}(1984)}]{shen1984principles}%
  \BibitemOpen
  \bibfield  {author} {\bibinfo {author} {\bibfnamefont {Y.~R.}\ \bibnamefont
  {Shen}},\ }\href@noop {} {\emph {\bibinfo {title} {Principles of nonlinear
  optics}}}\ (\bibinfo  {publisher} {Wiley-Interscience, New York, NY, USA},\
  \bibinfo {year} {1984})\BibitemShut {NoStop}%
\bibitem [{\citenamefont {Shen}\ and\ \citenamefont
  {Bloembergen}(1965)}]{shen1965theory}%
  \BibitemOpen
  \bibfield  {author} {\bibinfo {author} {\bibfnamefont {Y.~R.}\ \bibnamefont
  {Shen}}\ and\ \bibinfo {author} {\bibfnamefont {N.}~\bibnamefont
  {Bloembergen}},\ }\bibfield  {title} {\bibinfo {title} {Theory of stimulated
  brillouin and raman scattering},\ }\href@noop {} {\bibfield  {journal}
  {\bibinfo  {journal} {Physical Review}\ }\textbf {\bibinfo {volume} {137}},\
  \bibinfo {pages} {A1787} (\bibinfo {year} {1965})}\BibitemShut {NoStop}%
\bibitem [{\citenamefont {Giordmaine}\ and\ \citenamefont
  {Kaiser}(1966)}]{giordmaine1966light}%
  \BibitemOpen
  \bibfield  {author} {\bibinfo {author} {\bibfnamefont {J.~A.}\ \bibnamefont
  {Giordmaine}}\ and\ \bibinfo {author} {\bibfnamefont {W.}~\bibnamefont
  {Kaiser}},\ }\bibfield  {title} {\bibinfo {title} {Light scattering by
  coherently driven lattice vibrations},\ }\href@noop {} {\bibfield  {journal}
  {\bibinfo  {journal} {Physical Review}\ }\textbf {\bibinfo {volume} {144}},\
  \bibinfo {pages} {676} (\bibinfo {year} {1966})}\BibitemShut {NoStop}%
\bibitem [{\citenamefont {Wang}\ \emph
  {et~al.}(2021{\natexlab{a}})\citenamefont {Wang}, \citenamefont {Wu},
  \citenamefont {Yin}, \citenamefont {Gong}, \citenamefont {Tu}, \citenamefont
  {Lin}, \citenamefont {Liu}, \citenamefont {Shi}, \citenamefont {Zhang},
  \citenamefont {Wu}, \citenamefont {Lei}, \citenamefont {Dong},\ and\
  \citenamefont {Wang}}]{PhysRevB.104.165110}%
  \BibitemOpen
  \bibfield  {author} {\bibinfo {author} {\bibfnamefont {Z.~X.}\ \bibnamefont
  {Wang}}, \bibinfo {author} {\bibfnamefont {Q.}~\bibnamefont {Wu}}, \bibinfo
  {author} {\bibfnamefont {Q.~W.}\ \bibnamefont {Yin}}, \bibinfo {author}
  {\bibfnamefont {C.~S.}\ \bibnamefont {Gong}}, \bibinfo {author}
  {\bibfnamefont {Z.~J.}\ \bibnamefont {Tu}}, \bibinfo {author} {\bibfnamefont
  {T.}~\bibnamefont {Lin}}, \bibinfo {author} {\bibfnamefont {Q.~M.}\
  \bibnamefont {Liu}}, \bibinfo {author} {\bibfnamefont {L.~Y.}\ \bibnamefont
  {Shi}}, \bibinfo {author} {\bibfnamefont {S.~J.}\ \bibnamefont {Zhang}},
  \bibinfo {author} {\bibfnamefont {D.}~\bibnamefont {Wu}}, \bibinfo {author}
  {\bibfnamefont {H.~C.}\ \bibnamefont {Lei}}, \bibinfo {author} {\bibfnamefont
  {T.}~\bibnamefont {Dong}},\ and\ \bibinfo {author} {\bibfnamefont {N.~L.}\
  \bibnamefont {Wang}},\ }\bibfield  {title} {\bibinfo {title} {Unconventional
  charge density wave and photoinduced lattice symmetry change in the kagome
  metal ${\mathrm{csv}}_{3}{\mathrm{sb}}_{5}$ probed by time-resolved
  spectroscopy},\ }\href {https://doi.org/10.1103/PhysRevB.104.165110}
  {\bibfield  {journal} {\bibinfo  {journal} {Phys. Rev. B}\ }\textbf {\bibinfo
  {volume} {104}},\ \bibinfo {pages} {165110} (\bibinfo {year}
  {2021}{\natexlab{a}})}\BibitemShut {NoStop}%
\bibitem [{\citenamefont {Uykur}\ \emph {et~al.}(2021)\citenamefont {Uykur},
  \citenamefont {Ortiz}, \citenamefont {Iakutkina}, \citenamefont {Wenzel},
  \citenamefont {Wilson}, \citenamefont {Dressel},\ and\ \citenamefont
  {Tsirlin}}]{uykur2021low}%
  \BibitemOpen
  \bibfield  {author} {\bibinfo {author} {\bibfnamefont {E.}~\bibnamefont
  {Uykur}}, \bibinfo {author} {\bibfnamefont {B.}~\bibnamefont {Ortiz}},
  \bibinfo {author} {\bibfnamefont {O.}~\bibnamefont {Iakutkina}}, \bibinfo
  {author} {\bibfnamefont {M.}~\bibnamefont {Wenzel}}, \bibinfo {author}
  {\bibfnamefont {S.}~\bibnamefont {Wilson}}, \bibinfo {author} {\bibfnamefont
  {M.}~\bibnamefont {Dressel}},\ and\ \bibinfo {author} {\bibfnamefont
  {A.}~\bibnamefont {Tsirlin}},\ }\bibfield  {title} {\bibinfo {title}
  {Low-energy optical properties of the nonmagnetic kagome metal csv 3 sb 5},\
  }\href@noop {} {\bibfield  {journal} {\bibinfo  {journal} {Physical Review
  B}\ }\textbf {\bibinfo {volume} {104}},\ \bibinfo {pages} {045130} (\bibinfo
  {year} {2021})}\BibitemShut {NoStop}%
\bibitem [{\citenamefont {Joshi}\ \emph {et~al.}(2019)\citenamefont {Joshi},
  \citenamefont {Hill}, \citenamefont {Chowdhury}, \citenamefont {Malliakas},
  \citenamefont {Tavazza}, \citenamefont {Chatterjee}, \citenamefont
  {Hight~Walker},\ and\ \citenamefont {Vora}}]{joshi2019short}%
  \BibitemOpen
  \bibfield  {author} {\bibinfo {author} {\bibfnamefont {J.}~\bibnamefont
  {Joshi}}, \bibinfo {author} {\bibfnamefont {H.~M.}\ \bibnamefont {Hill}},
  \bibinfo {author} {\bibfnamefont {S.}~\bibnamefont {Chowdhury}}, \bibinfo
  {author} {\bibfnamefont {C.~D.}\ \bibnamefont {Malliakas}}, \bibinfo {author}
  {\bibfnamefont {F.}~\bibnamefont {Tavazza}}, \bibinfo {author} {\bibfnamefont
  {U.}~\bibnamefont {Chatterjee}}, \bibinfo {author} {\bibfnamefont {A.~R.}\
  \bibnamefont {Hight~Walker}},\ and\ \bibinfo {author} {\bibfnamefont {P.~M.}\
  \bibnamefont {Vora}},\ }\bibfield  {title} {\bibinfo {title} {Short-range
  charge density wave order in 2 h-t a s 2},\ }\href@noop {} {\bibfield
  {journal} {\bibinfo  {journal} {Physical Review B}\ }\textbf {\bibinfo
  {volume} {99}},\ \bibinfo {pages} {245144} (\bibinfo {year}
  {2019})}\BibitemShut {NoStop}%
\bibitem [{\citenamefont {Liu}\ \emph {et~al.}(2022)\citenamefont {Liu},
  \citenamefont {Ma}, \citenamefont {He}, \citenamefont {Li}, \citenamefont
  {Tan}, \citenamefont {Liu}, \citenamefont {Xu}, \citenamefont {Tang},
  \citenamefont {Watanabe}, \citenamefont {Taniguchi} \emph
  {et~al.}}]{liu2022observation}%
  \BibitemOpen
  \bibfield  {author} {\bibinfo {author} {\bibfnamefont {G.}~\bibnamefont
  {Liu}}, \bibinfo {author} {\bibfnamefont {X.}~\bibnamefont {Ma}}, \bibinfo
  {author} {\bibfnamefont {K.}~\bibnamefont {He}}, \bibinfo {author}
  {\bibfnamefont {Q.}~\bibnamefont {Li}}, \bibinfo {author} {\bibfnamefont
  {H.}~\bibnamefont {Tan}}, \bibinfo {author} {\bibfnamefont {Y.}~\bibnamefont
  {Liu}}, \bibinfo {author} {\bibfnamefont {J.}~\bibnamefont {Xu}}, \bibinfo
  {author} {\bibfnamefont {W.}~\bibnamefont {Tang}}, \bibinfo {author}
  {\bibfnamefont {K.}~\bibnamefont {Watanabe}}, \bibinfo {author}
  {\bibfnamefont {T.}~\bibnamefont {Taniguchi}}, \emph {et~al.},\ }\bibfield
  {title} {\bibinfo {title} {Observation of anomalous amplitude modes in the
  kagome metal csv3sb5},\ }\href@noop {} {\bibfield  {journal} {\bibinfo
  {journal} {Nature communications}\ }\textbf {\bibinfo {volume} {13}},\
  \bibinfo {pages} {3461} (\bibinfo {year} {2022})}\BibitemShut {NoStop}%
\bibitem [{\citenamefont {He}\ \emph {et~al.}(2024)\citenamefont {He},
  \citenamefont {Peis}, \citenamefont {Cuddy}, \citenamefont {Zhao},
  \citenamefont {Li}, \citenamefont {Zhang}, \citenamefont {Stumberger},
  \citenamefont {Moritz}, \citenamefont {Yang}, \citenamefont {Gao} \emph
  {et~al.}}]{he2024anharmonic}%
  \BibitemOpen
  \bibfield  {author} {\bibinfo {author} {\bibfnamefont {G.}~\bibnamefont
  {He}}, \bibinfo {author} {\bibfnamefont {L.}~\bibnamefont {Peis}}, \bibinfo
  {author} {\bibfnamefont {E.~F.}\ \bibnamefont {Cuddy}}, \bibinfo {author}
  {\bibfnamefont {Z.}~\bibnamefont {Zhao}}, \bibinfo {author} {\bibfnamefont
  {D.}~\bibnamefont {Li}}, \bibinfo {author} {\bibfnamefont {Y.}~\bibnamefont
  {Zhang}}, \bibinfo {author} {\bibfnamefont {R.}~\bibnamefont {Stumberger}},
  \bibinfo {author} {\bibfnamefont {B.}~\bibnamefont {Moritz}}, \bibinfo
  {author} {\bibfnamefont {H.}~\bibnamefont {Yang}}, \bibinfo {author}
  {\bibfnamefont {H.}~\bibnamefont {Gao}}, \emph {et~al.},\ }\bibfield  {title}
  {\bibinfo {title} {Anharmonic strong-coupling effects at the origin of the
  charge density wave in csv3sb5},\ }\href@noop {} {\bibfield  {journal}
  {\bibinfo  {journal} {Nature Communications}\ }\textbf {\bibinfo {volume}
  {15}},\ \bibinfo {pages} {1895} (\bibinfo {year} {2024})}\BibitemShut
  {NoStop}%
\bibitem [{\citenamefont {Stojchevska}\ \emph {et~al.}(2012)\citenamefont
  {Stojchevska}, \citenamefont {Mertelj}, \citenamefont {Chu}, \citenamefont
  {Fisher},\ and\ \citenamefont {Mihailovic}}]{stojchevska2012doping}%
  \BibitemOpen
  \bibfield  {author} {\bibinfo {author} {\bibfnamefont {L.}~\bibnamefont
  {Stojchevska}}, \bibinfo {author} {\bibfnamefont {T.}~\bibnamefont
  {Mertelj}}, \bibinfo {author} {\bibfnamefont {J.-H.}\ \bibnamefont {Chu}},
  \bibinfo {author} {\bibfnamefont {I.~R.}\ \bibnamefont {Fisher}},\ and\
  \bibinfo {author} {\bibfnamefont {D.}~\bibnamefont {Mihailovic}},\ }\bibfield
   {title} {\bibinfo {title} {Doping dependence of femtosecond quasiparticle
  relaxation dynamics in ba (fe, co) 2 as 2 single crystals: Evidence for
  normal-state nematic fluctuations},\ }\href@noop {} {\bibfield  {journal}
  {\bibinfo  {journal} {Physical Review B}\ }\textbf {\bibinfo {volume} {86}},\
  \bibinfo {pages} {024519} (\bibinfo {year} {2012})}\BibitemShut {NoStop}%
\bibitem [{\citenamefont {Thewalt}\ \emph {et~al.}(2018)\citenamefont
  {Thewalt}, \citenamefont {Hayes}, \citenamefont {Hinton}, \citenamefont
  {Little}, \citenamefont {Patankar}, \citenamefont {Wu}, \citenamefont {Helm},
  \citenamefont {Stan}, \citenamefont {Tamura}, \citenamefont {Analytis} \emph
  {et~al.}}]{thewalt2018imaging}%
  \BibitemOpen
  \bibfield  {author} {\bibinfo {author} {\bibfnamefont {E.}~\bibnamefont
  {Thewalt}}, \bibinfo {author} {\bibfnamefont {I.~M.}\ \bibnamefont {Hayes}},
  \bibinfo {author} {\bibfnamefont {J.~P.}\ \bibnamefont {Hinton}}, \bibinfo
  {author} {\bibfnamefont {A.}~\bibnamefont {Little}}, \bibinfo {author}
  {\bibfnamefont {S.}~\bibnamefont {Patankar}}, \bibinfo {author}
  {\bibfnamefont {L.}~\bibnamefont {Wu}}, \bibinfo {author} {\bibfnamefont
  {T.}~\bibnamefont {Helm}}, \bibinfo {author} {\bibfnamefont {C.~V.}\
  \bibnamefont {Stan}}, \bibinfo {author} {\bibfnamefont {N.}~\bibnamefont
  {Tamura}}, \bibinfo {author} {\bibfnamefont {J.~G.}\ \bibnamefont
  {Analytis}}, \emph {et~al.},\ }\bibfield  {title} {\bibinfo {title} {Imaging
  anomalous nematic order and strain in optimally doped bafe 2 (as, p) 2},\
  }\href@noop {} {\bibfield  {journal} {\bibinfo  {journal} {Physical review
  letters}\ }\textbf {\bibinfo {volume} {121}},\ \bibinfo {pages} {027001}
  (\bibinfo {year} {2018})}\BibitemShut {NoStop}%
\bibitem [{\citenamefont {Liu}\ \emph {et~al.}(2018)\citenamefont {Liu},
  \citenamefont {Zhang}, \citenamefont {Deng}, \citenamefont {Wen},
  \citenamefont {Li}, \citenamefont {Chia}, \citenamefont {Wang},\ and\
  \citenamefont {Xiao}}]{liu2018transient}%
  \BibitemOpen
  \bibfield  {author} {\bibinfo {author} {\bibfnamefont {S.}~\bibnamefont
  {Liu}}, \bibinfo {author} {\bibfnamefont {C.}~\bibnamefont {Zhang}}, \bibinfo
  {author} {\bibfnamefont {Q.}~\bibnamefont {Deng}}, \bibinfo {author}
  {\bibfnamefont {H.-h.}\ \bibnamefont {Wen}}, \bibinfo {author} {\bibfnamefont
  {J.-x.}\ \bibnamefont {Li}}, \bibinfo {author} {\bibfnamefont {E.~E.}\
  \bibnamefont {Chia}}, \bibinfo {author} {\bibfnamefont {X.}~\bibnamefont
  {Wang}},\ and\ \bibinfo {author} {\bibfnamefont {M.}~\bibnamefont {Xiao}},\
  }\bibfield  {title} {\bibinfo {title} {Transient electronic anisotropy in
  overdoped naf e 1- x c ox as superconductors},\ }\href@noop {} {\bibfield
  {journal} {\bibinfo  {journal} {Physical Review B}\ }\textbf {\bibinfo
  {volume} {97}},\ \bibinfo {pages} {020505} (\bibinfo {year}
  {2018})}\BibitemShut {NoStop}%
\bibitem [{\citenamefont {Liu}\ \emph {et~al.}(2024{\natexlab{b}})\citenamefont
  {Liu}, \citenamefont {Li}, \citenamefont {Tan}, \citenamefont {Liu},
  \citenamefont {Shi}, \citenamefont {Zhai}, \citenamefont {Lin}, \citenamefont
  {Cao}, \citenamefont {Yan}, \citenamefont {Zhang} \emph
  {et~al.}}]{liu2024charge}%
  \BibitemOpen
  \bibfield  {author} {\bibinfo {author} {\bibfnamefont {L.}~\bibnamefont
  {Liu}}, \bibinfo {author} {\bibfnamefont {Y.}~\bibnamefont {Li}}, \bibinfo
  {author} {\bibfnamefont {H.}~\bibnamefont {Tan}}, \bibinfo {author}
  {\bibfnamefont {Y.}~\bibnamefont {Liu}}, \bibinfo {author} {\bibfnamefont
  {Y.}~\bibnamefont {Shi}}, \bibinfo {author} {\bibfnamefont {Y.}~\bibnamefont
  {Zhai}}, \bibinfo {author} {\bibfnamefont {H.}~\bibnamefont {Lin}}, \bibinfo
  {author} {\bibfnamefont {G.}~\bibnamefont {Cao}}, \bibinfo {author}
  {\bibfnamefont {B.}~\bibnamefont {Yan}}, \bibinfo {author} {\bibfnamefont
  {G.-M.}\ \bibnamefont {Zhang}}, \emph {et~al.},\ }\bibfield  {title}
  {\bibinfo {title} {Charge density wave coexisting with amplified nematicity
  in the correlated kagome metal cscr3sb5},\ }\href@noop {} {\bibfield
  {journal} {\bibinfo  {journal} {arXiv preprint arXiv:2411.06778}\ } (\bibinfo
  {year} {2024}{\natexlab{b}})}\BibitemShut {NoStop}%
\bibitem [{\citenamefont {Merlin}(1997)}]{merlin1997generating}%
  \BibitemOpen
  \bibfield  {author} {\bibinfo {author} {\bibfnamefont {R.}~\bibnamefont
  {Merlin}},\ }\bibfield  {title} {\bibinfo {title} {Generating coherent thz
  phonons with light pulses},\ }\href@noop {} {\bibfield  {journal} {\bibinfo
  {journal} {Solid state communications}\ }\textbf {\bibinfo {volume} {102}},\
  \bibinfo {pages} {207} (\bibinfo {year} {1997})}\BibitemShut {NoStop}%
\bibitem [{\citenamefont {Stevens}\ \emph {et~al.}(2002)\citenamefont
  {Stevens}, \citenamefont {Kuhl},\ and\ \citenamefont
  {Merlin}}]{stevens2002coherent}%
  \BibitemOpen
  \bibfield  {author} {\bibinfo {author} {\bibfnamefont {T.}~\bibnamefont
  {Stevens}}, \bibinfo {author} {\bibfnamefont {J.}~\bibnamefont {Kuhl}},\ and\
  \bibinfo {author} {\bibfnamefont {R.}~\bibnamefont {Merlin}},\ }\bibfield
  {title} {\bibinfo {title} {Coherent phonon generation and the two stimulated
  raman tensors},\ }\href@noop {} {\bibfield  {journal} {\bibinfo  {journal}
  {Physical Review B}\ }\textbf {\bibinfo {volume} {65}},\ \bibinfo {pages}
  {144304} (\bibinfo {year} {2002})}\BibitemShut {NoStop}%
\bibitem [{\citenamefont {Dhar}\ \emph {et~al.}(1994)\citenamefont {Dhar},
  \citenamefont {Rogers},\ and\ \citenamefont {Nelson}}]{dhar1994time}%
  \BibitemOpen
  \bibfield  {author} {\bibinfo {author} {\bibfnamefont {L.}~\bibnamefont
  {Dhar}}, \bibinfo {author} {\bibfnamefont {J.~A.}\ \bibnamefont {Rogers}},\
  and\ \bibinfo {author} {\bibfnamefont {K.~A.}\ \bibnamefont {Nelson}},\
  }\bibfield  {title} {\bibinfo {title} {Time-resolved vibrational spectroscopy
  in the impulsive limit},\ }\href@noop {} {\bibfield  {journal} {\bibinfo
  {journal} {Chemical Reviews}\ }\textbf {\bibinfo {volume} {94}},\ \bibinfo
  {pages} {157} (\bibinfo {year} {1994})}\BibitemShut {NoStop}%
\bibitem [{\citenamefont {Yan}\ \emph {et~al.}(1985)\citenamefont {Yan},
  \citenamefont {Gamble~Jr},\ and\ \citenamefont {Nelson}}]{yan1985impulsive}%
  \BibitemOpen
  \bibfield  {author} {\bibinfo {author} {\bibfnamefont {Y.-X.}\ \bibnamefont
  {Yan}}, \bibinfo {author} {\bibfnamefont {E.~B.}\ \bibnamefont {Gamble~Jr}},\
  and\ \bibinfo {author} {\bibfnamefont {K.~A.}\ \bibnamefont {Nelson}},\
  }\bibfield  {title} {\bibinfo {title} {Impulsive stimulated scattering:
  General importance in femtosecond laser pulse interactions with matter, and
  spectroscopic applications},\ }\href@noop {} {\bibfield  {journal} {\bibinfo
  {journal} {The Journal of chemical physics}\ }\textbf {\bibinfo {volume}
  {83}},\ \bibinfo {pages} {5391} (\bibinfo {year} {1985})}\BibitemShut
  {NoStop}%
\bibitem [{\citenamefont {Gray}\ \emph {et~al.}(2024)\citenamefont {Gray},
  \citenamefont {Deng}, \citenamefont {Tian}, \citenamefont {Chilcote},
  \citenamefont {Dodge}, \citenamefont {Brahlek},\ and\ \citenamefont
  {Wu}}]{gray2024time}%
  \BibitemOpen
  \bibfield  {author} {\bibinfo {author} {\bibfnamefont {I.}~\bibnamefont
  {Gray}}, \bibinfo {author} {\bibfnamefont {Q.}~\bibnamefont {Deng}}, \bibinfo
  {author} {\bibfnamefont {Q.}~\bibnamefont {Tian}}, \bibinfo {author}
  {\bibfnamefont {M.}~\bibnamefont {Chilcote}}, \bibinfo {author}
  {\bibfnamefont {J.~S.}\ \bibnamefont {Dodge}}, \bibinfo {author}
  {\bibfnamefont {M.}~\bibnamefont {Brahlek}},\ and\ \bibinfo {author}
  {\bibfnamefont {L.}~\bibnamefont {Wu}},\ }\bibfield  {title} {\bibinfo
  {title} {Time-resolved magneto-optical effects in the altermagnet candidate
  mnte},\ }\href@noop {} {\bibfield  {journal} {\bibinfo  {journal} {Applied
  Physics Letters}\ }\textbf {\bibinfo {volume} {125}} (\bibinfo {year}
  {2024})}\BibitemShut {NoStop}%
\bibitem [{\citenamefont {Cheng}\ \emph {et~al.}(1991)\citenamefont {Cheng},
  \citenamefont {Vidal}, \citenamefont {Zeiger}, \citenamefont {Dresselhaus},
  \citenamefont {Dresselhaus},\ and\ \citenamefont
  {Ippen}}]{cheng1991mechanism}%
  \BibitemOpen
  \bibfield  {author} {\bibinfo {author} {\bibfnamefont {T.}~\bibnamefont
  {Cheng}}, \bibinfo {author} {\bibfnamefont {J.}~\bibnamefont {Vidal}},
  \bibinfo {author} {\bibfnamefont {H.}~\bibnamefont {Zeiger}}, \bibinfo
  {author} {\bibfnamefont {G.}~\bibnamefont {Dresselhaus}}, \bibinfo {author}
  {\bibfnamefont {M.}~\bibnamefont {Dresselhaus}},\ and\ \bibinfo {author}
  {\bibfnamefont {E.}~\bibnamefont {Ippen}},\ }\bibfield  {title} {\bibinfo
  {title} {Mechanism for displacive excitation of coherent phonons in sb, bi,
  te, and ti2o3},\ }\href@noop {} {\bibfield  {journal} {\bibinfo  {journal}
  {Applied Physics Letters}\ }\textbf {\bibinfo {volume} {59}},\ \bibinfo
  {pages} {1923} (\bibinfo {year} {1991})}\BibitemShut {NoStop}%
\bibitem [{\citenamefont {Zeiger}\ \emph {et~al.}(1992)\citenamefont {Zeiger},
  \citenamefont {Vidal}, \citenamefont {Cheng}, \citenamefont {Ippen},
  \citenamefont {Dresselhaus},\ and\ \citenamefont
  {Dresselhaus}}]{zeiger1992theory}%
  \BibitemOpen
  \bibfield  {author} {\bibinfo {author} {\bibfnamefont {H.}~\bibnamefont
  {Zeiger}}, \bibinfo {author} {\bibfnamefont {J.}~\bibnamefont {Vidal}},
  \bibinfo {author} {\bibfnamefont {T.}~\bibnamefont {Cheng}}, \bibinfo
  {author} {\bibfnamefont {E.}~\bibnamefont {Ippen}}, \bibinfo {author}
  {\bibfnamefont {G.}~\bibnamefont {Dresselhaus}},\ and\ \bibinfo {author}
  {\bibfnamefont {M.}~\bibnamefont {Dresselhaus}},\ }\bibfield  {title}
  {\bibinfo {title} {Theory for displacive excitation of coherent phonons},\
  }\href@noop {} {\bibfield  {journal} {\bibinfo  {journal} {Physical Review
  B}\ }\textbf {\bibinfo {volume} {45}},\ \bibinfo {pages} {768} (\bibinfo
  {year} {1992})}\BibitemShut {NoStop}%
\bibitem [{\citenamefont {De~Silvestri}\ \emph {et~al.}(1985)\citenamefont
  {De~Silvestri}, \citenamefont {Fujimoto}, \citenamefont {Ippen},
  \citenamefont {Gamble~Jr}, \citenamefont {Williams},\ and\ \citenamefont
  {Nelson}}]{de1985femtosecond}%
  \BibitemOpen
  \bibfield  {author} {\bibinfo {author} {\bibfnamefont {S.}~\bibnamefont
  {De~Silvestri}}, \bibinfo {author} {\bibfnamefont {J.}~\bibnamefont
  {Fujimoto}}, \bibinfo {author} {\bibfnamefont {E.}~\bibnamefont {Ippen}},
  \bibinfo {author} {\bibfnamefont {E.~B.}\ \bibnamefont {Gamble~Jr}}, \bibinfo
  {author} {\bibfnamefont {L.~R.}\ \bibnamefont {Williams}},\ and\ \bibinfo
  {author} {\bibfnamefont {K.~A.}\ \bibnamefont {Nelson}},\ }\bibfield  {title}
  {\bibinfo {title} {Femtosecond time-resolved measurements of optic phonon
  dephasing by impulsive stimulated raman scattering in $\alpha$-perylene
  crystal from 20 to 300 k},\ }\href@noop {} {\bibfield  {journal} {\bibinfo
  {journal} {Chemical physics letters}\ }\textbf {\bibinfo {volume} {116}},\
  \bibinfo {pages} {146} (\bibinfo {year} {1985})}\BibitemShut {NoStop}%
\bibitem [{\citenamefont {Ruhman}\ \emph {et~al.}(1987)\citenamefont {Ruhman},
  \citenamefont {Joly},\ and\ \citenamefont {Nelson}}]{ruhman1987time}%
  \BibitemOpen
  \bibfield  {author} {\bibinfo {author} {\bibfnamefont {S.}~\bibnamefont
  {Ruhman}}, \bibinfo {author} {\bibfnamefont {A.~G.}\ \bibnamefont {Joly}},\
  and\ \bibinfo {author} {\bibfnamefont {K.~A.}\ \bibnamefont {Nelson}},\
  }\bibfield  {title} {\bibinfo {title} {Time-resolved observations of coherent
  molecular vibrational motion and the general occurrence of impulsive
  stimulated scattering},\ }\href@noop {} {\bibfield  {journal} {\bibinfo
  {journal} {The Journal of chemical physics}\ }\textbf {\bibinfo {volume}
  {86}},\ \bibinfo {pages} {6563} (\bibinfo {year} {1987})}\BibitemShut
  {NoStop}%
\bibitem [{\citenamefont {Wu}\ \emph {et~al.}(2022{\natexlab{b}})\citenamefont
  {Wu}, \citenamefont {Ortiz}, \citenamefont {Tan}, \citenamefont {Wilson},
  \citenamefont {Yan}, \citenamefont {Birol},\ and\ \citenamefont
  {Blumberg}}]{wu2022charge}%
  \BibitemOpen
  \bibfield  {author} {\bibinfo {author} {\bibfnamefont {S.}~\bibnamefont
  {Wu}}, \bibinfo {author} {\bibfnamefont {B.~R.}\ \bibnamefont {Ortiz}},
  \bibinfo {author} {\bibfnamefont {H.}~\bibnamefont {Tan}}, \bibinfo {author}
  {\bibfnamefont {S.~D.}\ \bibnamefont {Wilson}}, \bibinfo {author}
  {\bibfnamefont {B.}~\bibnamefont {Yan}}, \bibinfo {author} {\bibfnamefont
  {T.}~\bibnamefont {Birol}},\ and\ \bibinfo {author} {\bibfnamefont
  {G.}~\bibnamefont {Blumberg}},\ }\bibfield  {title} {\bibinfo {title} {Charge
  density wave order in the kagome metal a v 3 sb 5 (a= cs, rb, k)},\
  }\href@noop {} {\bibfield  {journal} {\bibinfo  {journal} {Physical Review
  B}\ }\textbf {\bibinfo {volume} {105}},\ \bibinfo {pages} {155106} (\bibinfo
  {year} {2022}{\natexlab{b}})}\BibitemShut {NoStop}%
\bibitem [{\citenamefont {Nakayama}\ \emph {et~al.}(2021)\citenamefont
  {Nakayama}, \citenamefont {Li}, \citenamefont {Kato}, \citenamefont {Liu},
  \citenamefont {Wang}, \citenamefont {Takahashi}, \citenamefont {Yao},\ and\
  \citenamefont {Sato}}]{nakayama2021multiple}%
  \BibitemOpen
  \bibfield  {author} {\bibinfo {author} {\bibfnamefont {K.}~\bibnamefont
  {Nakayama}}, \bibinfo {author} {\bibfnamefont {Y.}~\bibnamefont {Li}},
  \bibinfo {author} {\bibfnamefont {T.}~\bibnamefont {Kato}}, \bibinfo {author}
  {\bibfnamefont {M.}~\bibnamefont {Liu}}, \bibinfo {author} {\bibfnamefont
  {Z.}~\bibnamefont {Wang}}, \bibinfo {author} {\bibfnamefont {T.}~\bibnamefont
  {Takahashi}}, \bibinfo {author} {\bibfnamefont {Y.}~\bibnamefont {Yao}},\
  and\ \bibinfo {author} {\bibfnamefont {T.}~\bibnamefont {Sato}},\ }\bibfield
  {title} {\bibinfo {title} {Multiple energy scales and anisotropic energy gap
  in the charge-density-wave phase of the kagome superconductor csv 3 sb 5},\
  }\href@noop {} {\bibfield  {journal} {\bibinfo  {journal} {Physical Review
  B}\ }\textbf {\bibinfo {volume} {104}},\ \bibinfo {pages} {L161112} (\bibinfo
  {year} {2021})}\BibitemShut {NoStop}%
\bibitem [{\citenamefont {Zhou}\ \emph {et~al.}(2021)\citenamefont {Zhou},
  \citenamefont {Li}, \citenamefont {Fan}, \citenamefont {Hao}, \citenamefont
  {Dai}, \citenamefont {Wang}, \citenamefont {Yao},\ and\ \citenamefont
  {Wen}}]{zhou2021origin}%
  \BibitemOpen
  \bibfield  {author} {\bibinfo {author} {\bibfnamefont {X.}~\bibnamefont
  {Zhou}}, \bibinfo {author} {\bibfnamefont {Y.}~\bibnamefont {Li}}, \bibinfo
  {author} {\bibfnamefont {X.}~\bibnamefont {Fan}}, \bibinfo {author}
  {\bibfnamefont {J.}~\bibnamefont {Hao}}, \bibinfo {author} {\bibfnamefont
  {Y.}~\bibnamefont {Dai}}, \bibinfo {author} {\bibfnamefont {Z.}~\bibnamefont
  {Wang}}, \bibinfo {author} {\bibfnamefont {Y.}~\bibnamefont {Yao}},\ and\
  \bibinfo {author} {\bibfnamefont {H.-H.}\ \bibnamefont {Wen}},\ }\bibfield
  {title} {\bibinfo {title} {Origin of charge density wave in the kagome metal
  csv 3 sb 5 as revealed by optical spectroscopy},\ }\href@noop {} {\bibfield
  {journal} {\bibinfo  {journal} {Physical Review B}\ }\textbf {\bibinfo
  {volume} {104}},\ \bibinfo {pages} {L041101} (\bibinfo {year}
  {2021})}\BibitemShut {NoStop}%
\bibitem [{\citenamefont {Wang}\ \emph
  {et~al.}(2021{\natexlab{b}})\citenamefont {Wang}, \citenamefont {Ma},
  \citenamefont {Zhang}, \citenamefont {Yang}, \citenamefont {Zhao},
  \citenamefont {Ou}, \citenamefont {Zhu}, \citenamefont {Ni}, \citenamefont
  {Lu}, \citenamefont {Chen} \emph {et~al.}}]{wang2021distinctive}%
  \BibitemOpen
  \bibfield  {author} {\bibinfo {author} {\bibfnamefont {Z.}~\bibnamefont
  {Wang}}, \bibinfo {author} {\bibfnamefont {S.}~\bibnamefont {Ma}}, \bibinfo
  {author} {\bibfnamefont {Y.}~\bibnamefont {Zhang}}, \bibinfo {author}
  {\bibfnamefont {H.}~\bibnamefont {Yang}}, \bibinfo {author} {\bibfnamefont
  {Z.}~\bibnamefont {Zhao}}, \bibinfo {author} {\bibfnamefont {Y.}~\bibnamefont
  {Ou}}, \bibinfo {author} {\bibfnamefont {Y.}~\bibnamefont {Zhu}}, \bibinfo
  {author} {\bibfnamefont {S.}~\bibnamefont {Ni}}, \bibinfo {author}
  {\bibfnamefont {Z.}~\bibnamefont {Lu}}, \bibinfo {author} {\bibfnamefont
  {H.}~\bibnamefont {Chen}}, \emph {et~al.},\ }\bibfield  {title} {\bibinfo
  {title} {Distinctive momentum dependent charge-density-wave gap observed in
  csv $ \_3 $ sb $ \_5 $ superconductor with topological kagome lattice},\
  }\href@noop {} {\bibfield  {journal} {\bibinfo  {journal} {arXiv preprint
  arXiv:2104.05556}\ } (\bibinfo {year} {2021}{\natexlab{b}})}\BibitemShut
  {NoStop}%
\bibitem [{\citenamefont {Ning}\ \emph {et~al.}(2024)\citenamefont {Ning},
  \citenamefont {Oh}, \citenamefont {Su}, \citenamefont {von Hoegen},
  \citenamefont {Porter}, \citenamefont {Capa~Salinas}, \citenamefont {Nguyen},
  \citenamefont {Chollet}, \citenamefont {Sato}, \citenamefont {Esposito} \emph
  {et~al.}}]{ning2024dynamical}%
  \BibitemOpen
  \bibfield  {author} {\bibinfo {author} {\bibfnamefont {H.}~\bibnamefont
  {Ning}}, \bibinfo {author} {\bibfnamefont {K.~H.}\ \bibnamefont {Oh}},
  \bibinfo {author} {\bibfnamefont {Y.}~\bibnamefont {Su}}, \bibinfo {author}
  {\bibfnamefont {A.}~\bibnamefont {von Hoegen}}, \bibinfo {author}
  {\bibfnamefont {Z.}~\bibnamefont {Porter}}, \bibinfo {author} {\bibfnamefont
  {A.}~\bibnamefont {Capa~Salinas}}, \bibinfo {author} {\bibfnamefont {Q.~L.}\
  \bibnamefont {Nguyen}}, \bibinfo {author} {\bibfnamefont {M.}~\bibnamefont
  {Chollet}}, \bibinfo {author} {\bibfnamefont {T.}~\bibnamefont {Sato}},
  \bibinfo {author} {\bibfnamefont {V.}~\bibnamefont {Esposito}}, \emph
  {et~al.},\ }\bibfield  {title} {\bibinfo {title} {Dynamical decoding of the
  competition between charge density waves in a kagome superconductor},\
  }\href@noop {} {\bibfield  {journal} {\bibinfo  {journal} {Nature
  Communications}\ }\textbf {\bibinfo {volume} {15}},\ \bibinfo {pages} {7286}
  (\bibinfo {year} {2024})}\BibitemShut {NoStop}%
\bibitem [{\citenamefont {Kautzsch}\ \emph
  {et~al.}(2023{\natexlab{b}})\citenamefont {Kautzsch}, \citenamefont {Oey},
  \citenamefont {Li}, \citenamefont {Ren}, \citenamefont {Ortiz}, \citenamefont
  {Pokharel}, \citenamefont {Seshadri}, \citenamefont {Ruff}, \citenamefont
  {Kongruengkit}, \citenamefont {Harter} \emph
  {et~al.}}]{kautzsch2023incommensurate}%
  \BibitemOpen
  \bibfield  {author} {\bibinfo {author} {\bibfnamefont {L.}~\bibnamefont
  {Kautzsch}}, \bibinfo {author} {\bibfnamefont {Y.~M.}\ \bibnamefont {Oey}},
  \bibinfo {author} {\bibfnamefont {H.}~\bibnamefont {Li}}, \bibinfo {author}
  {\bibfnamefont {Z.}~\bibnamefont {Ren}}, \bibinfo {author} {\bibfnamefont
  {B.~R.}\ \bibnamefont {Ortiz}}, \bibinfo {author} {\bibfnamefont
  {G.}~\bibnamefont {Pokharel}}, \bibinfo {author} {\bibfnamefont
  {R.}~\bibnamefont {Seshadri}}, \bibinfo {author} {\bibfnamefont
  {J.}~\bibnamefont {Ruff}}, \bibinfo {author} {\bibfnamefont {T.}~\bibnamefont
  {Kongruengkit}}, \bibinfo {author} {\bibfnamefont {J.~W.}\ \bibnamefont
  {Harter}}, \emph {et~al.},\ }\bibfield  {title} {\bibinfo {title}
  {Incommensurate charge-stripe correlations in the kagome superconductor
  csv3sb5- x sn x},\ }\href@noop {} {\bibfield  {journal} {\bibinfo  {journal}
  {npj Quantum Materials}\ }\textbf {\bibinfo {volume} {8}},\ \bibinfo {pages}
  {37} (\bibinfo {year} {2023}{\natexlab{b}})}\BibitemShut {NoStop}%
\bibitem [{\citenamefont {Wang}\ \emph
  {et~al.}(2021{\natexlab{c}})\citenamefont {Wang}, \citenamefont {Jiang},
  \citenamefont {Yin}, \citenamefont {Li}, \citenamefont {Wang}, \citenamefont
  {Huang}, \citenamefont {Shao}, \citenamefont {Liu}, \citenamefont {Zhu},
  \citenamefont {Shumiya} \emph {et~al.}}]{wang2021electronic}%
  \BibitemOpen
  \bibfield  {author} {\bibinfo {author} {\bibfnamefont {Z.}~\bibnamefont
  {Wang}}, \bibinfo {author} {\bibfnamefont {Y.-X.}\ \bibnamefont {Jiang}},
  \bibinfo {author} {\bibfnamefont {J.-X.}\ \bibnamefont {Yin}}, \bibinfo
  {author} {\bibfnamefont {Y.}~\bibnamefont {Li}}, \bibinfo {author}
  {\bibfnamefont {G.-Y.}\ \bibnamefont {Wang}}, \bibinfo {author}
  {\bibfnamefont {H.-L.}\ \bibnamefont {Huang}}, \bibinfo {author}
  {\bibfnamefont {S.}~\bibnamefont {Shao}}, \bibinfo {author} {\bibfnamefont
  {J.}~\bibnamefont {Liu}}, \bibinfo {author} {\bibfnamefont {P.}~\bibnamefont
  {Zhu}}, \bibinfo {author} {\bibfnamefont {N.}~\bibnamefont {Shumiya}}, \emph
  {et~al.},\ }\bibfield  {title} {\bibinfo {title} {Electronic nature of chiral
  charge order in the kagome superconductor cs v 3 sb 5},\ }\href@noop {}
  {\bibfield  {journal} {\bibinfo  {journal} {Physical Review B}\ }\textbf
  {\bibinfo {volume} {104}},\ \bibinfo {pages} {075148} (\bibinfo {year}
  {2021}{\natexlab{c}})}\BibitemShut {NoStop}%
\bibitem [{\citenamefont {Oey}\ \emph {et~al.}(2022)\citenamefont {Oey},
  \citenamefont {Ortiz}, \citenamefont {Kaboudvand}, \citenamefont
  {Frassineti}, \citenamefont {Garcia}, \citenamefont {Cong}, \citenamefont
  {Sanna}, \citenamefont {Mitrovi{\'c}}, \citenamefont {Seshadri},\ and\
  \citenamefont {Wilson}}]{oey2022fermi}%
  \BibitemOpen
  \bibfield  {author} {\bibinfo {author} {\bibfnamefont {Y.~M.}\ \bibnamefont
  {Oey}}, \bibinfo {author} {\bibfnamefont {B.~R.}\ \bibnamefont {Ortiz}},
  \bibinfo {author} {\bibfnamefont {F.}~\bibnamefont {Kaboudvand}}, \bibinfo
  {author} {\bibfnamefont {J.}~\bibnamefont {Frassineti}}, \bibinfo {author}
  {\bibfnamefont {E.}~\bibnamefont {Garcia}}, \bibinfo {author} {\bibfnamefont
  {R.}~\bibnamefont {Cong}}, \bibinfo {author} {\bibfnamefont {S.}~\bibnamefont
  {Sanna}}, \bibinfo {author} {\bibfnamefont {V.~F.}\ \bibnamefont
  {Mitrovi{\'c}}}, \bibinfo {author} {\bibfnamefont {R.}~\bibnamefont
  {Seshadri}},\ and\ \bibinfo {author} {\bibfnamefont {S.~D.}\ \bibnamefont
  {Wilson}},\ }\bibfield  {title} {\bibinfo {title} {Fermi level tuning and
  double-dome superconductivity in the kagome metal csv 3 sb 5- x sn x},\
  }\href@noop {} {\bibfield  {journal} {\bibinfo  {journal} {Physical Review
  Materials}\ }\textbf {\bibinfo {volume} {6}},\ \bibinfo {pages} {L041801}
  (\bibinfo {year} {2022})}\BibitemShut {NoStop}%
\bibitem [{\citenamefont {Zhong}\ \emph {et~al.}(2023)\citenamefont {Zhong},
  \citenamefont {Liu}, \citenamefont {Wu}, \citenamefont {Guguchia},
  \citenamefont {Yin}, \citenamefont {Mine}, \citenamefont {Li}, \citenamefont
  {Najafzadeh}, \citenamefont {Das}, \citenamefont {Mielke~III} \emph
  {et~al.}}]{zhong2023nodeless}%
  \BibitemOpen
  \bibfield  {author} {\bibinfo {author} {\bibfnamefont {Y.}~\bibnamefont
  {Zhong}}, \bibinfo {author} {\bibfnamefont {J.}~\bibnamefont {Liu}}, \bibinfo
  {author} {\bibfnamefont {X.}~\bibnamefont {Wu}}, \bibinfo {author}
  {\bibfnamefont {Z.}~\bibnamefont {Guguchia}}, \bibinfo {author}
  {\bibfnamefont {J.-X.}\ \bibnamefont {Yin}}, \bibinfo {author} {\bibfnamefont
  {A.}~\bibnamefont {Mine}}, \bibinfo {author} {\bibfnamefont {Y.}~\bibnamefont
  {Li}}, \bibinfo {author} {\bibfnamefont {S.}~\bibnamefont {Najafzadeh}},
  \bibinfo {author} {\bibfnamefont {D.}~\bibnamefont {Das}}, \bibinfo {author}
  {\bibfnamefont {C.}~\bibnamefont {Mielke~III}}, \emph {et~al.},\ }\bibfield
  {title} {\bibinfo {title} {Nodeless electron pairing in csv3sb5-derived
  kagome superconductors},\ }\href@noop {} {\bibfield  {journal} {\bibinfo
  {journal} {Nature}\ }\textbf {\bibinfo {volume} {617}},\ \bibinfo {pages}
  {488} (\bibinfo {year} {2023})}\BibitemShut {NoStop}%
\bibitem [{\citenamefont {Teng}\ \emph {et~al.}(2022)\citenamefont {Teng},
  \citenamefont {Chen}, \citenamefont {Ye}, \citenamefont {Rosenberg},
  \citenamefont {Liu}, \citenamefont {Yin}, \citenamefont {Jiang},
  \citenamefont {Oh}, \citenamefont {Hasan}, \citenamefont {Neubauer} \emph
  {et~al.}}]{teng2022discovery}%
  \BibitemOpen
  \bibfield  {author} {\bibinfo {author} {\bibfnamefont {X.}~\bibnamefont
  {Teng}}, \bibinfo {author} {\bibfnamefont {L.}~\bibnamefont {Chen}}, \bibinfo
  {author} {\bibfnamefont {F.}~\bibnamefont {Ye}}, \bibinfo {author}
  {\bibfnamefont {E.}~\bibnamefont {Rosenberg}}, \bibinfo {author}
  {\bibfnamefont {Z.}~\bibnamefont {Liu}}, \bibinfo {author} {\bibfnamefont
  {J.-X.}\ \bibnamefont {Yin}}, \bibinfo {author} {\bibfnamefont {Y.-X.}\
  \bibnamefont {Jiang}}, \bibinfo {author} {\bibfnamefont {J.~S.}\ \bibnamefont
  {Oh}}, \bibinfo {author} {\bibfnamefont {M.~Z.}\ \bibnamefont {Hasan}},
  \bibinfo {author} {\bibfnamefont {K.~J.}\ \bibnamefont {Neubauer}}, \emph
  {et~al.},\ }\bibfield  {title} {\bibinfo {title} {Discovery of charge density
  wave in a kagome lattice antiferromagnet},\ }\href@noop {} {\bibfield
  {journal} {\bibinfo  {journal} {Nature}\ }\textbf {\bibinfo {volume} {609}},\
  \bibinfo {pages} {490} (\bibinfo {year} {2022})}\BibitemShut {NoStop}%
\bibitem [{\citenamefont {Oh}\ \emph {et~al.}(2024)\citenamefont {Oh},
  \citenamefont {Biswas}, \citenamefont {Klemm}, \citenamefont {Tan},
  \citenamefont {Hashimoto}, \citenamefont {Lu}, \citenamefont {Yan},
  \citenamefont {Dai}, \citenamefont {Birgeneau},\ and\ \citenamefont
  {Yi}}]{oh2024tunability}%
  \BibitemOpen
  \bibfield  {author} {\bibinfo {author} {\bibfnamefont {J.~S.}\ \bibnamefont
  {Oh}}, \bibinfo {author} {\bibfnamefont {A.}~\bibnamefont {Biswas}}, \bibinfo
  {author} {\bibfnamefont {M.}~\bibnamefont {Klemm}}, \bibinfo {author}
  {\bibfnamefont {H.}~\bibnamefont {Tan}}, \bibinfo {author} {\bibfnamefont
  {M.}~\bibnamefont {Hashimoto}}, \bibinfo {author} {\bibfnamefont
  {D.}~\bibnamefont {Lu}}, \bibinfo {author} {\bibfnamefont {B.}~\bibnamefont
  {Yan}}, \bibinfo {author} {\bibfnamefont {P.}~\bibnamefont {Dai}}, \bibinfo
  {author} {\bibfnamefont {R.~J.}\ \bibnamefont {Birgeneau}},\ and\ \bibinfo
  {author} {\bibfnamefont {M.}~\bibnamefont {Yi}},\ }\bibfield  {title}
  {\bibinfo {title} {Tunability of charge density wave in a magnetic kagome
  metal},\ }\href@noop {} {\bibfield  {journal} {\bibinfo  {journal} {arXiv
  preprint arXiv:2404.02231}\ } (\bibinfo {year} {2024})}\BibitemShut {NoStop}%
\bibitem [{\citenamefont {Kresse}\ and\ \citenamefont
  {Furthm{\"u}ller}(1996{\natexlab{a}})}]{kresse1996efficiency}%
  \BibitemOpen
  \bibfield  {author} {\bibinfo {author} {\bibfnamefont {G.}~\bibnamefont
  {Kresse}}\ and\ \bibinfo {author} {\bibfnamefont {J.}~\bibnamefont
  {Furthm{\"u}ller}},\ }\bibfield  {title} {\bibinfo {title} {Efficiency of
  ab-initio total energy calculations for metals and semiconductors using a
  plane-wave basis set},\ }\href@noop {} {\bibfield  {journal} {\bibinfo
  {journal} {Computational materials science}\ }\textbf {\bibinfo {volume}
  {6}},\ \bibinfo {pages} {15} (\bibinfo {year}
  {1996}{\natexlab{a}})}\BibitemShut {NoStop}%
\bibitem [{\citenamefont {Kresse}\ and\ \citenamefont
  {Furthm{\"u}ller}(1996{\natexlab{b}})}]{kresse1996efficient}%
  \BibitemOpen
  \bibfield  {author} {\bibinfo {author} {\bibfnamefont {G.}~\bibnamefont
  {Kresse}}\ and\ \bibinfo {author} {\bibfnamefont {J.}~\bibnamefont
  {Furthm{\"u}ller}},\ }\bibfield  {title} {\bibinfo {title} {Efficient
  iterative schemes for ab initio total-energy calculations using a plane-wave
  basis set},\ }\href@noop {} {\bibfield  {journal} {\bibinfo  {journal}
  {Physical review B}\ }\textbf {\bibinfo {volume} {54}},\ \bibinfo {pages}
  {11169} (\bibinfo {year} {1996}{\natexlab{b}})}\BibitemShut {NoStop}%
\bibitem [{\citenamefont {Perdew}\ \emph {et~al.}(1996)\citenamefont {Perdew},
  \citenamefont {Burke},\ and\ \citenamefont
  {Ernzerhof}}]{perdew1996generalized}%
  \BibitemOpen
  \bibfield  {author} {\bibinfo {author} {\bibfnamefont {J.~P.}\ \bibnamefont
  {Perdew}}, \bibinfo {author} {\bibfnamefont {K.}~\bibnamefont {Burke}},\ and\
  \bibinfo {author} {\bibfnamefont {M.}~\bibnamefont {Ernzerhof}},\ }\bibfield
  {title} {\bibinfo {title} {Generalized gradient approximation made simple},\
  }\href@noop {} {\bibfield  {journal} {\bibinfo  {journal} {Physical review
  letters}\ }\textbf {\bibinfo {volume} {77}},\ \bibinfo {pages} {3865}
  (\bibinfo {year} {1996})}\BibitemShut {NoStop}%
\bibitem [{\citenamefont {Grimme}\ \emph {et~al.}(2010)\citenamefont {Grimme},
  \citenamefont {Antony}, \citenamefont {Ehrlich},\ and\ \citenamefont
  {Krieg}}]{grimme2010consistent}%
  \BibitemOpen
  \bibfield  {author} {\bibinfo {author} {\bibfnamefont {S.}~\bibnamefont
  {Grimme}}, \bibinfo {author} {\bibfnamefont {J.}~\bibnamefont {Antony}},
  \bibinfo {author} {\bibfnamefont {S.}~\bibnamefont {Ehrlich}},\ and\ \bibinfo
  {author} {\bibfnamefont {H.}~\bibnamefont {Krieg}},\ }\bibfield  {title}
  {\bibinfo {title} {A consistent and accurate ab initio parametrization of
  density functional dispersion correction (dft-d) for the 94 elements h-pu},\
  }\href@noop {} {\bibfield  {journal} {\bibinfo  {journal} {The Journal of
  chemical physics}\ }\textbf {\bibinfo {volume} {132}} (\bibinfo {year}
  {2010})}\BibitemShut {NoStop}%
\bibitem [{\citenamefont {Togo}\ and\ \citenamefont
  {Tanaka}(2015)}]{togo2015first}%
  \BibitemOpen
  \bibfield  {author} {\bibinfo {author} {\bibfnamefont {A.}~\bibnamefont
  {Togo}}\ and\ \bibinfo {author} {\bibfnamefont {I.}~\bibnamefont {Tanaka}},\
  }\bibfield  {title} {\bibinfo {title} {First principles phonon calculations
  in materials science},\ }\href@noop {} {\bibfield  {journal} {\bibinfo
  {journal} {Scripta Materialia}\ }\textbf {\bibinfo {volume} {108}},\ \bibinfo
  {pages} {1} (\bibinfo {year} {2015})}\BibitemShut {NoStop}%
\bibitem [{\citenamefont {Garrett}(2001)}]{garrett2001femtosecond}%
  \BibitemOpen
  \bibfield  {author} {\bibinfo {author} {\bibfnamefont {G.~A.}\ \bibnamefont
  {Garrett}},\ }\href@noop {} {\emph {\bibinfo {title} {Femtosecond pulsed
  laser excitation of coherent and squeezed phonon fields}}}\ (\bibinfo
  {publisher} {University of Michigan},\ \bibinfo {year} {2001})\BibitemShut
  {NoStop}%
\bibitem [{\citenamefont {Garrett}\ \emph {et~al.}(1996)\citenamefont
  {Garrett}, \citenamefont {Albrecht}, \citenamefont {Whitaker},\ and\
  \citenamefont {Merlin}}]{garrett1996coherent}%
  \BibitemOpen
  \bibfield  {author} {\bibinfo {author} {\bibfnamefont {G.}~\bibnamefont
  {Garrett}}, \bibinfo {author} {\bibfnamefont {T.}~\bibnamefont {Albrecht}},
  \bibinfo {author} {\bibfnamefont {J.}~\bibnamefont {Whitaker}},\ and\
  \bibinfo {author} {\bibfnamefont {R.}~\bibnamefont {Merlin}},\ }\bibfield
  {title} {\bibinfo {title} {Coherent thz phonons driven by light pulses and
  the sb problem: what is the mechanism?},\ }\href@noop {} {\bibfield
  {journal} {\bibinfo  {journal} {Physical review letters}\ }\textbf {\bibinfo
  {volume} {77}},\ \bibinfo {pages} {3661} (\bibinfo {year}
  {1996})}\BibitemShut {NoStop}%
\bibitem [{\citenamefont {Aroyo}\ \emph
  {et~al.}(2006{\natexlab{a}})\citenamefont {Aroyo}, \citenamefont
  {Perez-Mato}, \citenamefont {Capillas}, \citenamefont {Kroumova},
  \citenamefont {Ivantchev}, \citenamefont {Madariaga}, \citenamefont {Kirov},\
  and\ \citenamefont {Wondratschek}}]{aroyo2006bilbao}%
  \BibitemOpen
  \bibfield  {author} {\bibinfo {author} {\bibfnamefont {M.~I.}\ \bibnamefont
  {Aroyo}}, \bibinfo {author} {\bibfnamefont {J.~M.}\ \bibnamefont
  {Perez-Mato}}, \bibinfo {author} {\bibfnamefont {C.}~\bibnamefont
  {Capillas}}, \bibinfo {author} {\bibfnamefont {E.}~\bibnamefont {Kroumova}},
  \bibinfo {author} {\bibfnamefont {S.}~\bibnamefont {Ivantchev}}, \bibinfo
  {author} {\bibfnamefont {G.}~\bibnamefont {Madariaga}}, \bibinfo {author}
  {\bibfnamefont {A.}~\bibnamefont {Kirov}},\ and\ \bibinfo {author}
  {\bibfnamefont {H.}~\bibnamefont {Wondratschek}},\ }\bibfield  {title}
  {\bibinfo {title} {Bilbao crystallographic server: I. databases and
  crystallographic computing programs},\ }\href@noop {} {\bibfield  {journal}
  {\bibinfo  {journal} {Zeitschrift f{\"u}r Kristallographie-Crystalline
  Materials}\ }\textbf {\bibinfo {volume} {221}},\ \bibinfo {pages} {15}
  (\bibinfo {year} {2006}{\natexlab{a}})}\BibitemShut {NoStop}%
\bibitem [{\citenamefont {Aroyo}\ \emph
  {et~al.}(2006{\natexlab{b}})\citenamefont {Aroyo}, \citenamefont {Kirov},
  \citenamefont {Capillas}, \citenamefont {Perez-Mato},\ and\ \citenamefont
  {Wondratschek}}]{aroyo2006bilbao2}%
  \BibitemOpen
  \bibfield  {author} {\bibinfo {author} {\bibfnamefont {M.~I.}\ \bibnamefont
  {Aroyo}}, \bibinfo {author} {\bibfnamefont {A.}~\bibnamefont {Kirov}},
  \bibinfo {author} {\bibfnamefont {C.}~\bibnamefont {Capillas}}, \bibinfo
  {author} {\bibfnamefont {J.}~\bibnamefont {Perez-Mato}},\ and\ \bibinfo
  {author} {\bibfnamefont {H.}~\bibnamefont {Wondratschek}},\ }\bibfield
  {title} {\bibinfo {title} {Bilbao crystallographic server. ii.
  representations of crystallographic point groups and space groups},\
  }\href@noop {} {\bibfield  {journal} {\bibinfo  {journal} {Foundations of
  Crystallography}\ }\textbf {\bibinfo {volume} {62}},\ \bibinfo {pages} {115}
  (\bibinfo {year} {2006}{\natexlab{b}})}\BibitemShut {NoStop}%
\bibitem [{\citenamefont {Aroyo}\ \emph {et~al.}(2011)\citenamefont {Aroyo},
  \citenamefont {Perez-Mato}, \citenamefont {Orobengoa}, \citenamefont {Tasci},
  \citenamefont {de~la Flor},\ and\ \citenamefont
  {Kirov}}]{aroyo2011crystallography}%
  \BibitemOpen
  \bibfield  {author} {\bibinfo {author} {\bibfnamefont {M.~I.}\ \bibnamefont
  {Aroyo}}, \bibinfo {author} {\bibfnamefont {J.~M.}\ \bibnamefont
  {Perez-Mato}}, \bibinfo {author} {\bibfnamefont {D.}~\bibnamefont
  {Orobengoa}}, \bibinfo {author} {\bibfnamefont {E.}~\bibnamefont {Tasci}},
  \bibinfo {author} {\bibfnamefont {G.}~\bibnamefont {de~la Flor}},\ and\
  \bibinfo {author} {\bibfnamefont {A.}~\bibnamefont {Kirov}},\ }\bibfield
  {title} {\bibinfo {title} {Crystallography online: Bilbao crystallographic
  server},\ }\href@noop {} {\bibfield  {journal} {\bibinfo  {journal} {Bulg.
  Chem. Commun}\ }\textbf {\bibinfo {volume} {43}},\ \bibinfo {pages} {183}
  (\bibinfo {year} {2011})}\BibitemShut {NoStop}%
\bibitem [{\citenamefont {Ishioka}\ \emph {et~al.}(2006)\citenamefont
  {Ishioka}, \citenamefont {Kitajima},\ and\ \citenamefont
  {Misochko}}]{ishioka2006temperature}%
  \BibitemOpen
  \bibfield  {author} {\bibinfo {author} {\bibfnamefont {K.}~\bibnamefont
  {Ishioka}}, \bibinfo {author} {\bibfnamefont {M.}~\bibnamefont {Kitajima}},\
  and\ \bibinfo {author} {\bibfnamefont {O.~V.}\ \bibnamefont {Misochko}},\
  }\bibfield  {title} {\bibinfo {title} {Temperature dependence of coherent a1g
  and eg phonons of bismuth},\ }\href@noop {} {\bibfield  {journal} {\bibinfo
  {journal} {Journal of Applied Physics}\ }\textbf {\bibinfo {volume} {100}}
  (\bibinfo {year} {2006})}\BibitemShut {NoStop}%
\end{thebibliography}%

\clearpage



\section*{APPENDIX A: DFT calculation details}

\begin{figure*}[t]
    \centering
    \includegraphics[width=17.2cm]{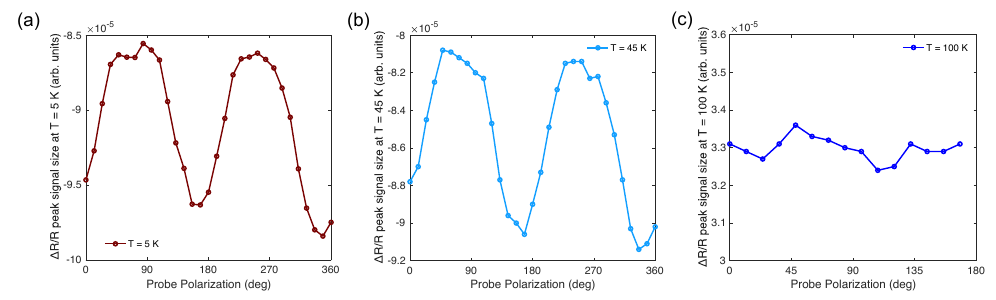}
    \caption{Probe polarization dependence of $\Delta$R/R peak signal size at T = 5 K, 45 K and 100 K in CsV$_3$Sb$_5$. Data is taken at the same spot position as Fig. \ref{fig5} of the main text. }
    \label{fig8}
\end{figure*}

\begin{figure*}[t]
    \centering
    \includegraphics[width=17.2cm]{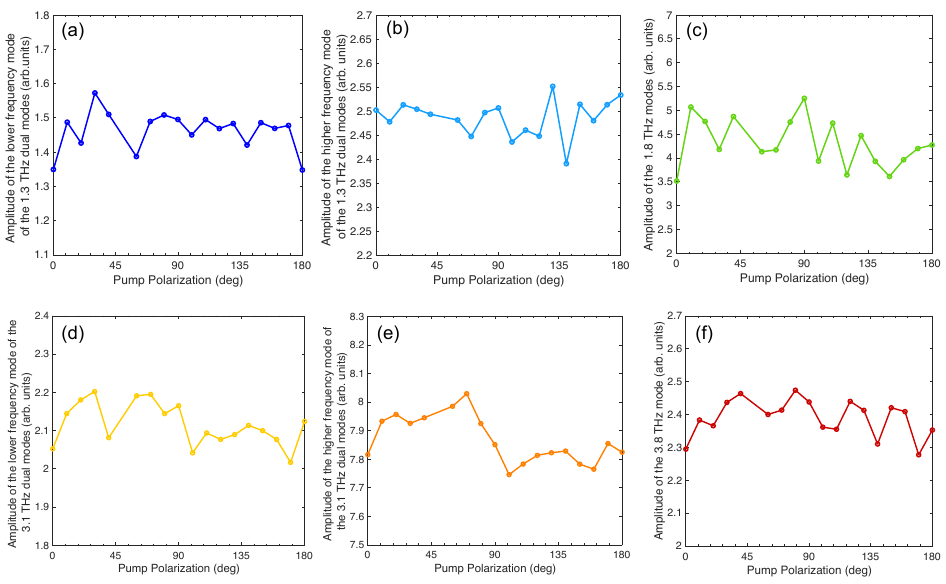}
    \caption{Pump polarization dependence of the amplitude of coherent phonon modes taken at T = 5 K in CsV$_3$Sb$_5$ including the 1.3 THz dual modes, 3.1 THz dual modes, 1.8 THz mode and 3.8 THz mode. }
    \label{fig9}
\end{figure*}

All density functional theory (DFT) calculations are performed using the Vienna Ab-initio Simulation Package\cite{kresse1996efficiency, kresse1996efficient}. The exchange–correlation interactions between electrons are treated using the generalized gradient approximation (GGA) of the Perdew–Burke–Ernzerhof (PBE) type\cite{perdew1996generalized}. A plane-wave energy cutoff of 300 eV is employed throughout. Structural optimizations with the DFT-D3 method of Grimme with zero-damping function\cite{grimme2010consistent} are carried out until the residual forces on all atoms are less than 1 meV/Å. The stable Star-of-David (SD) and inverse Star-of-David (ISD) structures are obtained from previous calculations\cite{tan2021charge}. For the three-dimensional charge density wave phases, we construct initial structures by stacking the 2D SD and ISD configurations along the c-axis, followed by full structural relaxation. As a result, two energetically favorable phases are identified: an ISD + ISD configuration with an interlayer $\pi$-phase shift (point group $D_{2h}$) and an ISD + SD configuration without a $\pi$-phase shift (point group $D_{6h}$). Phonon dispersions and vibrational modes are calculated using the Phonopy package within the finite-displacement approach\cite{togo2015first}. For the $D_{2h}$ structure, a 2 $\times$ 2 $\times$ 2 supercell is used with a 2 $\times$ 2 $\times$ 2 k-point mesh for reciprocal space sampling. For the $D_{6h}$ structure, a 2 $\times$ 2 $\times$ 1 supercell is adopted, along with the same 2 $\times$ 2 $\times$ 2 k-point mesh. 

\section*{APPENDIX B: Anisotropy shown by varying the probe polarization below T$_{\text{CDW}}$}

We provide here the supplementary data of probe polarization dependence of our TR-reflectivity signal below and above T$_{\text{CDW}}$ in CsV$_3$Sb$_5$ (Fig. \ref{fig8}).  

Clearly, at T = 5 K and 45 K, the TR-reflectivity signal size shows two-fold probe polarization angle dependence. While at T = 100 K, above T$_{\text{CDW}}$, the TR-reflectivity signal size is independent of probe polarization angle. Note for any probe polarization angle $\varphi$, $\varphi$ + 180$^\circ$ is equivalent to $\varphi$. And for T = 100 K data, only $\varphi$ = 0 - 180$^\circ$ is shown. 

This probe polarization dependent measurements have also been successfully used to probe the breaking of $C_6$ in Kagome CsCr$_3$Sb$_5$\cite{liu2024charge}. CsCr$_3$Sb$_5$ has space group P6/mmm in the high temperature phase. The authors in \cite{liu2024charge} also performed TR-reflectivity at various probe polarization directions at different temperatures. Below $T^* = 55$ K, TR-reflectivity shows anisotropy and two-fold dependence on probe polarization directions, while the TRR signals show no probe polarization dependence above $T^*$ in the $C_6$-preserved high temperature phase. They attribute the anisotropic TR-reflectivity signal at various probe polarization directions to nematicity which breaks $C_6$ to $C_2$. From their results, TR-reflectivity signal shows no probe polarization dependence in the $C_6$-preserved state and shows anisotropic two-fold dependence on probe polarization directions in the $C_2$ state. Our results are essentially the same as in \cite{liu2024charge}, the only difference is the $C_6$ breaking in our CsV$_3$Sb$_5$ is from the orthorhombic CDW structure induced by interlayer $\pi$ phase shift which breaks $C_6$ in the high temperature phase of CsV$_3$Sb$_5$ down to $C_2$.

\section*{APPENDIX C: Confirmation of DECP via pump polarization dependent study}

DECP requires absorption at the pump frequency in order to disturb the electronic energy distribution in the material, and only fully symmetric modes can be observed\cite{cheng1991mechanism, zeiger1992theory} in DECP mechanism. Fully symmetric phonon modes host $\Gamma_1^+$ symmetry, such as $A_{1g}$ modes in $D_{6h}$ point group and $A_g$ modes in $D_{2h}$ point group. We acknowledge that there are reports of the observation of non-fully-symmetric coherent phonon modes in literature also in absorbing media, and thus provide a more careful explanation below to confirm all detected modes are fully-symmetric modes. 

In the process of impulsive stimulated Raman scattering (ISRS), the amplitude of the coherent phonon is proportional to\cite{garrett2001femtosecond, garrett1996coherent}: 
$$\left[E_{pu}^i \left(\frac{\partial \chi^{ij}}{\partial Q}\right)_0 E_{pu}^j \right]\left[E_{pr}^i \left(\frac{\partial \chi^{ij}}{\partial Q}\right)_0 E_{pr}^j\right]$$
where $E_{pu}$ indicates the pump light polarization and $E_{pr}$ indicates the probe polarization. $Q$ is the normal mode amplitude of the phonon, $\chi$ is the electric susceptibility, and $\left(\partial \chi^{ij}/\partial Q\right)_0$ is the Raman tensor element. The Raman scattering amplitude is proportional to the square of the modulus of $\left(\partial \chi^{ij}/\partial Q\right)_0$. The Raman tensor form of $E_{2g}$ mode is $\left( \begin{smallmatrix} 0 & -d \\ -d & 0 \end{smallmatrix} \right)$ and $\left( \begin{smallmatrix} d & 0 \\ 0 & -d \end{smallmatrix} \right)$ with only one nonzero independent parameter\cite{aroyo2006bilbao, aroyo2006bilbao2, aroyo2011crystallography}. 

Above T$_{\text{CDW}}$, previous Raman spectroscopy detected one $A_{1g}$ mode at 4.1 THz and one $E_{2g}$ mode at 3.6 THz with comparable amplitudes\cite{wu2022charge, liu2022observation}. This main 3.6 THz $E_{2g}$ mode persists across T$_{\text{CDW}}$ at all temperatures with minimal frequency change of less than 1 cm$^{-1}$ (corresponding to 0.030 THz) and increasing amplitude as temperature increases\cite{wu2022charge, liu2022observation}. However, as seen from Fig. 2b from the main text, in our TR-reflectivity measurements, we only detect the 4.1 THz $A_{1g}$ mode above T$_{\text{CDW}}$. Below T$_{\text{CDW}}$, we do not observe this main 3.6 THz mode either. We do not observe such a mode that matches with the features of Raman-detected 3.6 THz $E_{2g}$ mode, i.e. there is no such a mode near 3.6 THz that persists at all temperatures with increasing amplitude and minimal frequency change ($<$ 0.030 THz) as temperature increases. The closest detected mode is the 3.86 THz mode, but it quickly weakens and softens as temperature increases, in contrast to the 3.6 THz $E_{2g}$ mode. Since the 3.6 THz non-fully symmetric $E_{2g}$ mode has the dominantly maximum Raman scattering amplitude among all non-fully-symmetric Raman-active modes\cite{wu2022charge}, without detecting this mode, we can rule out the detection of non-fully-symmetric Raman modes in our TR-reflectivity measurements. 


We also performed additional pump polarization dependent measurements as used in K. Ishioka et al., J. Appl. Phys. 100, 093501 (2006) \cite{ishioka2006temperature} to determine the symmetry of our observed coherent phonons. Suppose the actual CDW state is $C_6$-breaking with point group $D_{2h}$(e.g. ISD + ISD with interlayer $\pi$ phase shift, SD + ISD with interlayer $\pi$ phase shift). For non-fully-symmetric Raman modes, when the light is normal incident on (001) surface, one only need to consider $B_{1g}$ mode in $D_{2h}$ point group, with the Raman tensor form of $\left( \begin{smallmatrix} 0 & d \\ d & 0 \end{smallmatrix} \right)$ \cite{aroyo2006bilbao, aroyo2006bilbao2, aroyo2011crystallography}. We denote $E_{pr} = (\cos\theta, \sin\theta)$, where $\theta$ is the angle between the probe polarization direction and in-plane axis, and $\theta$ is fixed since we fix the probe polarization direction. We denote $E_{pu} = (\cos\varphi, \sin\varphi)$ where $\varphi$ is the angle of pump polarization direction with respect to in-plane axis. By varying pump polarization direction, $\varphi$ is varied. If we consider excitation of non-fully-symmetric Raman active phonons via ISRS, using the above equation, one gets the amplitude of $B_{1g}$ mode via ISRS is: 
$I_{B_{1g}}\propto \sin2\varphi$ showing two-fold dependence on pump polarization direction. Here we fix the probe polarization direction and measure the coherent phonon amplitude vs pump polarization direction of our observed 1.3 THz dual modes, 1.8 THz mode, 3.1 THz dual modes, and 3.8 THz mode in the CDW phase, as shown in Fig. \ref{fig9}. The 4.1 THz mode that shows up at all temperature is determined to be the main lattice $A_{1g}$ mode that is present in the high temperature P6/mmm phase. The pump polarization direction is varied between 0 and 180$^\circ$ and we note pump polarization direction between 180$^\circ$ and 360$^\circ$ should be equivalent. 

For these observed modes, if they were $B_{1g}$ modes generated via ISRS, $I_{B_{1g}}\propto\sin2\varphi$, their amplitude should show obvious two-fold dependence on pump polarization direction with nodes where the amplitude goes to zero. However, no observed coherent phonon modes obey this pump polarization dependence. We thus confirm these observed coherent phonon modes are not $B_{1g}$ modes, and they can only be fully-symmetric $A_g$ modes since the pump and probe light are normal incident on (001) surface. 

Or, if the actual CDW structure has $D_{6h}$ point group (e.g. 2 $\times$ 2 $\times$ 1 SD, 2 $\times$ 2 $\times$ 1 ISD, and SD + ISD without interlayer $\pi$ phase shift), for non-fully-symmetric Raman-active modes, when the light is normal incident on (001) surface, one will only need to consider $E_{2g}$ mode. Plugging in the Raman tensor of $E_{2g}$ mode gives the amplitude of $E_{2g}$ mode via ISRS as $I_{E_{2g}}\propto\cos2\varphi$ or $I_{E_{2g}}\propto\sin2\varphi$. Similarly, the pump polarization dependence of the amplitude of these observed coherent phonon modes does not match with $E_{2g}$ mode, thus the observed coherent phonons can only be fully-symmetric $A_{1g}$ modes. No matter which symmetry we assign to the CDW state, we confirm the observed coherent phonons in the CDW state are fully symmetric. Since all the coherent phonons generated are fully-symmetric mode, our results thus match better with DECP mechanism, same as previous pump-probe studies on CsV$_3$Sb$_5$\cite{ratcliff2021coherent}. 

\section*{APPENDIX D: Pump fluence dependence of coherent phonon spectrum}

\begin{figure}[t]
    \centering
    \includegraphics[width=8.6cm]{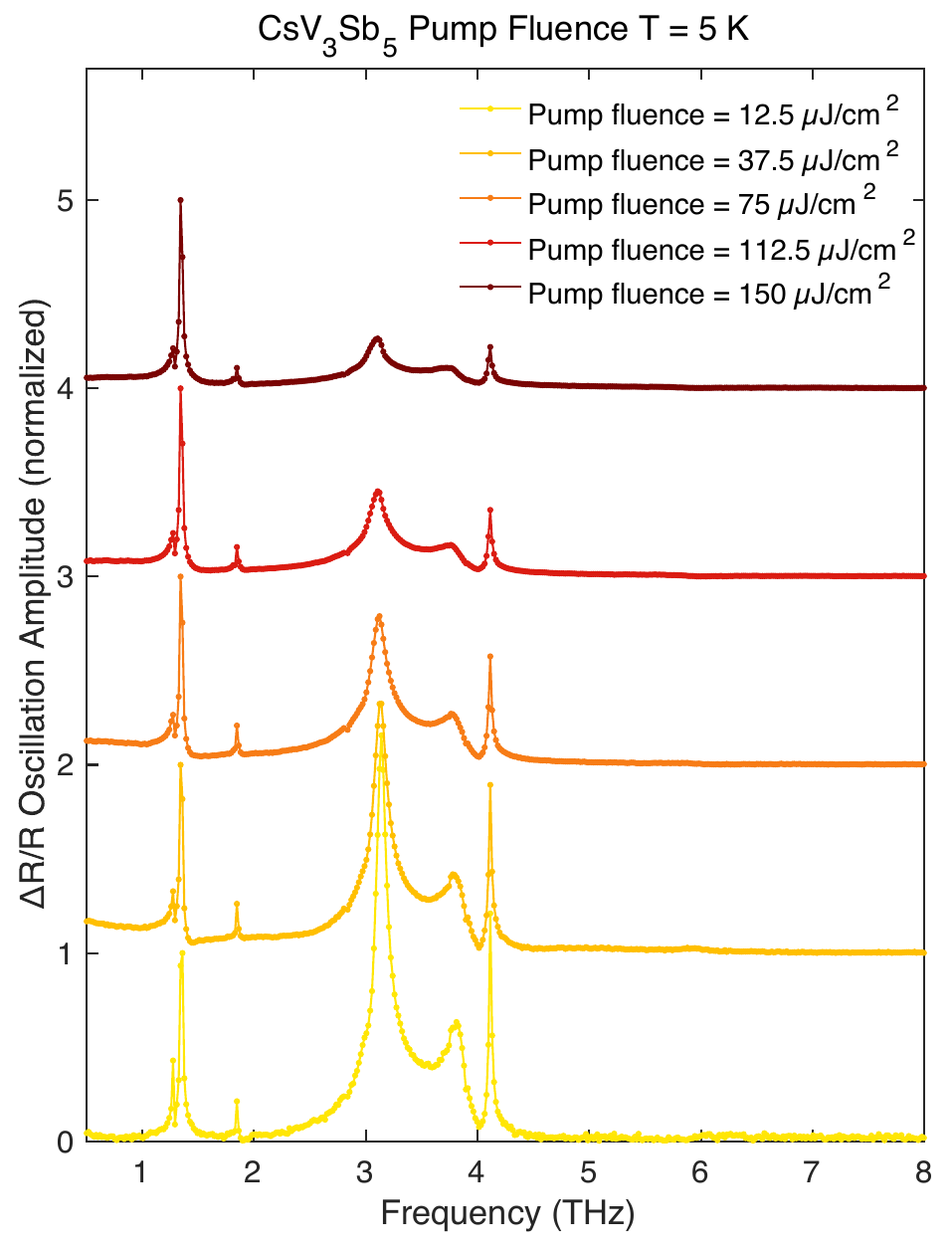}
    \caption{Pump fluence dependence of the coherent phonon spectrum taken at T = 5 K in CsV$_3$Sb$_5$. All datasets are normalized by the higher frequency peak of the 1.3 THz dual modes. }
    \label{fig10}
\end{figure}

Here, we also show the coherent phonon spectrum of CsV$_3$Sb$_5$ at T = 5 K with varying pump fluence F, up to the maximum pump fluence of 150 $\mu$J/cm$^2$ of our laser (Fig. \ref{fig10}). All coherent phonon spectra are normalized by the amplitude of the higher frequency mode of the 1.3 THz dual modes. At the lowest pump fluence, we reproduce the 1.3 THz dual modes, 1.8 THz mode, 3.1 THz modes, 3.8 THz mode and 4.1 THz mode reported in our main text, with the 3.1 THz mode having the strongest amplitude. As pump fluence is increased, the 3.1 THz, 3.8 THz and 4.1 THz modes become relatively weaker and the higher frequency mode of the 1.3 THz dual modes dominates at high pump fluences. 

\begin{figure*}[t]
    \centering
    \includegraphics[width=17.2cm]{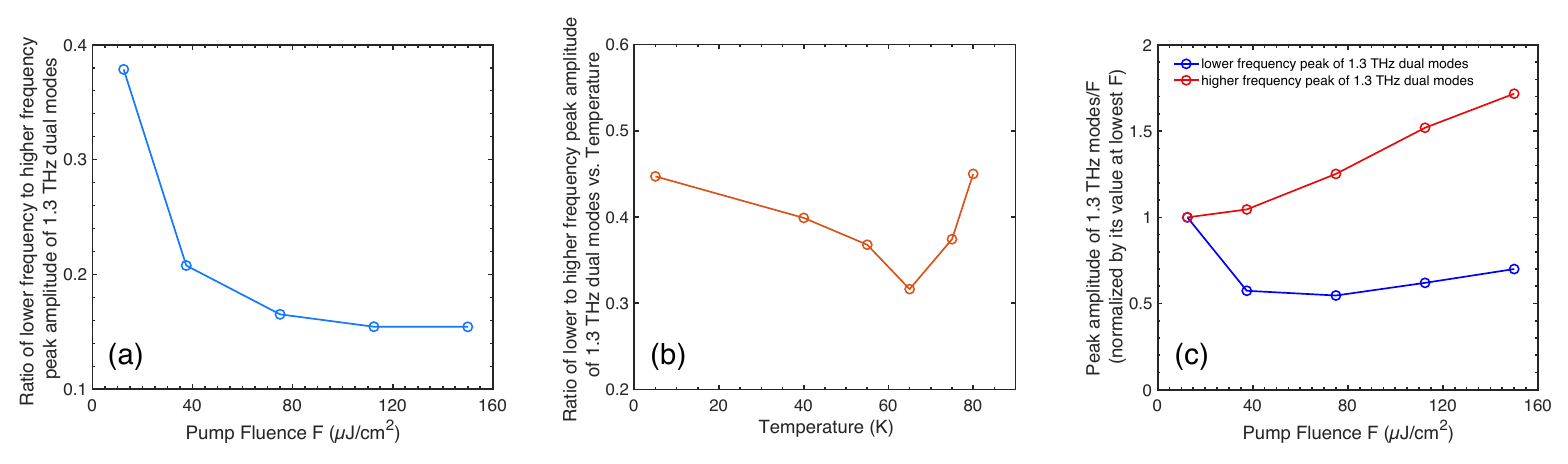}
    \caption{\textbf{Peak amplitudes of 1.3 THz dual modes as pump fluence changes. }(a) The ratio of lower frequency to higher frequency peak amplitude of 1.3 THz dual modes vs. pump fluence. Data extracted from Fig. \ref{fig10}. (b) The ratio of lower frequency to higher frequency peak amplitude of 1.3 THz dual modes vs. temperature. (c) Peak amplitude of 1.3 THz modes/F (normalized by its value at lowest F) as a function of pump fluence. Data extracted from Fig. \ref{fig10}. }
    \label{fig11}
\end{figure*}

We also notice the change of the amplitude ratio between the lower frequency and higher frequency mode of the 1.3 THz dual modes as a function of pump fluence. From Fig. \ref{fig10}, one can find the lower frequency peak of the 1.3 THz dual modes becomes relatively weaker with respect to the higher frequency peak of the 1.3 THz dual modes as the pump fluence increases. For these two peaks of the 1.3 THz dual modes, we plot the ratio of the lower frequency peak amplitude to the higher frequency peak amplitude as a function of pump fluence in Fig. \ref{fig11}a, showing the decrease of this ratio as pump fluence is increased. This indicates the amplitude of the lower frequency peak of the 1.3 THz dual modes gets suppressed with respect to the amplitude of the higher frequency peak of the 1.3 THz dual modes. We rule out the effect of increased laser heating at higher fluences by plotting the temperature dependence of this peak amplitude ratio with fixed pump fluence of $\sim 10 \mu$J/cm$^2$ (Fig. \ref{fig11}b). This peak amplitude ratio does not show monotonic decrease as temperature increases, and its value stays near 0.4, compared to $\sim 0.15$ at high fluences. 

In DECP mechanism, the coherent phonon amplitude in $\Delta$R/R is proportional to pump fluence\cite{zeiger1992theory}. Thus, for an independent coherent phonon, its amplitude normalized by pump fluence will be a constant. In Fig. \ref{fig11}c, we show the peak amplitude divided by pump fluence of the lower and higher frequency peak of the 1.3 THz dual modes as a function of pump fluence. Note we normalize each curve by its value at lowest pump fluence to better compare these two curves. The peak amplitude/pump fluence of the higher frequency peak shows increasing behavior, but the peak amplitude/pump fluence of the lower frequency peak got suppressed. The opposite trend of these two peaks suggests a competition behavior between them. 


These observations match with a recent TR-XRD study demonstrating the coexistence of MLL and LLL phases and competition between these two coexisting CDW phases in CsV$_3$Sb$_5$\cite{ning2024dynamical}. Herein, upon light excitation, the LLL phase gets more suppressed with respect to MLL, along with an expansion of the MLL domain and a contraction of the LLL domain. Thus, the lower frequency peak of the 1.3 THz dual modes will come from the more suppressed LLL phase and the higher frequency peak of the 1.3 THz dual modes will come from MLL phase. Indeed, as seen from the calculated fully-symmetric phonon spectrum of the coexistence of MLL and LLL state in Fig. \ref{fig6}c and Table \ref{table-1} of the main text, we can indeed find a 1.04 THz fully-symmetric mode from LLL and a 1.21 THz fully-symmetric mode from MLL that can match with our detected 1.3 THz dual modes in TR-reflectivity. Thus, this picture with the coexistence of MLL and LLL CDW phase matches with our TR-reflectivity results, and the fluence dependent ultrafast TR-reflectivity study reveals the competition of these two CDW phases. Our non-equilibrium dynamical study of this system thus provides more details of the exact CDW configuration and decodes the interplay between coexisting CDW phases.

\clearpage

\renewcommand{\tablename}{{\bf{Appendix Table}}}

\end{document}